\documentclass[twocolumn]{aastex62}
\usepackage{natbib}
\usepackage{amsmath, mathtools}
\usepackage{hyperref}
\usepackage{color}
\usepackage{comment}
\usepackage[T1]{fontenc}
\usepackage{gensymb}
\usepackage[caption=false]{subfig}

\graphicspath{{./}}

\shorttitle{$\alpha$-abundances of low-metallicity stars in the Galactic Center.}
\shortauthors{BENTLEY ET AL.}

\begin{document}

\author{Rory O. Bentley}
\affiliation{UCLA Galactic Center Group, Physics and Astronomy Department, UCLA, Los Angeles, CA 90095-1547, USA}

\author{Tuan Do}
\affiliation{UCLA Galactic Center Group, Physics and Astronomy Department, UCLA, Los Angeles, CA 90095-1547, USA}

\author{Wolfgang Kerzendorf}
\affiliation{Department of Physics and Astronomy, Michigan State University, East Lansing, MI 48824, USA}
\affiliation{Department of Computational Mathematics, Science, and Engineering, Michigan State University, East Lansing, MI 48824, USA}

\author{Devin S. Chu}
\affiliation{UCLA Galactic Center Group, Physics and Astronomy Department, UCLA, Los Angeles, CA 90095-1547, USA}

\author{Zhuo Chen}
\affiliation{UCLA Galactic Center Group, Physics and Astronomy Department, UCLA, Los Angeles, CA 90095-1547, USA}

\author{Quinn Konopacky}
\affiliation{Center for Astrophysics and Space Science, University of California San Diego, La Jolla, CA 92093, USA}

\author{Andrea Ghez}
\affiliation{UCLA Galactic Center Group, Physics and Astronomy Department, UCLA, Los Angeles, CA 90095-1547, USA}

\title{Measuring the $\alpha$-abundance of subsolar-metallicity stars in the Milky Way’s central half-parsec: testing globluar cluster and dwarf galaxy infall scenarios}

\begin{abstract}

While the Milky Way Nuclear star cluster has been studied extensively, how it formed is uncertain. Studies have shown it contains a solar and supersolar metallicity population that may have formed in-situ, along with a subsolar metallicity population that may have formed via mergers of globular clusters and dwarf galaxies. Stellar abundance measurements are critical to differentiate between formation scenarios. We present new measurements of [$M/H$] and $\alpha$-element abundances [$\alpha/Fe$] of two subsolar-metallicity stars in the Galactic Center. These observations were taken with the adaptive-optics assisted high-resolution (R=24,000) spectrograph NIRSPEC in the K-band (1.8 - 2.6 micron). These are the first $\alpha$-element abundance measurements of sub-solar metallicity stars in the Milky Way nuclear star cluster. We measure [$M/H$]=$-0.59\pm 0.11$, [$\alpha/Fe$]=$0.05\pm 0.15$ and [$M/H$]= $-0.81\pm 0.12$, [$\alpha/Fe$]= $0.15\pm 0.16$ for the two stars at the Galactic center; the uncertainties are dominated by systematic uncertainties in the spectral templates. The stars have an [$\alpha/Fe$] in-between the [$\alpha/Fe$] of globular clusters and dwarf galaxies at similar [$M/H$] values. Their abundances are very different than the bulk of the stars in the nuclear star cluster. These results indicate that the sub-solar metallicity population in the Milky Way nuclear star cluster likely originated from infalling dwarf galaxies or globular clusters and are unlikely to have formed in-situ.

\end{abstract}

\section{Introduction}    
\label{sect:intro}
The Milky Way nuclear star cluster (MW NSC) is a massive ($\sim10^{7} M_{\odot}$) and very dense collection of stars that surrounds the supermassive black hole at the very center of our Galaxy \citep{Schodel2014b, Chatzopoulos2015, FeldmeierKrause2017}. How the MW NSC formed is not well understood. The two primary scenarios are 'in-situ' formation \citep[e.g.][]{Milosavljevic2004, Pflamm-Altenburg2009}, and 'migration and merger' \citep[e.g.][]{Tremaine1975, Capuzzo-Dolcetta1993, Antonini2012, Antonini2013}. In in-situ formation, the stars formed roughly in their present positions. In migration and merger, the stars formed much further from the center, in globular clusters and dwarf galaxies. These then migrated inwards through dynamical friction, before being disrupted and mixed into the present-day MW NSC. However, identifying the origins of stars and stellar populations is not always simple, as the distinguishing observational tracers can be difficult to measure.

An important key to understanding the formation history of the stars in the nuclear star cluster is their metallicity and elemental abundances. Through determination of stellar metallicities, one can identify when the stars formed, and at what rate \citep[e.g.][]{Fuhrmann1998, Anders2014, Bensby2014}. Elemental abundances are very useful tools in disentangling the different stellar components of the Milky Way. Of particular interest here are the bulk metallicity [$M/H$] (often used interchangeably with the iron abundance [$Fe/H$]), and the $\alpha$-element (O, Mg, Si, S, Ca, and Ti) abundance [$\alpha/Fe$]. These two parameters change dramatically in a star depending on the environment in which a star formed. Comparisons between the stellar populations of the Galactic disk, bulge, halo, and surrounding dwarf galaxies find very different [$M/H$] and [$\alpha/Fe$] values from population to population \citep[e.g.][]{Bonifacio2004, Monaco2005, Letarte2010, Hayden2015, Norris2017, Hill2019, Zasowski2019, Schultheis2020}. 

Early spectroscopic studies of the MW NSC were hampered by the $\sim$30 magnitudes of extinction in optical bands (where abundance studies usually take place), and the intense stellar crowding in the Galactic center. Initial infrared studies in the K-band and H-band were able to observe a dozen stars in the Milky Way's central $\lesssim$10 pc, and found solar to supersolar iron abundances \citep{Carr2000, Ramirez2000, Blum2003, Cunha2007}.  The stellar crowding at the Galactic center meant that only the brightest stars could be observed (K < 9.5 mag). Reaching fainter magnitudes allows for more available targets (making it easier to study the stellar population in more detail), and allows one to study stars which are typically longer lived (making it possible to study a longer period of history, \citet{FeldmeierKrause2017}).

\begin{figure}[h]
\begin{center}
\includegraphics[width=3.5in]{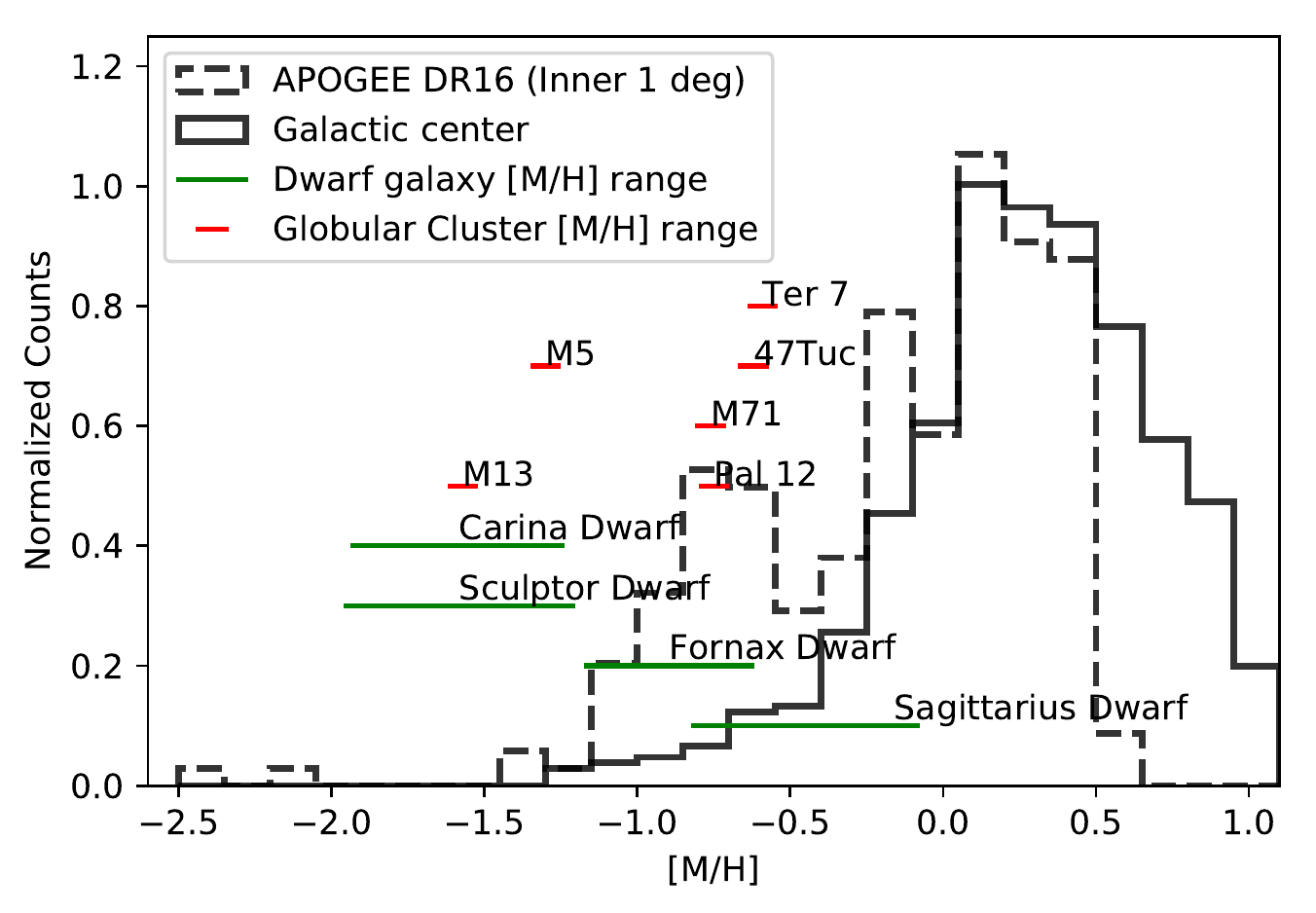}
\caption{\footnotesize The metallicity distributions of stars at the Galactic Center \citep[solid black histogram][]{FeldmeierKrause2017}, inner Milky Way bulge \citep[dashed black histogram][]{Ahumada2019}, and several globular clusters (red line) and dwarf galaxies (green lines). The Galactic center metallicity distribution is very wide, spanning a factor of $\sim100$ in metal abundance. Differences in the origins of these stars may be reason for this large variation in [$M/H$]. In this work, we test the hypothesis that the low-metallicity stars ([$M/H$] < -0.5) may have been brought to the Galactic center by infalling globular clusters or dwarf galaxies by using both [$M/H$] and [$\alpha/Fe$].}
\label{fig:mh_dist}
\end{center}
\end{figure}

Medium spectral-resolution (R $\sim$ 5,000) integral-field spectroscopy in recent years have found that the stars have a broad range of [$M/H$] values, spanning from -1.0 to above 0.5 dex \citep[Figure \ref{fig:mh_dist},][]{Do2015,FeldmeierKrause2017}. These findings are consistent with smaller samples observed at higher spectral resolution \citep{Schultheis2015,Ryde2016,Rich2017, Do2018, Thorsbro2020}.  This variation indicates a complex metal enrichment history of the MW NSC. The high metallicity stars likely formed in-situ or in the nearby bulge from gas enriched by metals from past generations of Milky Way stars. However, the subsolar metallicity population may have had a different origin, potentially being the remnant of a globular cluster or dwarf galaxy infall. The recent discovery of dynamical distinctions between the subsolar and supersolar metallicity populations in the central 10 pc suggests that the subsolar metallicity stars may be the remnants of an infall \citep{Feldmeier-Krause2020, Do2020}. Additionally, the discovery of RR Lyrae variables in the bulge in \citet{Minniti2016}, provides additional tracers of past infall events. An infall origin hypothesis could be tested by higher spectral resolution studies of the stars probing [$\alpha/Fe$], which has a small effect on the spectra and cannot be measured at medium resolution.

Although there have been studies of [$\alpha/Fe$] within the MW NSC \citep{Rich2017, Do2018, Thorsbro2020}, the small sample sizes ($\lesssim$ 10 stars) have missed the subsolar metallicity population, which comprises $\lesssim$10$\%$ of the total population \citep{Do2015, FeldmeierKrause2017}. \citet{Schultheis2015} and \citet{Ryde2016} analyzed an adaptive optics assisted high resolution (R $\sim$ 24,000) spectrum of an inner bulge red giant, and found [$\alpha/Fe$] = 0.4 and [$Fe/H$] = -1.0. This star is located 58 pc from the MW NSC, within the Milky Way bulge but outside the MW NSC, which has an effective radius $r_{eff}$ = 4.2$\pm$0.4 \citep{Schodel2014a}.

In this paper, we measure for the first time both the metallicity and alpha-abundance of sub-solar metallicity stars within 1 pc of the supermassive black hole at the Galactic center with new high-resolution K-band spectra ($R \sim 24,000$). Our aim is to constrain the physical origins of this population by comparing their composition to globular clusters and dwarf galaxies to assess the infall and merger hypothesis. If these stars are in fact mergers, we may expect the relationship between [$M/H$] and [$\alpha/Fe$] to be comparable to these objects \citep{Searle1978}. Similarly, these measurements could also show whether they are formed in-situ like most other stars in the inner bulge \citet{Feltzing2013}.

In this work, we use full spectrum fitting to obtain our [$\alpha/Fe$] and [$M/H$] measurements, and provide the first full spectrum fitting calibrations at high spectral resolution in K-band. In full-spectrum fitting, a large portion of the observed spectrum is fit to a synthetic spectrum over a range of model-dependant physical parameters. This method can measure stellar parameters (effective temperature $T_{eff}$, surface gravity log $g$, [$M/H$], and [$\alpha/Fe$]) and was used in in APOGEE \citep{Holtzman2015}, along with several of the Galactic Center studies discussed above \citep{Do2015, Do2018, FeldmeierKrause2017}. As part of this calibration, we also compare fitting with two spectral grids that cover the region of parameter space for our study: the BOSZ grid and the PHOENIX grid. The BOSZ grid \citep{Bohlin2017}, designed for flux calibration of standard stars for the James Webb Space Telescope, and the PHOENIX grid \citep{Husser2013}, a widely-used grid designed for a range of stellar astrophysics applications, are of particular interest for high-resolution near-infrared spectroscopy.

Section \ref{sect:obs_data} describes the observations and data. Section \ref{sect:specfit} describes the full-spectrum fitting, summarizes the experiments to test the quality and robustness of our fits, and our fit results for the Galactic center stars. Section \ref{sect:disc} provides discussion, and Section \ref{sect:conclu} is a conclusion. Section \ref{sect:appendix} contains more in-depth discussion of the experiments in Section \ref{sect:results}.

\section{Observations and Data}    
\label{sect:obs_data}

\subsection{NIRSPAO Observations}
\label{sect:observ}
The observations in this study were taken between April 2016 and July 2017 with NIRSPAO, which consists of the NIRSPEC spectrograph \citep{McLean1998, McLean2005}, behind laser guide star AO \citep{Wizinowich2006, vanDam2006} at the W. M. Keck Observatory. Observations were taken in echelle mode, yielding a spectral resolution of R $\sim$ 24,000 \citep{McLean1998} with a slit aperture of 0.041"x2.26". Laser guide star was used at the center of the field of view for each observation, and for lower order tip-tilt corrections, a R=13.7 mag star USNO 0600-28577051 (17:45:40.720–29:00:11.20) was used as a natural guide star. Details on the observations are listed in Table \ref{tab:obs_stars}. 

\subsection{Sample Selection}
\begin{figure}[h]
\includegraphics[width=3.5in]{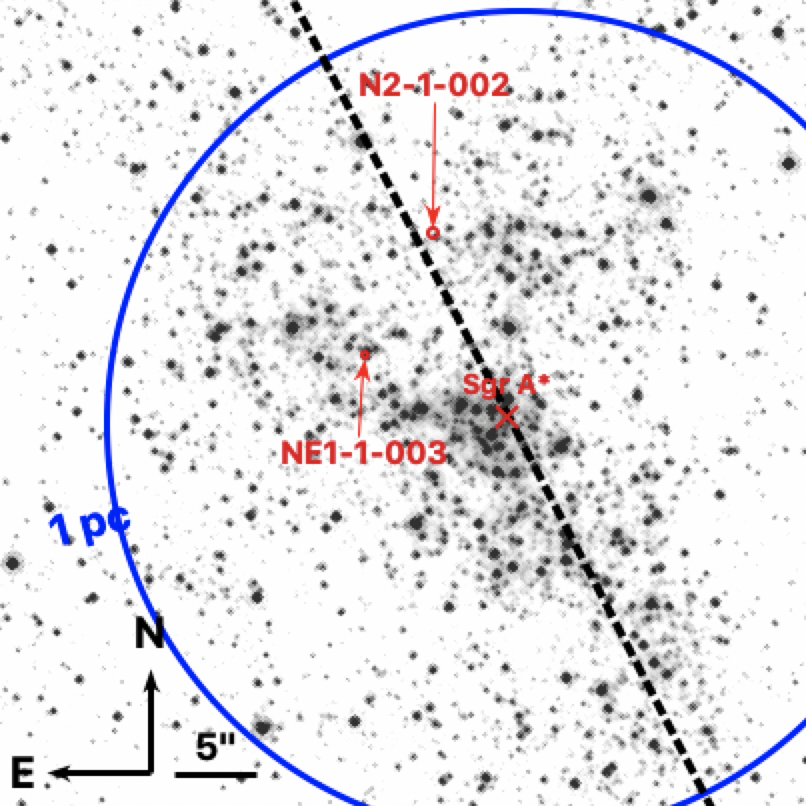}
\caption{\footnotesize An HST WFC3-IR image taken with the F153M filter of the Galactic center in the near-infrared from Hosek et al. (in prep), that shows the location of the 2 stars in our sample (red circles). These stars are within 1 pc (blue circle) in projection from Sgr A*. Black dashed line indicates the Galactic plane. }
\label{fig:gc_star_map}
\end{figure}

The two Galactic center stars are selected from the sample of stars studied in \citet{Do2015}. These stars are part of a program to followup observations of stars in \citet{Do2015} with K $<$ 13 mag at high spectral resolution in K-band. Early results from this program were reported in \citet{Do2018}. Here we focus on the two sub-solar metallicity stars with signal-to-noise ratio (SNR) $>$ 10 from this program (see Appendix \ref{sect:stat_err_vs_snr} for discussion about SNR). The properties of the two stars included in this study, N2-1-002 and NE1-1-003 are listed in Table \ref{tab:obs_stars}, and their positions are shown in Figure \ref{fig:gc_star_map}.

For calibrations, we observe six stars from open clusters that were used to calibrate the APOGEE abundance measurements \citep{Meszaros2013}. These stars have metallicities that span the expected range of metallicities based on the results of \citet{Do2015}. The star 2M18482584+4828027, which is one of the highest metallicity stars in the APOGEE catalog, was also observed to provide a calibrator at very high metallicity. The observations of the calibrators obtained by us are in Table \ref{tab:obs_stars}, and their physical parameters from the APOGEE DR16 data release \citep{Ahumada2019} are listed in \ref{tab:cal_stars}. As the APOGEE physical parameter values for the cluster stars are consistent with previous measurements using different methods, we taking the APOGEE values as the true physical parameter values.

\subsection{Calibrators from Archival Data}
\label{sect:koa_cals}
We use the the Keck Observatory Archive to search for NIRSPEC K-band observations of stars in open clusters that have been observed in the APOGEE calibration sample \citep{Meszaros2013}, to be used as additional calibrators. Three calibrators in the cluster M71, and another observation of a calibrator we observed (2M19534827+1848021) matched our criteria. These spectra are from program C118NS (PI: Cohen, J.). The observations were taken in a very similar manner to our observations outlined in Table \ref{tab:obs_stars}, except with a slit aperture of 0.288"x24".  The raw spectra were downloaded and reduced using the method in Section \ref{sect:reduc}. These archival observations are also listed in Table \ref{tab:obs_stars}, and physical parameters are listed in Table \ref{tab:cal_stars}.

\begin{table*}
\begin{center}
\setlength{\tabcolsep}{1pt}
\caption{Observed Stars$^{\dagger\dagger}$}
\label{tab:obs_stars}
\begin{tabular}{ccccccccc}
\hline\hline
Name & K & RA & Dec & Date Observed & Integration Time & Position Angle & Purpose & SN \\
& (mag) &  &  & (UT) & (s) & & of Observation & \\
\hline
NE1-1-003 & 11.45 & 8.61$^{\dagger}$ & 3.762$^{\dagger}$ & 2016 April 30 & 2x900 & 356.5 & Science Target & 25 \\
N2-1-002 & 12.12 & 4.398$^{\dagger}$ & 10.997$^{\dagger}$ & 2016 May 16 & 2x900 & 121 & Science Target & 10 \\
2M19205338+3748282 & 9.77 & 19:20:53.38 & 37:48:28.2 & 2016 May 16 & 4x480 & 121 & Calibrator & 45 \\
2M19411705+4010517 & 7.82 & 19:41:17.1 & 40:10:51.8 & 2016 May 16 & 4x180 & 121 & Calibrator & 40 \\
2M19213390+3750202 & 8.80 & 19:21:33.9 & 37:50:20.2 & 2016 May 20 & 4x360 & 155 & Calibrator & 35 \\
2M19413439+4017482 & 7.99 & 19:41:34.4 & 40:17:48.2 & 2016 May 20 & 4x120 & 155 & Calibrator & 40 \\
2M19534827+1848021 & 8.09 & 19:53:48.3 & 18:48:02.3 & 2016 May 20 & 4x60 & 155 & Calibrator & 55 \\
2M15190324+0208032 & 8.50 & 15:19:03.2 & 02:08:03.3 & 2017 July 13 & 4x120 & 0 & Calibrator & 85 \\
2M18482584+4828027 & 9.5 & 18:48:25.9 & 48:28:02.7 & 2017 July 13 & 4x300 & 0 & Calibrator & 60 \\
2M19534827+1848021 & 8.09 & 19:53:48.3 & 18:48:02.3 & 2008 Oct 07 & 2x600 & 81 & Calibrator$^{\ast}$ & 80  \\
2M19535325+1846471 & 8.04 & 19:53:53.3 & 18:46:47.1 & 2008 Oct 07 & 4x600 & 81 & Calibrator$^{\ast}$ & 100 \\
2M19533757+1847286 & 8.42 & 19:53:37.6 & 18:47:28.6 & 2008 Oct 08 & 4x600 & 50 & Calibrator$^{\ast}$ & 90 \\
2M19534525+1846553 & 9.42 & 19:53:45.3 & 18:46:55.4 & 2008 Oct 08 & 2x600 & 50 & Calibrator$^{\ast}$ & 75 \\
\hline \hline
\end{tabular}
\end{center}
$^{\ast}$ Spectrum from Keck Observatory Archive.

$^{\dagger}$ Projected offset in " from the central supermassive black hole at 17:45:40.0409 -29:00:28.118 \citep{Schodel2010}.

$^{\dagger\dagger}$ All observations taken at echelle/cross-disperser setting 63.85/35.65, with filter Nirspec-7 (K-band). Slit aperture of 0.041"x2.26" used for our observations, 0.288"x26" for observations from Keck Observatory Archive.
\end{table*}

\begin{table*}
\begin{center}
\caption{Calibrator Stars for this paper, with APOGEE DR16 physical parameters}
\label{tab:cal_stars}
\begin{tabular}{ccccccc}
\hline\hline
Name & Cluster & H & $T_{eff}$ & log g & [$M/H$] & [$\alpha/Fe$] \\
& & (mag) & (K) & & & \\
\hline
2M19205338+3748282 & NGC 6791 & 9.96$\pm$0.02 & 4032$\pm$65 & 1.50$\pm$0.05 & 0.36$\pm$0.01 & 0.02$\pm$0.01  \\
2M19411705+4010517 & NGC 6819 & 8.01$\pm$0.02 & 4103$\pm$68 & 1.50$\pm$0.05 & 0.00$\pm$0.01 & 0.02$\pm$0.01  \\
2M19213390+3750202 & NGC 6791 & 9.05$\pm$0.02 & 3750$\pm$59 & 1.10$\pm$0.04 & 0.32$\pm$0.01 & 0.04$\pm$0.01  \\
2M19413439+4017482 & NGC 6819 & 8.20$\pm$0.02 & 4185$\pm$69 & 1.75$\pm$0.05 & 0.06$\pm$0.01 & 0.00$\pm$0.01  \\
2M19534827+1848021 & M71 & 8.27$\pm$0.03 & 4077$\pm$73 & 1.24$\pm$0.06 & -0.74$\pm$0.01 & 0.22$\pm$0.01 \\
2M15190324+0208032 & M5 & 8.63$\pm$0.03 & 4022$\pm$76 & 0.70$\pm$0.07 & -1.27$\pm$0.01 & 0.17$\pm$0.01 \\
2M18482584+4828027 & & 9.60$\pm$0.02 & 5572$\pm$122 & 3.81$\pm$0.07 & 0.44$\pm$0.01 & -0.04$\pm$0.01 \\
2M19535325+1846471 & M71 & 8.20$\pm$0.02 & 4002$\pm$72 & 1.04$\pm$0.06 & -0.82$\pm$0.01 & 0.23$\pm$0.01  \\
2M19533757+1847286 & M71 & 8.68$\pm$0.02 & 4004$\pm$71 & 1.14$\pm$0.06 & -0.74$\pm$0.01 & 0.26$\pm$0.01  \\
2M19534525+1846553 & M71 & 9.56$\pm$0.02 & 4305$\pm$78 & 1.75$\pm$0.07 & -0.73$\pm$0.01 & 0.19$\pm$0.01  \\
\hline \hline

\hline
\end{tabular}
\end{center}
\end{table*}

\subsection{Data Reduction}
\label{sect:reduc}
\begin{figure*}[h]
\begin{center}
\includegraphics[width=7in]{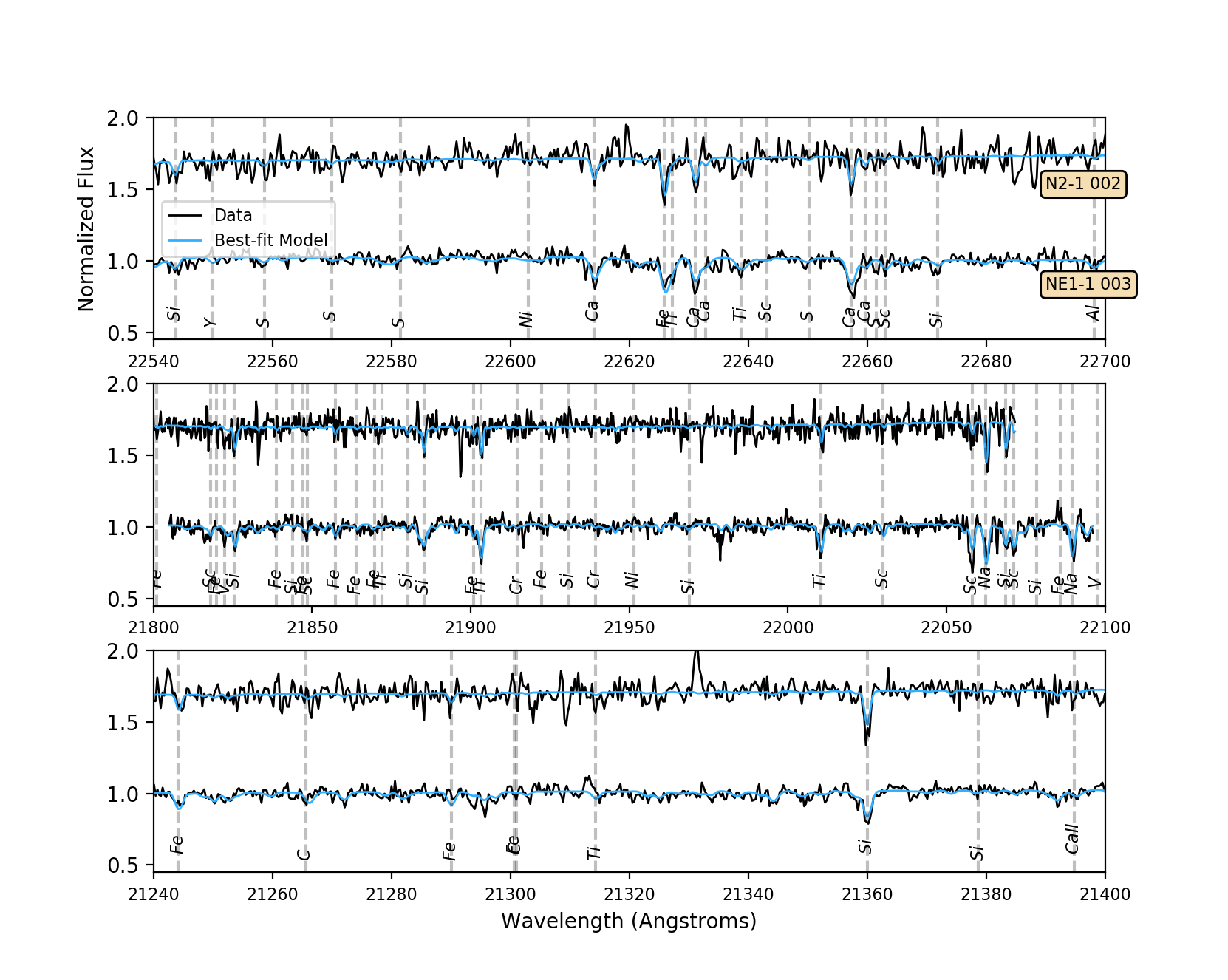}
\caption{\footnotesize NIRSPAO spectra (black) and best-fit models (blue) of the two Galactic center stars in our sample, shifted to rest wavelengths. These are the first high spectral resolution observations of low-metallicity stars ([Fe/H] < -0.5) in the central 1 pc from the supermassive black hole. See Appendix \label{sect:grid_results} for all spectral orders of all calibrator stars and Galactic center stars.}
\label{fig:all_gc_specs}
\end{center}
\end{figure*}

Observations were reduced using the REDSPEC package\footnote{\url{https://www2.keck.hawaii.edu/inst/nirspec/redspec}}. This is the standard data reduction package for pre-2019 NIRSPEC observations. Etalon lamps were used for wavelength solutions, and spatial rectification was performed with standard stars. The many features of the etalon lamp gives excellent relative wavelength calibration in each order, but there is a systematic offset different spectral orders. To account for this, we simultaneously fit for a relative velocity offset between orders when fitting the data. The REDSPEC spectral extraction routine was used to obtain the final spectra. Orders 33 through 37 were reduced and inspected, although only orders 34 through 36 are used in the fitting. Orders 34-36 (covering 2.1-2.3 $\mu$m) have the highest SN and and best telluric feature subtractions. A portion of the reduced Galactic center star spectra are shown in Figure \ref{fig:all_gc_specs}. Appendix \ref{sec:atlas} shows all of the spectra. 

\subsection{Additional Data}
\label{sect:additional_data}

We use Ks photometry for N2-1-002 and NE1-1-003 from \citet{Schodel2010}. Measurements in \citet{Schodel2010} were made from images taken with the NACO instrument on the ESO VLT unit telescope 4. We use the known offsets of the two stars from Sgr A* to identify them in the \citet{Schodel2010} database. The $Ks$ magnitudes are $Ks$=12.12$\pm$0.02 for N2-1-002 and $Ks$=11.45$\pm$0.02 for NE1-1-003.

Radial velocity measurements of N2-1-002 and NE1-1-003 used in this work are from \citet{Do2015}. They used medium resolution (R=5,400) spectra from NIFS spectrograph on the Gemini North telescope. We apply a correction for the motion of the the local standard of rest to the velocities reported in \citet{Do2015}. We find $v_{rad}$=250 km/s for N2-1-002 and $v_{rad}$=146 km/s for NE1-1-003.

We use proper motion measurements for N2-1-002 and NE1-1-003 from Hosek et al. (in prep). These measurements were made using HST WFC3-IR observations in the F153M filter from 2010 to 2020. N2-1-002 has a proper motion of $\mu$=2.2 mas/yr (corresponding to 87 km/s at the Galactic center) and NE1-1-003 has a proper motion of $\mu$=8.3 mas/yr (corresponding to 265 km/s at the Galactic center). We use the proper motion and radial velocity information in section \ref{sect:results_location} to constrain the location of these stars to the innermost 0.75 pc of the MW NSC.

\section{Methods}    
\label{sect:specfit}
In this section, we describe the methodology we employ for parameter estimation, calibrations, and resulting [$M/H$] and [$\alpha/Fe$] measurements for stars in our sample. Section \ref{sect:method} describes the Bayesian inference model and full-spectral fitting method. Section \ref{sect:calibration} describes the calibration of our method for our K-band spectra, and our choice of spectral grid.

\label{sect:method}
\subsection{Full-Spectral Fitting with StarKit}    
\label{sect:starkit}

We use full-spectral fitting tool StarKit \citep{Kerzendorf2015} to fit our spectra. 
In this method we compare the observed spectrum to a synthetic model to determine the best fitting model parameters using Bayes' Theorem:

\begin{equation}
P(\theta| D) = \frac{P(D|\theta)P(\theta)}{P(D)}
\end{equation}
where $D$ is the observed spectrum, and the model parameters $ \theta = (T_{eff}, ~\log g, ~\textrm{[M/H]}, \textrm{[}\alpha\textrm{/Fe]}, v_{rad},v_{rot},\sigma_{\rm add,order})$, where $v_{rad}$ is the radial velocity and $v_{rot}$ is the rotational velocity, and $\sigma_{\rm add,order}$ is the additive uncertainty to the flux. The priors on the model parameters are $P(\theta)$ and $P(D)$ is the evidence, which acts as the normalization. The combined likelihood for an observed spectrum is:
\begin{equation}
P(D|\theta) = \prod^{\lambda_n}_{\lambda=\lambda_0}\frac{1}{\epsilon_{\lambda,obs}\sqrt{2\pi}}\exp{(-(F_{\lambda,obs} - F_\lambda(\theta))^2/2\epsilon_{\lambda,obs}^{2})},
\end{equation}
where $F_{\lambda,obs}$ is the observed spectrum, $F_{\lambda}(\theta)$ is the model spectrum evaluated with a given set of model parameters, and $\epsilon_{\lambda,obs}$ is the 1 $\sigma$ total uncertainty for each wavelength. This likelihood assumes that the uncertainty for each flux point is approximately Gaussian. For computational efficiency, we use the log-likelihood in place of the likelihood:
\begin{equation}
\ln P(D|\theta) \propto -\frac{1}{2} \sum^{\lambda_n}_{\lambda=\lambda_0}((F_{\lambda,obs} - F_\lambda(\theta))^2/\epsilon_{\lambda,obs}^{2}).
\end{equation}

The total flux uncertainty $\epsilon_{\lambda,obs}$ at each wavelength includes the observed flux uncertainty, $\sigma_{\lambda,obs}$ and an additional error component, $\sigma_{\rm add,order}$ to account for systematic errors in the model grid: $\epsilon_{\lambda,obs}^{2} = \sigma_{\rm add,order}^2 + \sigma_{\lambda,obs}^2$. This additive error ($\sigma_{\rm add,order}$) is fitted as part of the likelihood calculation simultaneously with the model parameters. A single $\sigma_{\rm add,order}$ value is fitted for each order, which can range between 0 to 10$\%$ of the median flux. 

\begin{table*}
\begin{center}
\caption{Grid Parameter Ranges (full extent and ranges used in fits)}
\label{tab:grid_span}
\begin{tabular}{ccccc}
\hline\hline
Parameter & BOSZ (Step size) & PHOENIX (Step size) & Full BOSZ (Step size) & Full PHOENIX (Step size) \\
\hline
    Wavelength range$^{\ast}$ ($\mu$m) & 2.0-2.4 & 2.0-2.4 & 0.1-32 & 0.05-5.5 \\
    $T_{eff}$ (K) & 3500 to 7000 (250) & 2300 to 6000 (100) & 3500 to 30000 (250-1000) & 2300 to 12000 (100-250)\\
    log $g$ & 0.0 to 4.5 (0.5) & 0.0 to 4.5 (0.5) & 0.0 to 5.0 (0.5) &  0.0 to 6.0 (0.5) \\
    $[M/H]$ & -2.0 to 0.75 (0.25) & -2.0 to 1.0 (0.5) & -2.0 to 0.75 (0.25) & -4.0 to 1.0 (0.5-1.0) \\
    $[\alpha/Fe$$]$ & -0.25 to 0.5 (0.25) & -0.2 to 1.0 (0.2) & -0.25 to 0.5 (0.25) & -0.2 to 1.0 (0.2) \\
    R & 50000 & 50000 & 300000 & 500000 \\
\hline \hline
\end{tabular}
\end{center}
\end{table*}

StarKit has a modular structure that breaks the steps of constructing the model from the grid into different steps. Given a set of physical parameters, StarKit uses a linear interpolator to produce model spectra at arbitrary values from the spectral grid points. The spectrum is convolved with a rotational broadening profile, then shifted to the specified radial velocity. Afterwards, it is convolved with a Gaussian to the resolution of the NIRSPEC spectrograph (R $\sim$ 24,000) and resampled at the wavelength of observations. Finally, the model spectrum is normalized using a polynomial to fit the continuum of the observed spectrum.

We use uniform priors for the Bayesian inference for all fitted parameters (Table \ref{tab:prior_range}). The priors on $T_{eff}$ and $v_{rot}$ were chosen based on the range of values observed for red giant stars. Prior ranges in [$M/H$] range from -2.0 to 0.75 when using the BOSZ grid and from -2.0 to +1.0 for the PHOENIX grid. For [$\alpha/Fe$], the prior ranges from -0.25 to 0.5 for the for the BOSZ grid and -0.2 to +1.0 for the PHOENIX grid. The upper limits for the choice of priors on [$M/H$] and [$\alpha/Fe$] are based on limits in the pre-computed template spectral grids. We investigate both fitting for log $g$ and using photometric information to fix the value of log g for spectral fitting (Sections \ref{sec:system_logg}). We also use flat priors to fit for the wavelength offset between orders.

\begin{table}
\begin{center}
\caption{Prior ranges used in fits}
\label{tab:prior_range}
\begin{tabular}{cc}
\hline\hline
Parameter (unit) & Range \\
\hline
    $T_{eff}$ (K) & 3500 to 6000 \\
    log $g$$^{\ast}$ & from photometry \\
    $[M/H]$ & -2.0 to 0.75 \\
    $[\alpha/Fe$$]$ & -0.25 to 0.5 \\
    $v_{rot}$ (km/s) & 0 to 20 \\
    $v_{rad}$ (km/s) & -600 to 600 \\
    R & 15000 to 40000 \\
    Additive flux error (normalized flux) & 0 to 0.1 \\
\hline \hline
\end{tabular}
\end{center}
$^{\ast}$ log $g$ was fixed in the fits that best matched the reference physical parameters, and the fits to the Galactic center target stars.
\end{table}

We sample the posterior $P(\theta| D)$ using the MuliNest algorithm \citep{Feroz2008,Feroz2009} and report the central 68\% confidence interval for all parameters ($T_{eff}$, log $g$, [$M/H$], [$\alpha/Fe$], $v_{rot}$, $v_{rad}$). See \citet{Do2018} for additional details about the fitting.

\subsection{Calibration of K-band Spectra}    
\label{sect:calibration}

In this section, we present our investigation of the most likely sources of systematic errors in the parameter estimation for $T_{eff}$, log $g$, [$M/H$] and [$\alpha/Fe$]. We do this by comparing the StarKit fits to K-band spectra we obtained of open cluster stars previously used APOGEE survey for calibrations. These stars were used by the APOGEE survey to calibrate their parameter estimates with H-band spectra \citep{Meszaros2013}, so they have multiple previous measurements of their physical parameters. We need to do this cross calibration between K-band and H-band because there has been no large scale calibration done in K-band at the spectral resolution of our observations. 

\subsection{Choosing a Spectral Grid}
\label{sec:grid_choice}
To create the model spectra, StarKit requires a pre-computed grid of spectra, so the choice of the grid is important. The spectral grids consist of a large set of synthetic spectra computed from stellar atmosphere models over a range of model parameters. For this work, the most important parameters are $T_{eff}$, log $g$, [$M/H$], [$\alpha/Fe$]. For this work we use the BOSZ grid and PHOENIX grid, the 2 publicly available grid with the wavelength range and spectral resolution that allows us to measure [$\alpha/Fe$]. To choose between the two grids, we ran fits using the PHOENIX and BOSZ grids on the 10 calibrator stars. We then examined the fit residuals and the best-fit physical parameters to compare them with the APOGEE values for these stars \citep{Holtzman2015, Ahumada2019}. The parameter ranges of the BOSZ and PHOENIX grids that were used in the fits are shown in Table \ref{tab:grid_span}.

The two grids have important differences. They use different line lists, and line shape, profile, and depth calculations (see \citet{Husser2013}, \citet{Bohlin2017} for more information). The PHOENIX grid has a lower limit in $T_{eff}$ of $T_{eff}$=2300 K, compared to $T_{eff}$=3500 K for BOSZ. High metallicity stars at the Galactic center are likely to have temperatures below the 3500 K lower edge of the BOSZ grid. We therefore limit our current analysis to the low-metallicity stars at the Galactic center, which are more likely to have a temperature above 3500 K. The BOSZ grid allows for the C abundance [$C/M$] as a free parameter, but we chose to fix [$C/M$]= 0.0, in order to focus on the bulk properties of the star.

We find that the BOSZ grid is better able to reproduce the observed calibrator spectra than the PHOENIX grid (Appendix \ref{sect:grid_results}), showing smaller mean offsets and at least 1.6 times less scatter in the offsets (\ref{tab:mean_offsets}). The best-fit stellar parameters [$M/H$] has an average offset compared to APOGEE values of -0.18$\pm$0.16 with the BOSZ grid compared to -0.20$\pm$0.29 with the PHOENIX grid. Similarly, [$\alpha/Fe$] shows smaller mean offsets and scatter from the APOGEE values, with $\Delta$[$\alpha/Fe$] =-0.18$\pm$0.18 with the BOSZ grid and [$\alpha/Fe$]=-0.19$\pm$0.24 with the PHOENIX grid. Table \ref{tab:mean_offsets_small} summarizes the fits. More detail can be found in Appendix \label{sect:grid_results})

\subsection{Surface gravity}
\label{sec:surface_gravity}
We find in our fits of the calibrators that K-band spectra are insensitive to surface gravity, so additional information such as photometry are required to constrain log g. This lack of sensitivity is evident by the large scatter in the best-fit surface gravity for the calibrators compared to their reference values. This indicates that the spectral features in K-band are insensitive to changes in log $g$. We therefore fixed log $g$ in our fits to the value from APOGEE DR16 for the calibrators, or a value derived from stellar isochrones and photometry for the MW NSC stars. See Appendix \ref{sec:system_logg} for more details.

\subsection{Correcting for Systematic Uncertainties in Best-Parameters}
\label{sec:system_bestfit}
\begin{figure*}[h]
\begin{center}
\includegraphics[width=7in]{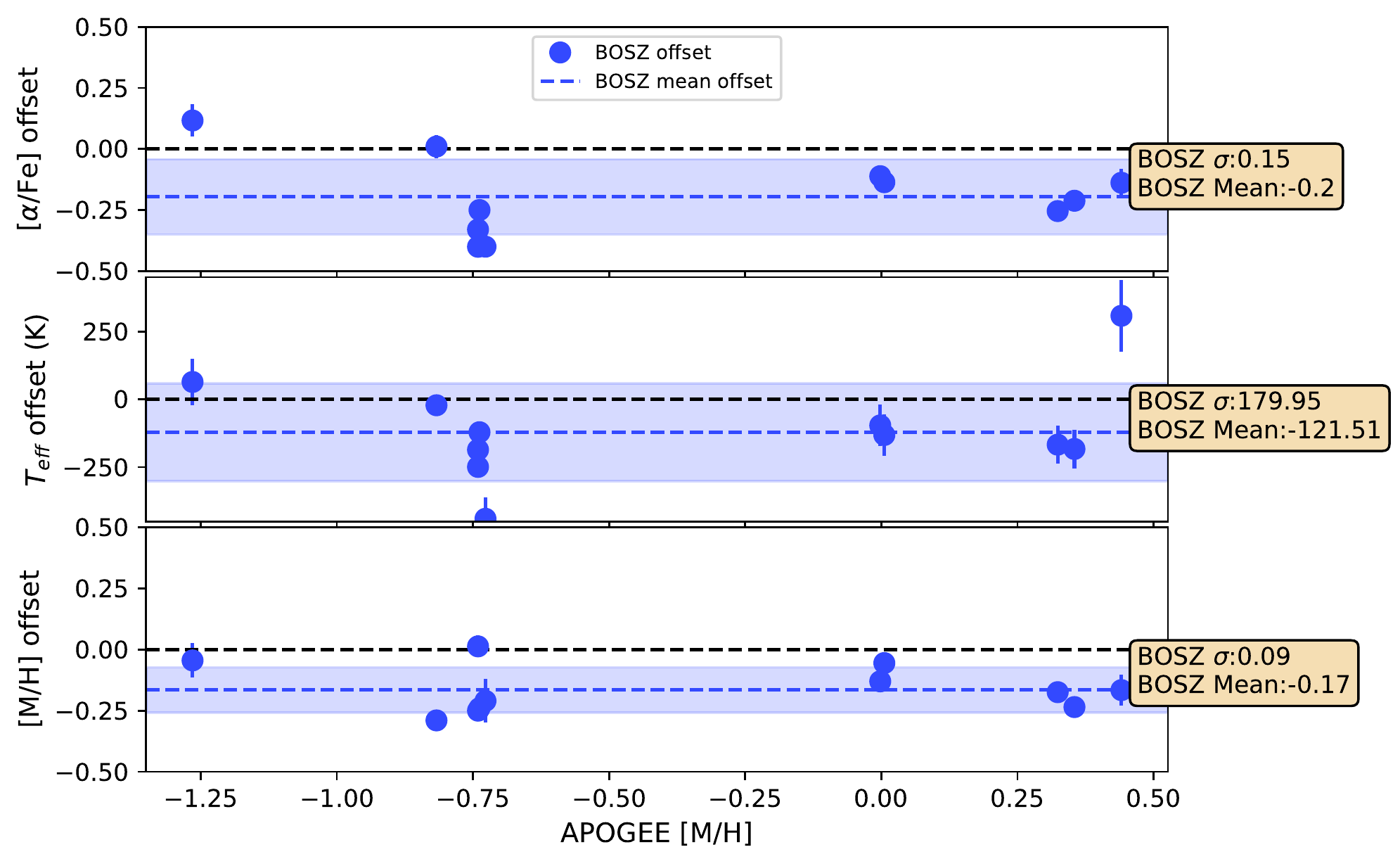}
\caption{\footnotesize We calibrate the K-band NIRSPAO spectra by using stars in star clusters that have also been observed by the APOGEE survey for their calibration in H-band. We use the offset between the best-fit StarKit model and the APOGEE value to determine the systematic offset and uncertainty in the StarKit model. The circular points are spectra taken for this work, and diamond points are previously observed and available in the Keck Observatory Archive. The blue dashed line is the mean offset and the shaded region is the 1$\sigma$ variation. The black dashed line is at an offset of zero. We find that the [$M/H$] and [$\alpha/Fe$] have a systematic offset of -0.17$\pm0.09$ and -0.2$\pm0.15$, respectively. We apply these offsets and their uncertainties to the fits of the Galactic center stars.}
\label{fig:grids_offsets}
\end{center}
\end{figure*}

We use comparisons with stellar calibrators discussed above to apply a correction to the best fit parameters using Starkit for fitting K-band spectra. This calibration step is similar to that applied by the APOGEE team in \citet{Meszaros2013} based on similar analyses with stars in open clusters. 
To correct for these offsets when fitting our spectra of stars in the Galactic Center, we use the mean systematic offset and its uncertainty between the StarKit fits of the calibrator stars and their reference values. We use the standard deviation in these offsets as an estimate of the uncertainty of this correction. We subtract this correction for each parameter and add the scatter in quadrature with the statistical uncertainties to estimate the final uncertainty for each parameter. Based on our experiments, we find that the most accurate way to measure stellar parameters in K-band with NIRSPAO and StarKit is to use the BOSZ grid and to fix log g with prior information (previous observations for the calibrators, photometrically derived values for Galactic center stars). The mean offsets and offset standard deviations used to account for systematic offsets are listed in Table \ref{tab:mean_offsets} row 3 ($v_{rad}$, log $g$ fixed), and shown in Figure \ref{fig:grids_offsets}.

\subsection{Investigating Additional Sources of Systematic Uncertainties}
\label{sect:exp_summary}

We investigate two additional sources of systematic uncertainty that have less impact on the best-fit parameters than those discussed above. Here, we summarize our findings for completeness. For details, see the Appendices.  

\begin{itemize}
  \item We try masking different portions of the spectra with large fitting residuals to see if that would improve the fit. We also experiment with masking regions that were most sensitive to the physical parameters such as [$M/H$]. None of the masks result in a significant improvement in the bringing the fits for the calibrator stars closer to the values from APOGEE. 
\item We test the effect of noise in our spectra by artificially adding random noise to our calibrator spectra to examine the SNR where the fits are robust. We add noise to simulate nine SNR values between 2 and 25 and then compare the fits to the case without extra noise. This experiment shows that the best-fit parameters begin to deviate from their values without extra noise at SNR<10. From this result, we exclude spectra with SNR<10 from further analysis as unreliable. 
  \item We also investigate how spectral resolution affects the best-fit parameters and find that generally, the best-fit parameters are consistent between R = 4000 and 24,000, though with larger uncertainties at lower resolution. We convolve the R$\sim$24,000 NIRSPAO spectra with a Gaussian filter to simulate lower spectral resolution. Reducing the resolution to R$\sim$4,000 produce offsets consistent with what we find at the R$\sim$24,000 spectra to within 1 $\sigma$. Additionally, we fit a sample of R$\sim$2,000 spectra from the SPEX stellar spectral library to compare the StarKit fits to literature values of $T_{eff}$, log $g$, and [$M/H$] at an even lower R. The mean offsets for the SPEX fits to their reference values were similar to the offset we see with APOGEE calibrators at higher resolutions, but the scatter is $\sim$30$\%$ greater at this lower resolution.
\end{itemize}

\begin{table*}[h]
\begin{center}
\caption{Mean offsets and standard deviations of the BOSZ and PHOENIX comparison fits, and the final offsets that are used in correcting fits.}
\label{tab:mean_offsets_small}
\begin{tabular}{cccccc}
\hline\hline
Experiment & Grid & $T_{eff}$ & log $g$ & [$M/H$] & [$\alpha/Fe$] \\
 & & K & & &  \\
\hline
R fixed & BOSZ & -145.0$\pm$178.0 & -0.09$\pm$0.45 & -0.18$\pm$0.16 & -0.18$\pm$0.18 \\
R fixed & PHOENIX & -147.0$\pm$250.0 & -1.22$\pm$1.07 & -0.2$\pm$0.29 & 0.19$\pm$0.24 \\
$v_{rad}$, log $g$ fixed$^{\ast}$ & BOSZ & -117.0$\pm$180.0 & 0.0$\pm$0.0 & -0.16$\pm$0.09 & -0.2$\pm$0.15 \\
\hline \hline
\end{tabular}
\end{center}
$^{\ast}$ These are the final values are used to correct for systematic offsets and error in the best-fit parameters for the Galactic center fits.
\end{table*}

\begin{table*}
\begin{center}
    \caption{Stellar physical parameters of Galactic Center star fits}
    \label{tab:gc_fits}
    \begin{tabular}{cccccccccc}
    \hline\hline
    Name & $T_{eff}$ & log g & [$M/H$] & [$\alpha/Fe$] & $T_{eff}$ & [$M/H$] & [$\alpha/Fe$] & $v_{rot}$ & Avg additive error \\
    & K & (fixed) & & & K (corrected) & (corrected) & (corrected) &  km/s & \\
    \hline
NE1-1-003 & 3830.0$\pm$193.0 & 0.6 & -0.75$\pm$0.11 & -0.15$\pm$0.15 & 3947.0$\pm$193.0 & -0.59$\pm$0.11 & 0.05$\pm$0.15 & 9.7$\pm$2.1 & 1.9$\pm$0.1\% \\
N2-1-002 & 4100.0$\pm$201.0 & 0.9 & -0.97$\pm$0.12 & -0.05$\pm$0.16 & 4217.0$\pm$201.0 & -0.81$\pm$0.12 & 0.15$\pm$0.16 & 5.0$\pm$2.6 & 5.8$\pm$0.1\% \\
\hline \hline
\end{tabular}
\end{center}
\end{table*}

\subsection{Results of calibrations}
\label{sect:grid_discuss}
Our work presents the most extensive testing, so far, of full-spectrum fitting in the K-band at R~24,000. We are able to recover the stellar parameters $T_{eff}$, [$M/H$], \& [$\alpha/Fe$] to an accuracy similar to what other surveys have found in other wavelengths. This shows that the K-band can be reliably used for bulk abundance determination with the full-spectrum fitting. $T_{eff}$, [$M/H$], and [$\alpha/Fe$] were measured to ~200 K, ~0.10 dex, and ~0.15 dex respectively. In comparison, typical APOGEE accuracy for these values are ~200 K, ~0.09 dex, and ~0.10 dex respectively. 
We find that the most difficult parameter to constrain with K-band spectra at R = 24,000 is the surface gravity. This is likely because of the lack of features in the spectra sensitive to log $g$ at our spectral resolution in this wavelength range. While the surface gravity is not well fit, we find that it also is not a strong source of bias because the spectra are not very sensitive to changes in the surface gravity. With additional information such as the distance and brightness of the stars, we also find that we can fix the surface gravity to values from photometry to further reduce the bias from this parameter. 

We find that the uncertainties in are dominated $T_{eff}$, [$M/H$], and [$\alpha/Fe$] by the systematic effects arising from mismatches in some lines between the template and observed spectra. In this work, we find that in the K-band, the lines from 22600-22670\r{A} (eg. 22631\r{A} Ca, 22657\r{A} Ca) and 22050-22090\r{A} (eg. 22056\r{A} Sc, 22068\r{A} Si), along with the Ti line at 22010\r{A} have larger systematic differences. The remaining regions of the spectra appear to be well-fit by the models.

In our current work, we focus on the subsolar metallicity stars because their physical parameters are within the pre-computed BOSZ grid. A large number of high metallicity stars could be studied in a similar way with additional grid points at lower temperatures. High metallicity stars are generally much cooler. Extending the BOSZ grid to lower temperatures ($T_{eff}$ < 3000 K) will allow a large number of stars to be fit using the full spectral fitting method.

\section{Results}
\label{sect:results}

\subsection{Chemical Abundances in the Galactic Center Stars}
\label{sect:results_abundances}

We find that the two Galactic center stars NE1-1-003 and N2-1-002 have sub-solar metallicities and have solar alpha-abundances. We fit the K-band spectra of these stars using the methodology outlined above. For this fit, we use a prior on value of log $g$ from Ks photometry from \citet{Schodel2010} and the SPISEA isochrone fitting package \citep{Hosek2020} with MIST isochrones \citep{Choi2016}.  We infer that log $g$ = 0.6 for NE1-1-003 and log $g$ = 0.9 for N2-1-002 (Appendix \ref{sec:system_logg}).  We apply the empirically determined corrections to $T_{eff}$, [$M/H$], [$\alpha/Fe$] found in Section \ref{sect:specfit} to the best-fit parameters and include the systematic uncertainties added in quadrature to the statistical uncertainties. We find that NE1-1-003 has $T_{eff}$ = 3947$\pm$193, [$M/H$] =  -0.59$\pm$0.11, [$\alpha/Fe$] = 0.05$\pm$0.15, and N2-1-002 has $T_{eff}$ = 4217$\pm$201, [$M/H$] =  -0.81$\pm$0.12, [$\alpha/Fe$] = 0.15$\pm$0.16 (for other parameters see Table \ref{tab:gc_fits}). 

\subsection{Constraining Location within the Nuclear Star Cluster}
\label{sect:results_location}

\begin{figure}[h]
\includegraphics[width=3.5in]{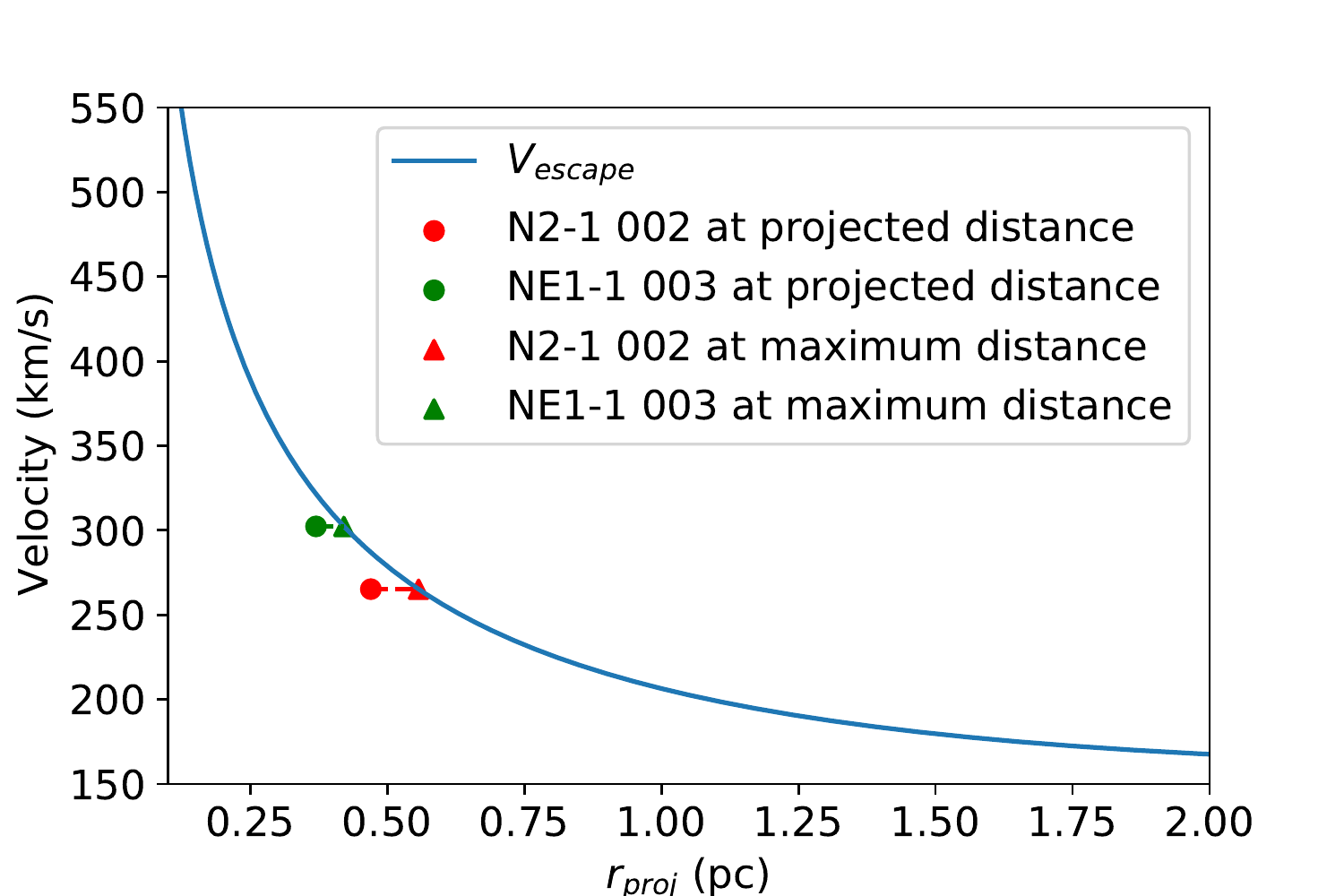}
\caption{\footnotesize The maximum escape velocity from the nuclear star cluster, $v_{escape}$, versus distance from Sgr A* $r$ (blue line). The green and red points indicate the two stars in the Galactic center studied in this work, with the dashed lines showing their 3-dimensional velocities. Assuming they are bound to the nuclear star cluster, their high 3D velocities place them within 0.6 pc from the supermassive black hole.}
\label{fig:v_esc}
\end{figure}

The two stars in our sample are consistent with being less than 0.75 pc from the supermassive black hole based on their 3D velocity. If we assume a star is bound to the supermassive black hole and the nuclear star cluster, the escape velocity at a given projected distance gives the maximum velocity that a star can have:
\begin{equation}
v_{esc} = \sqrt{\frac{2GM(r)}{r}},
\end{equation}
where $M(r)$ is the mass within a radius $r$ of the black hole, and $r = \sqrt{R_{2D}^2 + z^2}$, where $R_{2D}$ is the projected distance from the black hole and $z$ is the line of sight distance. 
A star's velocity and projected distance therefore gives an upper limit to the 3D distance that the star can be from the black hole. Here we use a black hole mass of $4\times10^6$ $M_\odot$ \citep{GravityCollaboration2018, Do2019}, and stellar mass distribution from \citet{Schodel2010}. Using the radial velocities from \citet{Do2015} and proper motion measurements from Hosek et al. (in prep.), NE1-1-003 likely is located between 0.37 pc and 0.48 pc and N2-1-002 is between 0.47 pc and 0.60 pc from the black hole (Figure \ref{fig:v_esc}). 

\section{Discussion}
\label{sect:disc}

Our present study offers new insight into the question of whether some parts of the Milky Way nuclear star cluster is the result of a merger with a globular cluster or dwarf galaxy. [$M/H$] and [$\alpha/Fe$] have been shown to vary between different regions of the Milky Way \citep{Hayden2015} and in globular clusters \citep{Pritzl2005} and dwarf galaxies \citep{Letarte2010, Norris2017, Hill2019}. The two stars in this study, N2-1-002 ( [$M/H$]=-0.81$\pm$0.12 and [$\alpha/Fe$]=0.15$\pm$0.16) and NE1-1-003 ([$M/H$]=-0.59$\pm$0.11 and [$\alpha/Fe$]=0.05$\pm$0.15) have sub-solar metallicity and approximately solar alpha-abundance. 

\begin{figure}[h]
\includegraphics[width=3.5in]{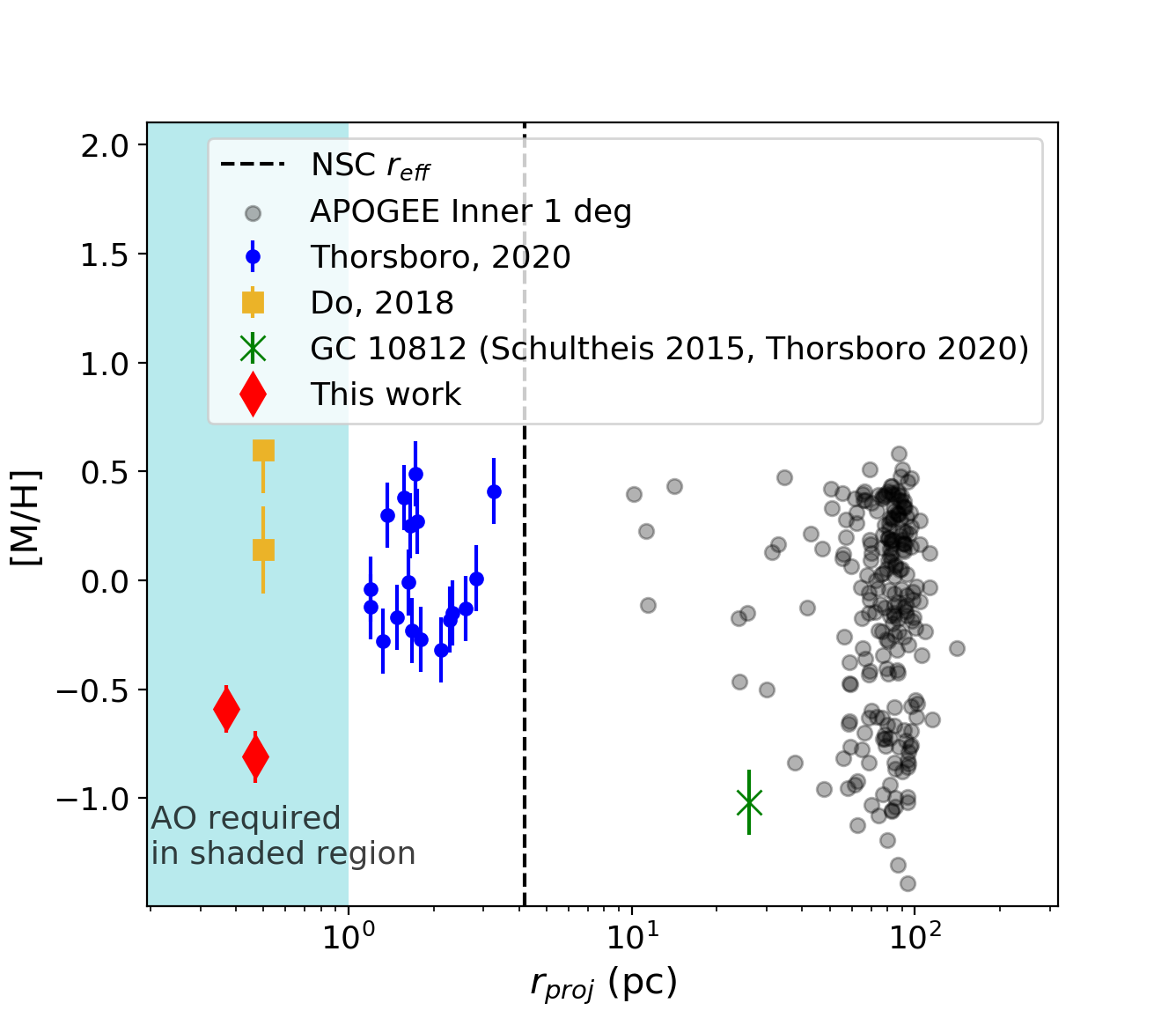}
\caption{\footnotesize Metallicity, [$M/H$], as a function of projected distance $r_{proj}$ for several samples of stars studied at high-spectral resolution at the  Galactic center. Red diamonds are the stars from this work, blue circles are stars from the main bulge population from \citet{Thorsbro2020}, the green circle is the sole metal-poor star from \citet{Thorsbro2020}, gold circles are from \citet{Do2018}, and the smaller grey points are inner bulge stars from APOGEE DR16. Our work explores the abundance of low-metallicity stars close the supermassive black hole. This work was enabled by adaptive optics (AO) which can spatially resolve this dense region and having a large sample of previous medium resolution spectra to find these rare stars. }
\label{fig:mh_rproj}
\end{figure}

\begin{figure*}
\includegraphics[width=7in]{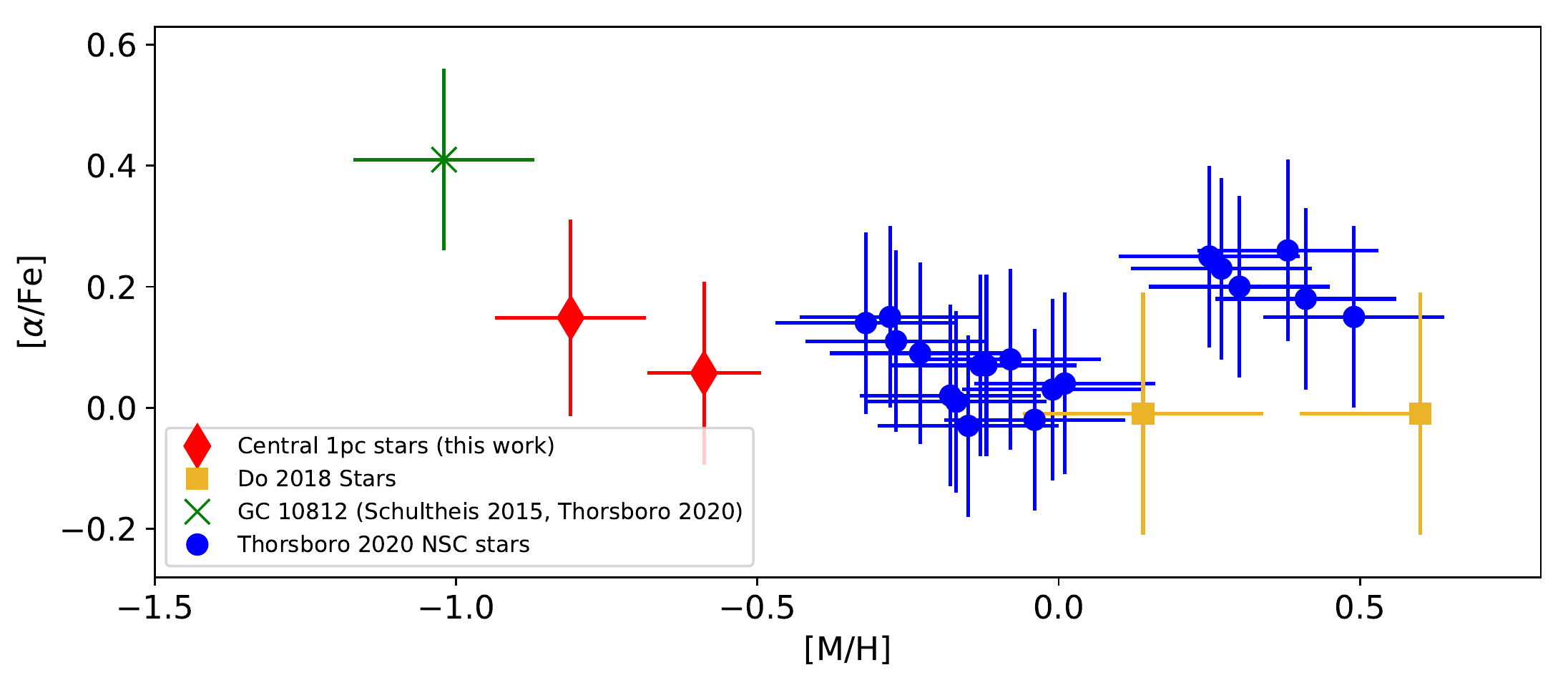}
\caption{\footnotesize [$\alpha/Fe$] vs [$M/H$] for stars in our sample (red diamonds), compared with previous observations inside the Milky Way Nuclear star cluster \citep[gold squares, blue circles][]{Do2018,Thorsbro2020}, and a metal-poor star at 40 pc from the Galactic center \citep[green x][]{Schultheis2015, Thorsbro2020}. The two stars in our sample occupy a different region in [$M/H$] and [$\alpha/Fe$] compared to the high metallicity population inside the nuclear star cluster and the low-metallicity stars outside the nuclear star cluster; this may indicate that low-metallicity stars inside the nuclear star cluster may have a different origin than the other previously studied populations. }
\label{fig:alpha_mh_thorsbro}
\end{figure*}

Our measurements for these stars provide the first $\alpha$-abundances for subsolar metallicity stars in the central 1-pc (Figure \ref{fig:mh_rproj}). There are no high resolution spectroscopic studies before our work because they represent $<10$\% of the total number of stars in the MW NSC. The small sample sizes (1 to $\sim10$) of previous high-resolution studies have missed these the sub-solar metallicity population of the MW NSC, and have measured alpha-abundances for the solar and super-solar metallicity population \citep{Carr2000, Ramirez2000, Cunha2007, Davies2009, Do2018, Thorsbro2020}. Previous high spectral resolution $\alpha$-abundance measurements of metal-poor stars have been made at larger Galactrocentric distances ($>25 pc$) than our study because of the difficulty in resolving stars without the use of adaptive optics (Figure \ref{fig:mh_rproj}, \citet{Schultheis2015, Ryde2016, Ahumada2019}). These studies outside the nuclear star clusters showed higher [$\alpha/Fe$] values \citet{Schultheis2015, Ahumada2019} compared to N2-1-002 and NE1-1-003 inside the nuclear star cluster (Figure \ref{fig:alpha_mh_thorsbro}). This suggests that the origin of the subsolar metallicity stars in the NSC may be different different than in the majority of the inner bulge.

\begin{figure*}
\includegraphics[width=7in]{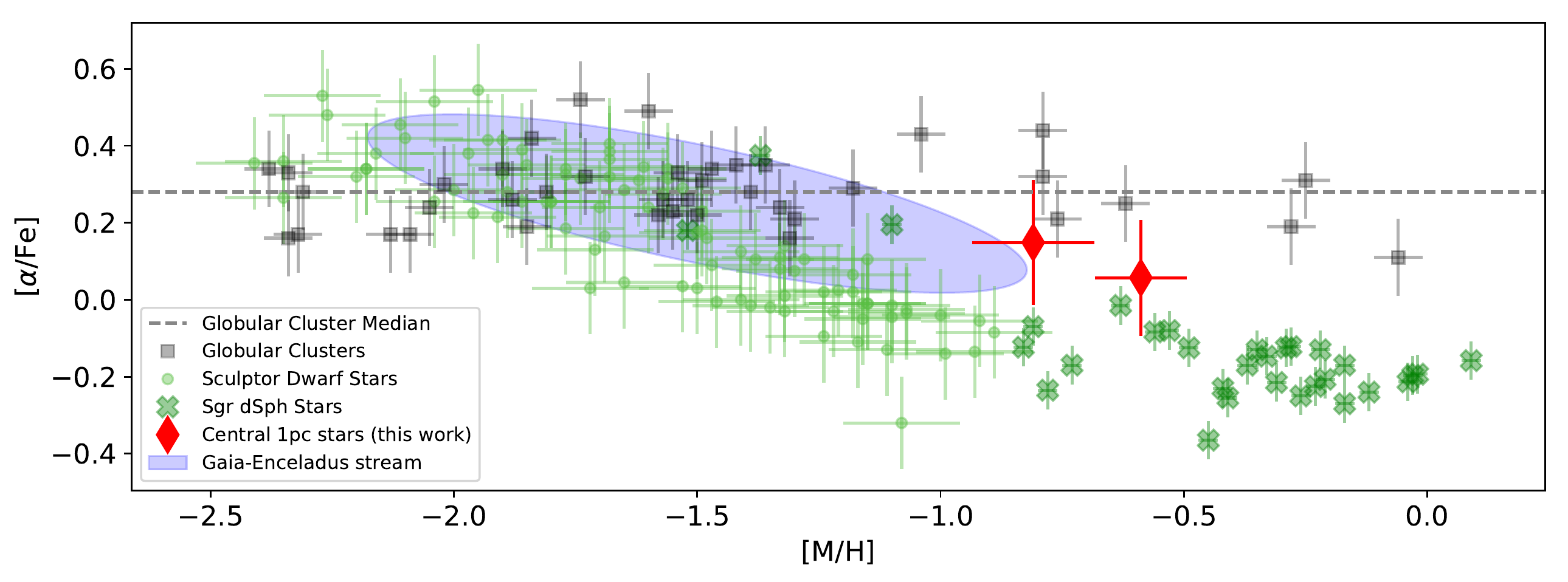}
\caption{\footnotesize The [$\alpha/Fe$] and [$M/H$] for the stars in our sample (red diamonds) are consistent with both globular clusters \citep[grey squares, grey dashed line marks median $\alpha/Fe$,][]{Pritzl2005} and dwarf galaxies \citep[green X and circles,][]{Monaco2005,Hill2019}, within their uncertainties. The low-metallicity nuclear star cluster stars are also close the observed properties of the Gaia-Enceladus stream \citep[blue ellipse,][]{Helmi2018}.
}
\label{fig:alpha_mh_dgal}
\end{figure*}

The [$M/H$] and [$\alpha/Fe$] of the two stars in this study fall between that of globular clusters and dwarf galaxies suggesting they may be the result of an infall event. These stars have metallicities much lower than 90$\%$ of stars in the nuclear star cluster, which suggest that these stars are not likely to be formed in the same way as the majority of stars in the cluster \citet{Do2015, Do2020, Ryde2016, Rich2017, Thorsbro2020} (Figures \ref{fig:mh_rproj}, \ref{fig:alpha_mh_thorsbro}). On the other hand, globular clusters have [$M/H$] ranging from -2.5 to -0.1, with [$\alpha/Fe$] ranging from 0.1 and 0.5. Dwarf galaxies tend to have large dispersion in both [$M/H$] and [$\alpha/Fe$], with [$\alpha/Fe$] generally lower than globular clusters at the same [$M/H$] (Figure \ref{fig:alpha_mh_dgal}, \citealt{Pritzl2005, Letarte2010, Norris2017, Hill2019}). The two stars in this study have similar metallicity to the Sgr dSph stars and the globular clusters Terzan 7, Palomar 12, NGC 6342, M71, and 47 Tuc \citep{Pritzl2005}. Given the current uncertainties, we cannot differentiate these two scenarios at this time. 

The presence of dwarf galaxy remnants in the MW NSC would significantly impact our ideas of how our NSC and supermassive black hole has evolved. \citet{Lang2013} proposed that a dwarf galaxy containing an intermediate mass black hole (IMBH) could reach the Galactic center, disrupting gas clouds in the central $\sim$100 pc, and dragging gas into the MW NSC. Some of this gas could have accreted onto Sgr A*, causing a burst of accretion activity, and forming the 'Fermi bubbles', a pair of gamma ray-emitting lobes of gas above and below the galactic plane \citep{Dobler2010, Su2010}. The remaining gas could initiate a burst of star formation in the central $\sim$1 pc, which results in the young stars we see today \citep{Paumard2006}. This hypothesis can be tested because the stars belonging to this infalling dwarf should still exist in the MW NSC today \citet{ArcaSedda2020}. The stars in our sample may have come from such a population, though additional stars are necessary to differentiate this scenario from an infalling star cluster. 

Other recent results also show evidence for infall into the Galactic center. \citet{Minniti2016} identified a dozen RR Lyrae variables, characteristic members of old and metal-poor populations, in the Galactic bulge, providing the first direct observational evidence for past infall events. \citet{FeldmeierKrause2020} found that there are two times as many metal-poor stars in the Galactic North portion of the MW NSC compared to elsewhere in the cluster. As infalling clusters are expected to form anisotropic structures that can persist for Gyr after infall \citep{PMB2014,ArcaSedda2020}, they suggest that this anisotropy could be evidence of a recent cluster infall which deposited our metal-poor stars within the MW NSC. \citet{Minniti2021} investigated the orbits of known globular clusters, noted a lack of clusters passing within 0.5$^{\circ}$ of the Galactic center. They discuss that clusters and possibly dwarf galaxy remnants passing into this region may be disrupted, and the constituent stars mixed into the inner bulge. In addition, \citet{Do2020} found different kinematic properties in the subsolar metallicity population compared to the super-solar population, which also consistent with an infall hypothesis. Our study offers additional evidence for this by showing that the alpha-abundances are not typical of the inner Milky Way, and therefore unlikely to be formed in-situ. Also detected from APOGEE data are signs that parts of the inner bulge was created from dwarf galaxy mergers \citep{Helmi2018, Horta2020}. Additional measurements bridging the APOGEE stars and the MW NSC stars could be used to test whether the stars in the MW NSC was part of this remnant galaxy.

While metal-poor stars could have formed in-situ at the Galactic center 13 Gyrs ago before significant metal enrichment, the stars in this study are unlikely to be part of this population. Due to the very fast metal-enrichment in the bulge, the vast majority of the earliest stars that may have formed in-situ would be expected to have [$M/H$] $\sim$ -1.0 and $\alpha$-enhanced with [$\alpha/Fe$] $\sim$ 0.5 \citep{Barbuy2009, Chiappini2011, Barbuy2014, Schiavon2017, Thorsbro2020}. While the stars in our study have similar metallicities, they are not alpha-enriched, which suggests formation in a system with longer star formation history. In addition, stellar isochrone fitting work in Chen et al., (in prep) suggests that the metal-poor component of the MW NSC has ages of 3-5 Gyr, far younger than the ~13 Gyr when the first stars in the bulge are believed to have formed \citep{Howes2015}. The results of \citet{Minniti2021}, showing a lack of globular clusters in the inner 0.5$^{\circ}$ (~100 pc) around the Galactic center, suggests that clusters which enter this region are disrupted. If the stars from the disrupted clusters remain in the inner 100 pc, some may migrate to the MW NSC, resulting in the metal-poor stars studied here. In the future, measurements of specific elemental abundances such as [$C/Fe$] and [$Al/Fe$] could be used to separate the in-situ versus the merger scenarios, as they have distinct chemical abundance patterns \citep{Schiavon2017, Fernandez-Trincado2019, Fernandez-Trincado2020}. 

At the present time, our understanding of the sub-solar metallicity stars is limited by the number of Galactic center stars observed at high spectral resolution and systematic uncertainties in the spectral grid. Increasing the sample size of metal-poor Galactic center stars will allow us to understand they lie in [$\alpha/Fe$]-[$M/H$] space, and allow us to differentiate between a potential globular cluster or in-situ origin, and a dwarf galaxy origin. We would expect a larger range of both metallicity and alpha-abundance for a dwarf galaxy origin compared to a globular cluster. A dwarf galaxy will also tend to have a [$\alpha/Fe$] vs. [$M/H$] relationship that depends on its star formation history. Measuring this relationship may also allow us to pinpoint which dissolved dwarf galaxy the Nuclear Star Cluster stars belong to. Improvements in the atomic data and radiative transfer models will help to reduce the uncertainties and provide a more precise comparison to other parts of the Milky Way. A dwarf galaxy infall may also deposit dark matter into the bulge, while a globular cluster infall would not. However, we cannot distinguish the effects of dark matter from luminous matter at the present time.

\section{Conclusions}
\label{sect:conclu}

We measure the alpha abundance for subsolar metallicity stars in the central parsec for the first time. We find that [$M/H$] = -0.59$\pm$0.11 and [$\alpha/Fe$] = 0.05$\pm$0.15 for the star NE1-1-003, and [$M/H$] = -0.81$\pm$0.12 and [$\alpha/Fe$] = 0.15$\pm$0.16 for the star N2-1-002.  We conducted extensive calibrations using K-band spectra of stars observed in H-band in the APOGEE survey, and show that our abundances for the calibrators are consistent with the APOGEE abundances within 0.1 dex in [$M/H$] and 0.15 dex in [$\alpha/Fe$].  The [$\alpha/Fe$] we find for the metal-poor stars are consistent with the hypothesis that the sub-solar metallicity population may be the result of a merger of a globular cluster or a dwarf galaxy with the nuclear star cluster. 

Our findings are consistent with other observations that suggest an infall origin for the sub-solar metallicity population of the Milky Way Nuclear Star Cluster. The next step in this work is to obtain additional measurements to differentiate between the infall of a dwarf galaxy and a globular cluster. The infall of a dwarf galaxy can significantly impact not just the composition of the nuclear star cluster, but also bring in gas for star formation and accretion onto the supermassive black hole, which might explain a number of observed phenomena such as the existence of young stars and the Fermi bubbles.

\section{Acknowledgements}
\label{sect:acknowl}
We would like to thank Jessica R. Lu, Mark Morris, Eric Becklin, and other members of the UCLA Galactic Center group for providing helpful comments and discussions. We also would like to thank Matthew Hosek for providing proper motion data on the two low-metallicity NSC stars, Jennifer Sobeck for providing information on potential calibrator stars from APOGEE. We also thank the staff of the Keck Observatory, especially Jim Lyke, Randy Campbell, Gary Puniwai, Heather Hershey, Hien Tran, Scott Dahm, Jason McIlroy, Joel Hicock, and Terry Stickel, for all their help in obtaining the new observations. 

The W. M. Keck Observatory is operated as a scientiﬁc partnership among the California Institute of Technology, the University of California, and the National Aeronautics and Space Administration. The authors wish to recognize that the summit of Maunakea has always held a very signiﬁcant cultural role for the indigenous Hawaiian community. We are most fortunate to have the opportunity to observe from this mountain. The Observatory was made possible by the generous ﬁnancial support of the W. M. Keck Foundation.

Support for this work was provided by NSF AAG grant NSF AAG AST-1909554.
\bibliography{StellarMetallicities}

\section{Appendices}
\label{sect:appendix}
In the appendices, we describe in greater detail the experiments to estimate systematic uncertainties in the full-spectral fitting method. Section \ref{sect:grid_results} describes the comparison of the BOSZ and PHOENIX spectral grids. Section \ref{sec:system_logg} describes the robustness of fitting surface gravity and improvements that can be made by using log $g$ estimates from photometry. Section \ref{sect:interpol} describes the uncertainty introduced by interpolation between spectra at different points in the spectral grid. Section \ref{sect:correlations} examines correlated uncertainties between the different physical parameters. Section \ref{sect:masking} describes our experiments with masking selected portions of our spectra based on several criteria to test if masking can reduced the systematic uncertainties due to template mismatch. Section \ref{sect:convolve} is our experiments with fitting spectra at a lower resolution. Section \ref{sect:stat_err_vs_snr} describes the relationship between signal-to-noise of the spectra and the statistical uncertainties in the fitted parameters. Section \ref{sec:atlas} contains the individual spectra and best fits for the stars in this work. 

\subsection{PHOENIX and BOSZ grid comparison.}
\label{sect:grid_results}

\begin{figure*}[h]
\begin{center}
\includegraphics[width=7in]{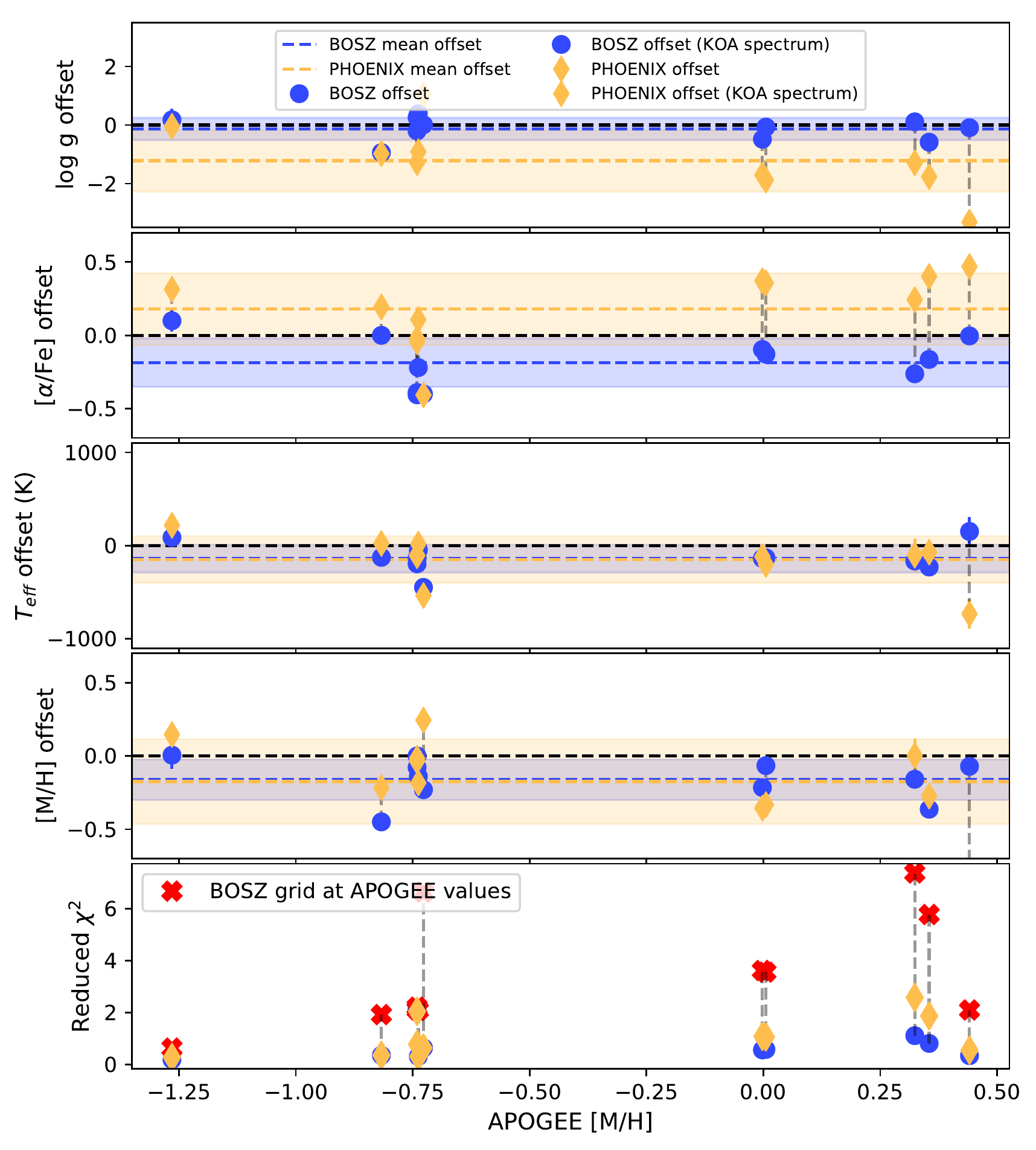}
\caption{\footnotesize We test the PHOENIX and BOSZ spectral grids by comparing the best fit parameters using these grids with their best fit values from the APOGEE survey. We find that in general the BOSZ grid has smaller offsets and scatter (blue) compared to the PHOENIX grid (orange) for log g (first panel), [$\alpha/$Fe] (second panel), $T_{eff}$ (3rd panel), and [M/H] (4th panel). This indicates that fits with the BOSZ grid are better able to reproduce the known properties of the calibrator stars. The red points in the bottom panel are the $\chi^{2}$ value with the BOSZ model set at the APOGEE values, showing that the best-fit parameters from StarKit are better matches to our spectra compared to the best-fit APOGEE values evaluated using the BOSZ and PHOENIX grids.
}
\label{fig:grids_offsets_both}
\end{center}
\end{figure*}

Here, we discuss the results of our experiment comparing the results of fitting the PHOENIX and BOSZ grids with full-spectrum fitting, and potential sources of the systematic uncertainties arising from the particular grids. 

We fit the spectra of calibrator stars in open clusters with both grids and compared them to the APOGEE DR16 values in order to determine which grid gives more robust results. These stars are also used as calibrators for the APOGEE pipelines and we use the APOGEE DR16 values (\citet{Ahumada2019} as the ground truth physical parameter values. The offsets between the measured values and APOGEE DR16 were calculated for each calibrator star along with the $\chi^{2}$ values for each fit for both grids (Figure \ref{fig:grids_offsets_both}). The BOSZ grid result in similar or better agreement with the APOGEE values than the PHOENIX grid. The fits from using the BOSZ also show smaller scatter in their offsets. 

\begin{figure}[h]
\includegraphics[width=7in]{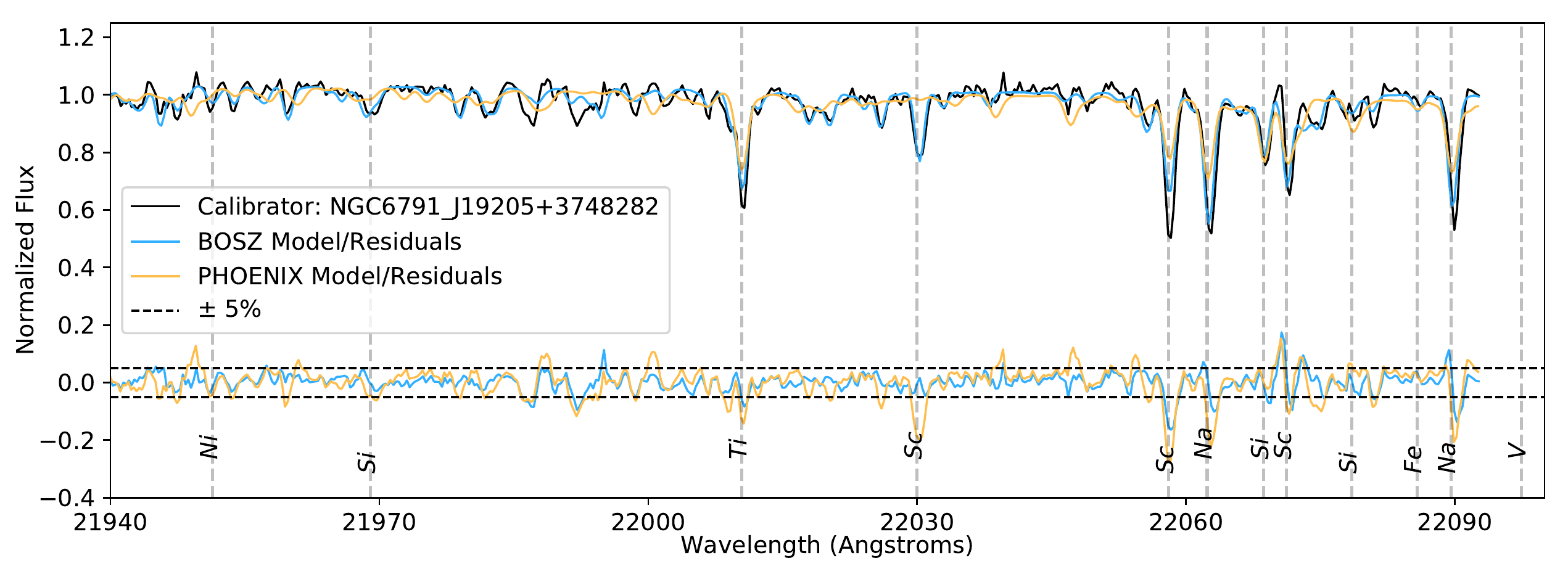}
\caption{\footnotesize Comparison of the best-fit spectrum from BOSZ (blue) and PHOENIX (orange) grid to the observed spectrum (black) of one of our calibrators (2M19205338+3748282). The bottom shows the residuals of the data minus the best-fit model; the black dashed line is $\pm$5$\%$ residual. The BOSZ grid produces smaller fitting residuals compared to the PHOENIX grid.}
\label{fig:phoenix_bosz_spec_comparsion}
\end{figure}

The BOSZ fit also shows smaller residuals than the fit using the PHOENIX grid for all calibrator stars (Figure \ref{fig:phoenix_bosz_spec_comparsion}). The relatively smaller residuals from the fits using the BOSZ grid is reflected in the $\chi^2$ values as well (Figure \ref{fig:grids_offsets_both}).

The scatter in the best-fit parameters for the BOSZ grid fits are similar or slightly larger than the APOGEE systematic uncertainties, except for surface gravity \citep{Meszaros2013}. The standard deviations in the offsets between the best-fit BOSZ grid values and APOGEE DR16 with all parameters fitted were found to be $\sigma_{T_{eff}}$=180K, $\sigma_{[M/H]}$=0.09, $\sigma_{[\alpha/Fe]}$=0.15, $\sigma_{\log g}=0.5$ (\ref{tab:mean_offsets_small}). \citet{Meszaros2013} stated the precision of the APOGEE values of these parameters as $\sigma_{T_{eff}}$=50 - 100K, $\sigma_{[M/H]}$=0.09, $\sigma_{[\alpha/Fe]}$=0.07, and $\sigma_{\log g}=0.17$. StarKit measurements with the BOSZ grid has similar scatter compared to the APOGEE \citet{Meszaros2013} results for [$M/H$], and has about twice the scatter for $T_{eff}$ and [$\alpha/Fe$]. We will discuss the $\log g$ measurements further in Section \ref{sec:system_logg}. Overall, our methodology can produce similar precision in metallicity and alpha-abundance using K-band spectra as APOGEE can in the H-band for these calibrators.

To test that our best-fit parameters produced the lowest residuals between the observed spectra and grid, the BOSZ model spectrum was evaluated at the APOGEE physical values for each calibrator, and the reduced $\chi^{2}$ calculated. The BOSZ model set to the best-fit parameters produced the smallest reduced $\chi^{2}$, as seen in Figure \ref{fig:grids_offsets}. These lower reduced $\chi^{2}$ values showed the APOGEE values were not the best match for our K-band spectra, and the StarKit fitting was not getting caught in a local reduced $\chi^{2}$ minimum.

 There likely multiple reasons why the BOSZ grid results in better fits than the PHOENIX grid. One explanation is that the PHOENIX grid uses older atomic data than BOSZ, which may have more updated transition probabilities \citep{Husser2013, Bohlin2017}. Different atmospheric assumptions in the calculations of line shapes and strengths could again introduce another source of error in the best-fit values. Our best-fitting models with PHOENIX appeared to have issues matching both strong and weak lines, as can be seen in Figure \ref{fig:phoenix_bosz_spec_comparsion} with the Si line at 21970. BOSZ also has issues matching some lines, such as the Na line at 22090\r{A} in Figure \ref{fig:phoenix_bosz_spec_comparsion}, although there are fewer mismatches, and they typically are not as large as the PHOENIX mismatches. Improvements in the grid template will have the greatest impact on our ability to measure abundances accurately.

\subsection{Testing for correlations between physical parameters and offsets}
\label{sect:correlations}

\begin{figure*}[h]
\begin{center}
\includegraphics[width=7in]{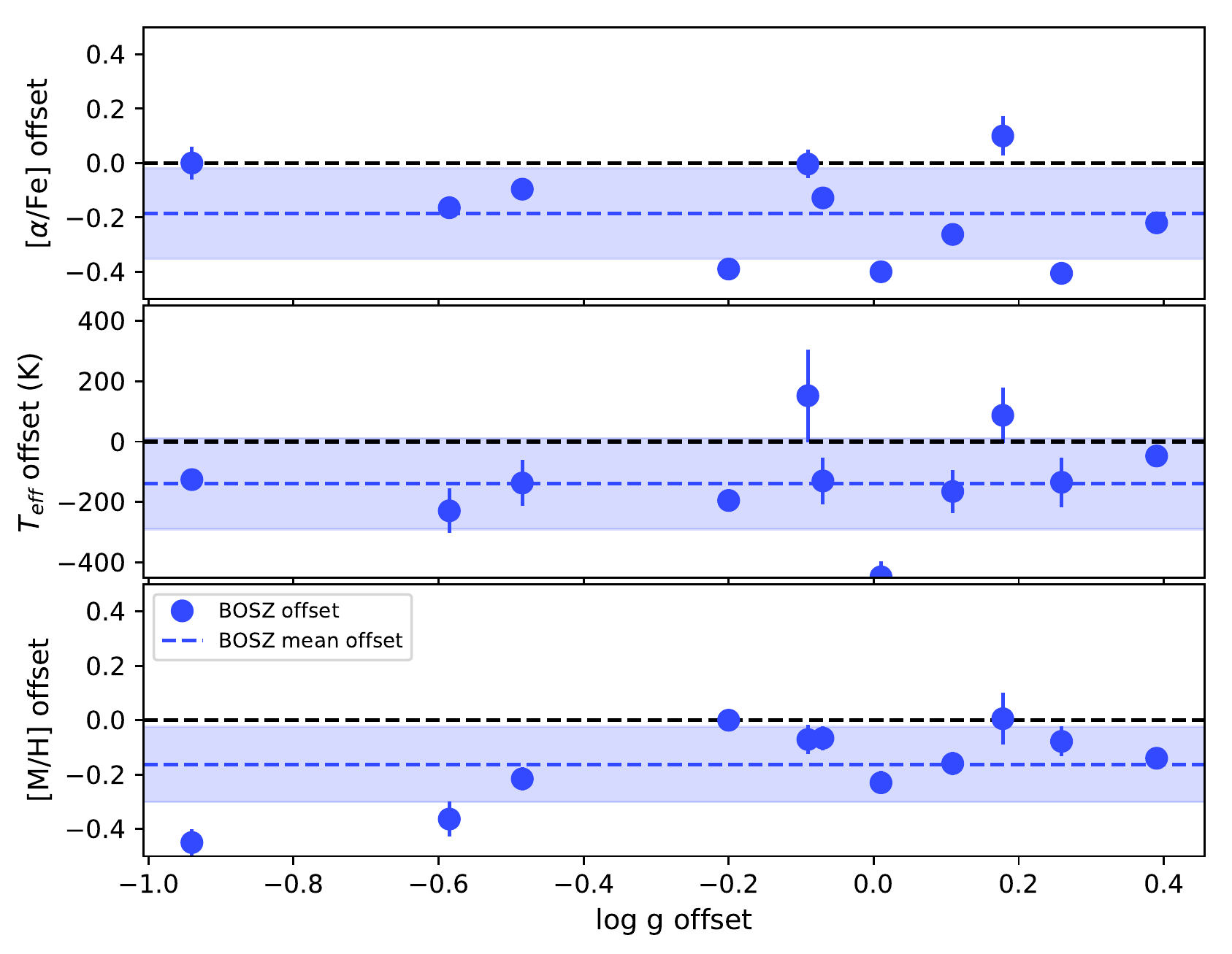}
\caption{\footnotesize We find log g does not have a very strong dependence on the physical parameters [M/H] and [$\alpha$/Fe] until it is offset by more than 0.4 dex. The figure shows the dependence of the offset between the reference APOGEE values for calibrator stars and the best-fit values using the BOSZ grid as a function of offset in log g from the APOGEE values ($\alpha$/Fe], $T_{eff}$, and [M/H]). We also compute the F-statistics and critical values are listed for each offset plot. If the F-statistic meets or exceeds 5$\%$ critical value, then there is at least a 5$\%$ significance to the correlation. There is a weak correlation between [$M/H$] and log $g$, but no significant correlations for the other parameters.}
\label{fig:nspec_offset_ftest}
\end{center}
\end{figure*}

We fit the relationship between the the BOSZ grid offset for each parameter ($T_{eff}$, log $g$, [$M/H$], and [$\alpha/Fe$]) from the calibration stars and their APOGEE values to examine correlations. 
To test for correlations, we use an F-test comparing a linear fit to the offsets to a constant fit; if the linear fit is a statistically better fit, then it is likely there is a correlation. 
We compute the F-statistic for the linear and constant fits at the 5$\%$ levels. If the F-statistic was lower than this critical value, we cannot reject the constant (uncorrelated fit) for the linear fit. The constant fit passes the F-test for all cases. There is a marginal correlation between the [$M/H$] offset and $\log$ $g$ offset, which was above the critical value for the model at a 10$\%$ significance level (Figure \ref{fig:nspec_offset_ftest}).

\subsection{Systematic Uncertainties in Surface Gravity}
\label{sec:system_logg}

We find that using photometry or other methods to estimate $\log g$ to be helpful to improve the accuracy of the abundance measurements in K-band. 
In Section \ref{sect:grid_results}, we found that the surface gravity, log $g$, has the largest offset between the best-fit value and the published APOGEE values for the calibrators; $\log g$ also shows slightly correlation with [$M/H$] offset (Section \ref{sect:correlations}). The average offset is log $g$ = -0.09$\pm$0.45 for the BOSZ grid. This is within the APOGEE log $g$ precision of 0.17 dex from \citet{Meszaros2013}, but the standard deviation of our log $g$ offsets are larger at 0.9 dex. Like the BOSZ grid, the PHOENIX grids physical parameter with the largest mean offset is the surface gravity. The average offset in log $g$ is -1.22$\pm$1.07 for the PHOENIX grid. We find that an offset of $\Delta$log $g$ = 0.4 dex is needed to change [$M/H$] and [$\alpha/Fe$] by greater than 0.1 dex, or within the systematic uncertainties for these parameters with log $g$ at the APOGEE or MIST isochrone values. $T_{eff}$ also remained within its systematic uncertainty at this offset.

\begin{figure}
\includegraphics[width=3.3in]{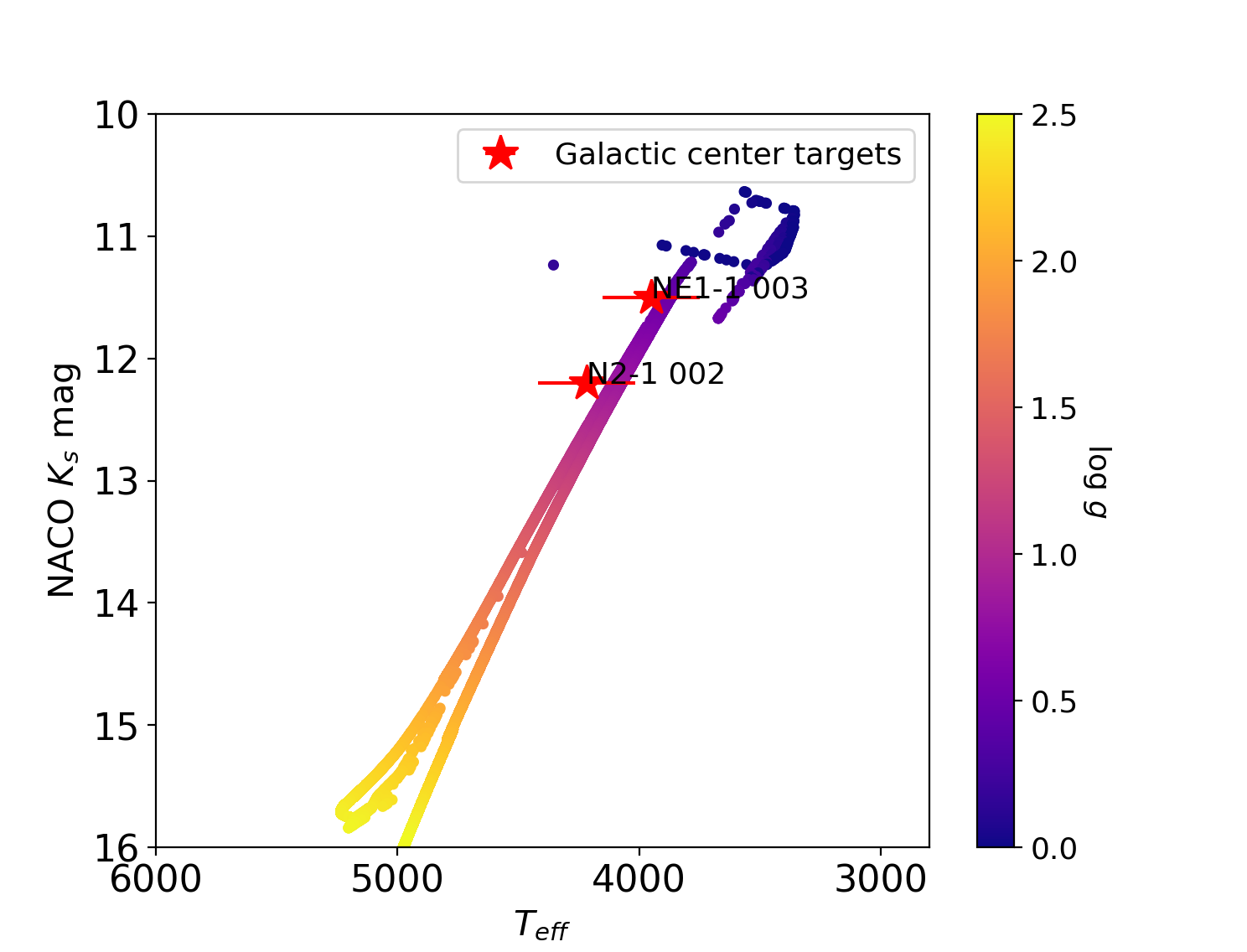}
\caption{\footnotesize The MIST isochrone \citep{Choi2016} with an [$Fe/H$] = -1.0 and an age of 5 Gyr is well matched to the photometry and temperature data for the two stars at the Galactic center (red stars). We use this isochrone to measure the surface gravity for NE1-1-003 (log $g$ = 0.6) and N2-1-002 (log $g$ = 0.9) using their $T_{eff}$ measurement and $K_{s}$ photometry.}
\label{fig:isochrome}
\end{figure}

The large offsets in the surface gravity suggests that the majority of the features in the section of K-band considered here are not very sensitive to surface gravity. We therefore use photometry and other sources of information to narrow the possible range of log $g$ for our fits. For the APOGEE stars, we fix the surface gravity to the best fit APOGEE DR16 values before fitting the K-band spectra. The surface gravity values for the Galactic center stars were determined using stellar evolutionary models for stars with similar ages and metallicities as the Galactic center. MIST evolutionary models \citep{Choi2016} with merged atmosphere models \citep[more details on merged atmosphere models in][]{Hosek2020} for [$M/H$]=0.5 and [$M/H$]=-1.0, at ages of 10 Gyr and 5 Gyr suggested log $g$=0.6 for NE1-1-003 and log $g$=0.9 for N2-1-002 would be appropriate for both the subsolar and supersolar metallicity populations at the two stars' respective magnitudes. A sample isochrone used is shown in Figure \ref{fig:isochrome}. For more details on the cluster modeling, see Chen et al. (in prep). 

By estimating the surface gravity to values from photometry, we can reduce this source of systematic uncertain below that of the spectral grid (see Section \ref{sect:results}). 

\subsection{Interpolation Uncertainty}
\label{sect:interpol}
Spectral grids are sampled at discreet values of physical parameters, and so in order to produce models at arbitrary parameter values, we must interpolate between grid points. This interpolation can introduce an additional source of uncertainty. To investigate how much the interpolation was contributing to the error, we remove a grid point, and then interpolated the spectrum at the parameters of the removed point. This is repeated at every point in the grid, and the value and wavelength of the maximum residual between the removed grid spectrum and the interpolated spectrum are recorded. This allows us examine the interpolation accuracy of different portions of the spectral grid. Our maps show that the worst interpolated regions were at the lower bounds of the grid's $T_{eff}$,  log $g$, and [$\alpha/Fe$] ranges, and at the upper bounds of the grid's [$M/H$] range. For example, the maximum interpolation residual at a grid point decreased as $T_{eff}$ increased from the lower bound. Overall, the interpolation errors are very small. Within the NIRPEC wavelength range used in this paper, the maximum interpolation residuals is $\lesssim0.5\%$.

Overall, our interpolation tests suggest that interpolation is not a major source of systematic uncertainty. The interpolated spectra generated when the grid points were removed were fitted with the BOSZ grid, and offsets between the best-fit parameter values and the actual grid point parameters were calculated. This gave us estimates of the maximum errors introduced by interpolation. These fits showed that the mean offsets and scatter in best-fit values about the grid points were $\Delta T_{eff}$ = -6$\pm$38, $\Delta$ log $g$ = 0.1$\pm$0.1, $\Delta$ [$M/H$] = -0.01$\pm$0.03 $\Delta [\alpha/Fe$] = -0.01$\pm$0.02. These additional sources of error were added in quadrature to the systematic and statistical errors. 

\subsection{Masking and Abundance Sensitivities}
\label{sect:masking}
We performed several experiments to examine whether including only regions where the lines are sensitive to changes in the physical parameters would produce more accurate fits. Overall, masking different portions of the spectra did not have a significant effect on the parameter estimates. 

One way to determine the changes in the line strengths is to measure how the flux scales with changes in the physical parameters at each wavelength. We measure this by fitting a linear trend to flux changes by varying one parameter at time by $\pm0.1$ dex, while fixing the others. We vary the parameters around the best-fit values for the spectra from 7 calibrator stars. We mask all points that do not vary by more than 5$\%$ when varying [$M/H$] by 0.1 dex, [$\alpha/Fe$] by 0.1 dex, or $T_{eff}$ by 400K, and re-fit these spectra. The best-fit values from this masking method did not change significantly compared to their non-masked best-fit values.

\begin{figure}[h]
\includegraphics[width=7in]{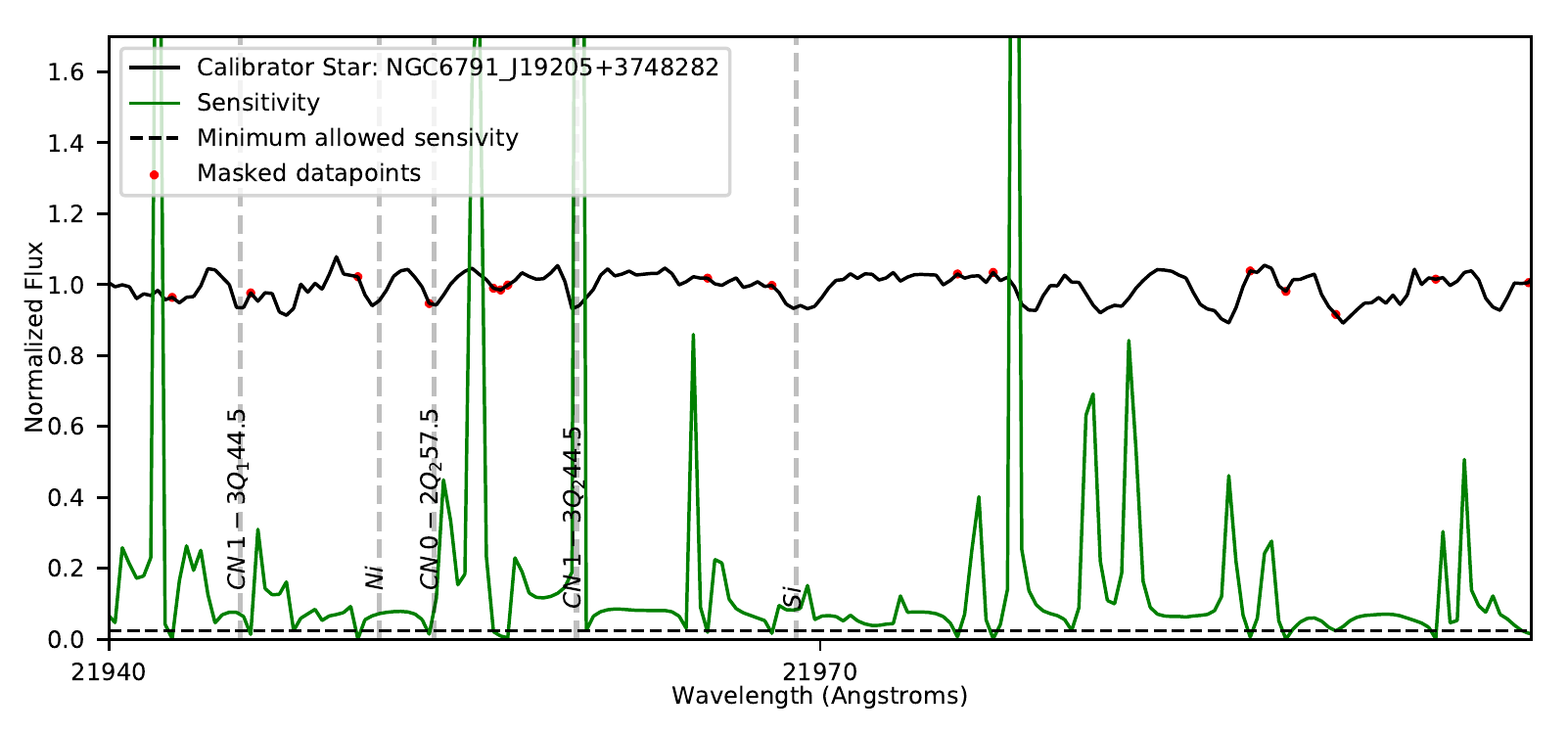}
\caption{\footnotesize An example of masking regions of a spectrum (2M19205338+3748282, black line) which respond the least to changes in physical parameters, as quantified by the $S_{\lambda}$ parameter from \citet{Sheminova2014}. We compute $S_{\lambda}$ (green line) and then mask values below 0.024 (dashed horizontal black line) . Red dots show masked datapoints which are not used in the fit. Vertical grey lines are spectral lines. We find that masking using this method does not significantly improve the quality of our fits.}
\label{fig:sl_spec}
\end{figure}

The next masking method tested used a parameter called the sensitivity function $S_{\lambda}$, described in \citet{Sheminova2014}. $S_{\lambda}$ is a parameter that describes how much the flux at a given wavelength changes as a physical parameter (in this case [$M/H$]) is increased and decreased by a specified increment. The expression for $S_{\lambda}$ is: 
\begin{equation}
S_{\lambda} = 100\frac{R_{\lambda}(A+\Delta A)-R_{\lambda}(A-\Delta A)}{R_{\lambda}(A))}
,     R_{\lambda}=1-F
\end{equation}
with A representing our physical parameter, and F our normalized flux. The $S_{\lambda}$ value at each wavelength was evaluated for changes of $\pm$0.1 dex about the unmasked best-fit [$M/H$], and regions with $S_{\lambda}$ less than a specified cut-off value were excluded from the fit. When $S_{\lambda}$ was compared to the model spectrum, the regions of highest $S_{\lambda}$ (or regions with the strongest response to physical parameter changes) were typically located in the wings of stronger lines, and in center of weaker lines. This is consistent with where \citet{Sheminova2014} found the greatest $S_{\lambda}$ values.  Figure \ref{fig:sl_spec} shows an example of how masking is applied with the spectrum of calibrator star 2M19205338+3748282.

\begin{figure}
\includegraphics[width=7in]{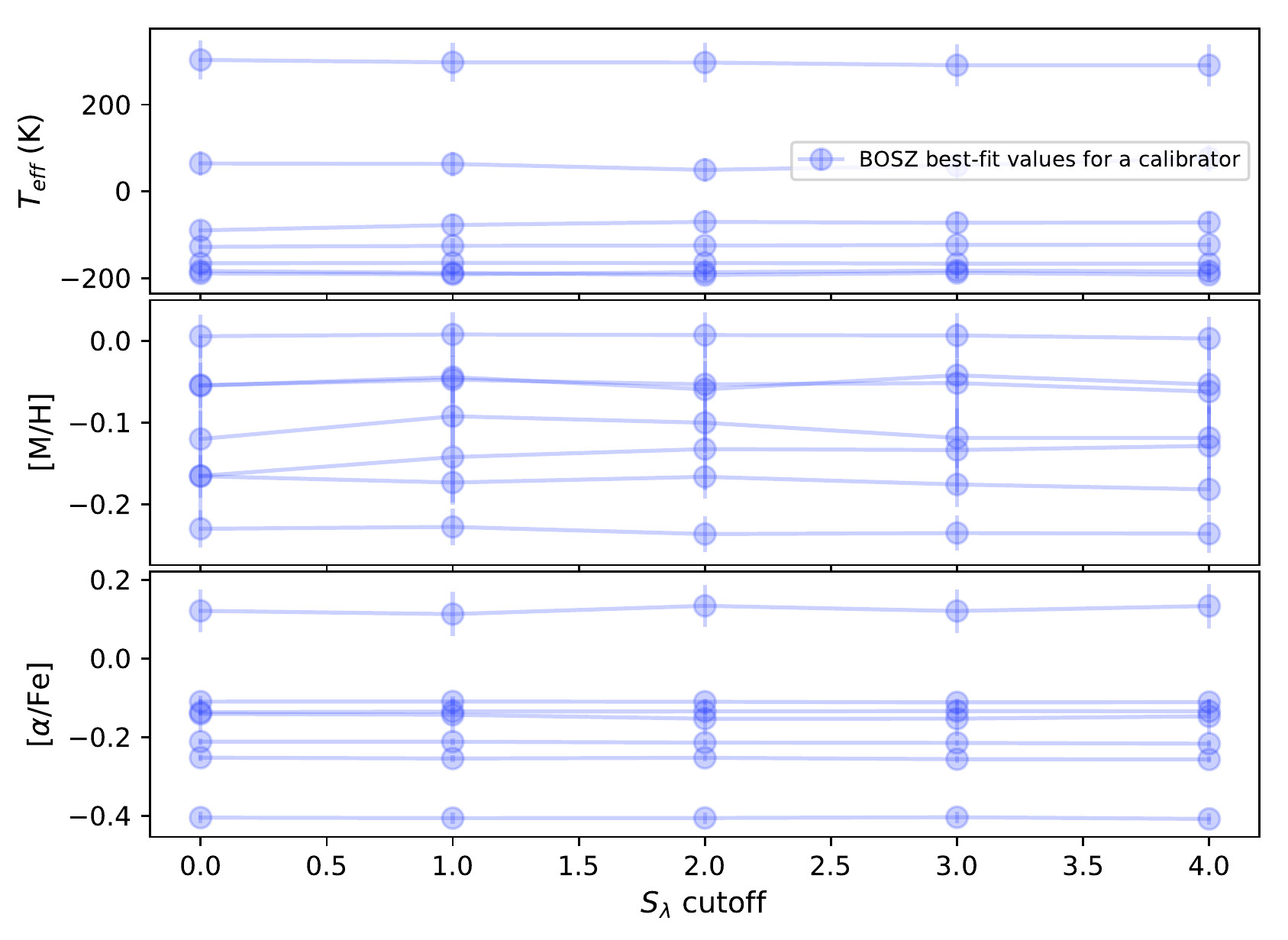}
\caption{\footnotesize The best-fit physical parameters for all calibrator stars for various threshold values of $S_{\lambda}$ used to mask and remove data points from the spectra before fitting. Higher thresholds of $S_{\lambda}$ remove more data points. log $g$ is fixed to the APOGEE DR16 value in these fits. Connected blue points are fits for the same calibrator star. Their is very little variance in the best-fit values for different $S_{\lambda}$ cutoffs, and no particular value improves the fit compared to no masking.}
\label{fig:sl_response}
\end{figure}

We run fits with different constraints on the cut-off value, and find that there is no significant improvement on fits with a fixed log $g$. We run fits with with $S_{\lambda}$ cutoff between 1 and 400 and log $g$ fixed to the reference value. The fraction of the data that is masked, and as a result excluded from the fits, ranges from 0$\%$ for the lowest $S_{\lambda}=1$ cutoff to 99$\%$ for the highest cutoff $S_{\lambda}=400$. We find no particular cut-off values significantly change the parameter estimates across all the calibrators (Figure \ref{fig:sl_response}). When log $g$ is allowed to be free, masking improves the estimation of log $g$ by reducing the scatter in the offsets from the reference values by 0.15 dex at a cut-off of $S_{\lambda}$=2.0 and higher. The quality of fits for [$M/H$], [$\alpha/Fe$], and $T_{eff}$ are not affected by masking. This indicates that $S_{\lambda}$-based masking may be helpful to explore if log $g$ values are important to derive from spectra in the future.

For the final masking method, the model was evaluated at the best-fit unmasked values, and the residuals between the observed spectrum and the unmasked model were found. The data points with residuals higher than a specified cut-off value were then masked. Nine residual cut-off values were tested, ranging from 17.5$\%$ to 2$\%$.  The quality of the fits does not improve with this masking method.

\subsection{Fitting with spectra convolved to a lower resolution}
\label{sect:convolve}
\begin{figure*}[h]
\begin{center}
\includegraphics[width=7in]{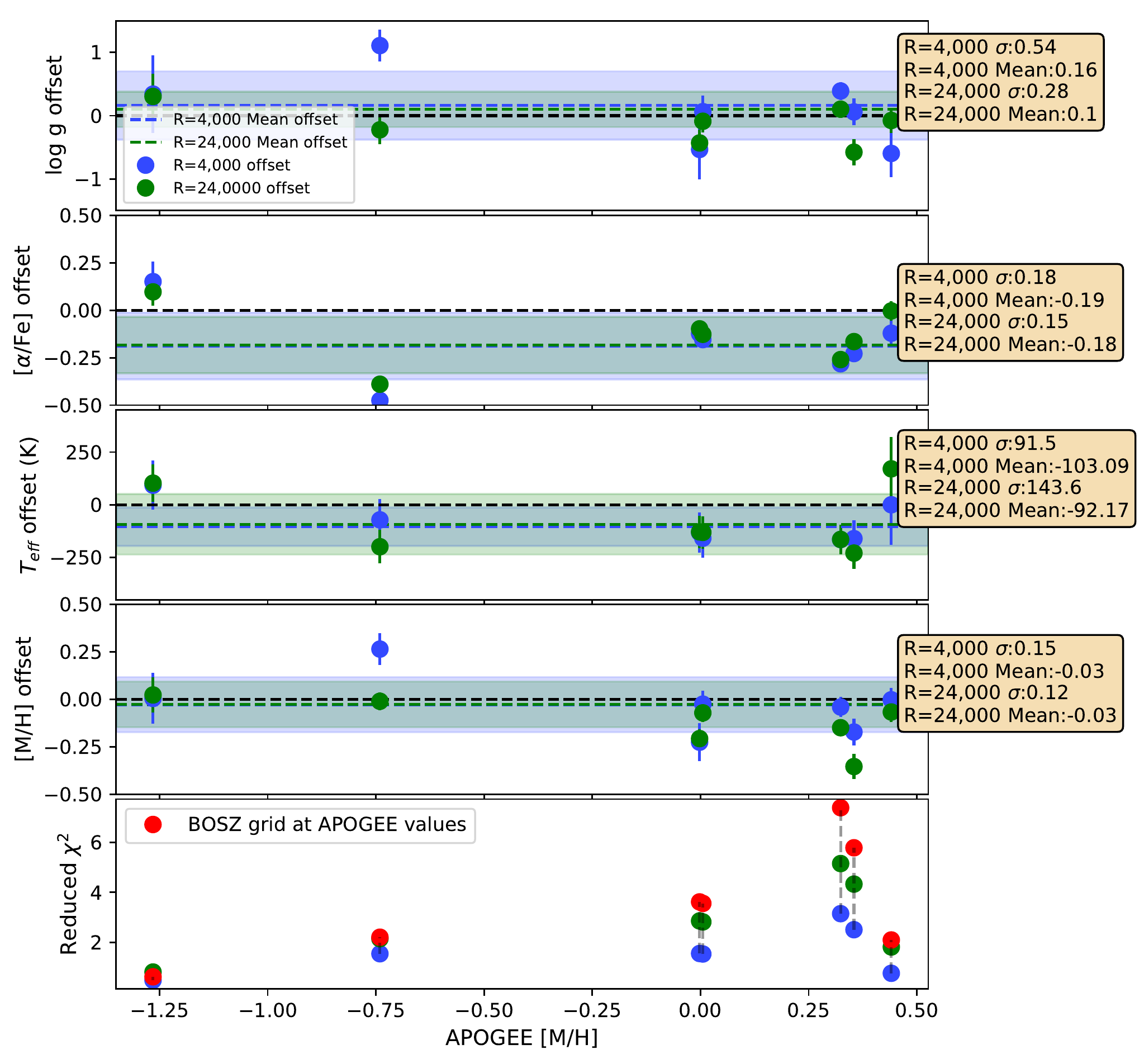}
\caption{\footnotesize We examine whether the best-fit physical parameters are biased at lower spectral resolution by convolving the R=24,000 NIRSPEC spectra to R=4,000. These mean offset and scatter at medium resolution (green) is similar to the mean and scatter at high resolution (blue). This indicates that medium resolution spectra and high resolution spectra have consistent results with out fitting method.}
\label{fig:r4000_offsets}
\end{center}
\end{figure*}

In order to ensure that the StarKit fitting produces similar best-fit parameters at different spectral resolutions, the calibrator star spectra are convolved with a Gaussian filter to R = 4,000, from R $\sim$ 24,000. The lower-resolution spectra is fit again with the BOSZ grid, and the results of the R = 4,000 calibrator star fits are compared with the R $\sim$ 24,000 fits to check for consistency. The lower-resolution fits are consistent to within ~1$\sigma$ of the higher-resolution fits using the BOSZ grid, as seen in Figure \ref{fig:r4000_offsets}.

\begin{figure*}[h]
\begin{center}
\includegraphics[width=7in]{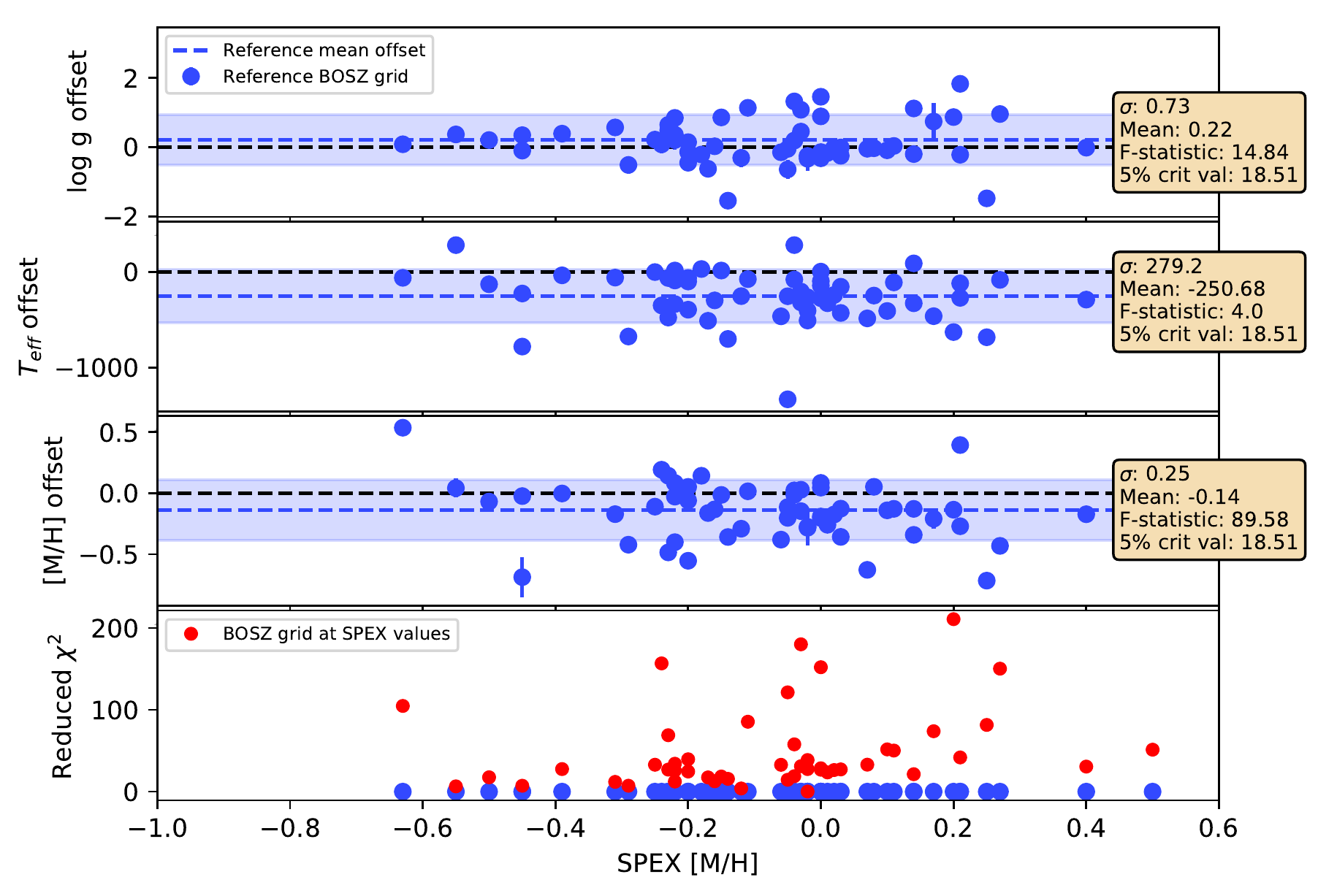}
\caption{\footnotesize We examine how well StarKit and the BOSZ grid fit stars at R~2,000 using the sample from  \citet{Rayner2009}. The fits at low resolution produce mean physical parameter offsets similar to the high resolution NIRSPEC sample, but with a factor of 2-3 times larger scatter. Reference physical values taken from \citet{Cesetti2013}.}
\label{fig:spex_offsets}
\end{center}
\end{figure*}

R$\sim$2,000 spectra from the SPEX stellar spectra library \citep{Rayner2009} are fit in the same manner as the NIRSPAO spectra, and offsets from the physical parameters from \citet{Cesetti2013} are calculated to provide an additional check for consistency at lower resolution. The offset plots for $T_{eff}$, log $g$, and [$M/H$] are shown in Figure \ref{fig:spex_offsets}. The mean offsets and scatter for $\Delta T_{eff}$ = -250$\pm$280 K, $\Delta \log g$ =  -0.22$\pm$0.73 dex, and [$\Delta M/H$] = -0.14$\pm$0.25 dex. The offset means are similar to what was found for the R $\sim$ 24,000 offsets, although with greater scatter. In summary, our fitting routine produces similar best-fit values at lower spectral resolution as at high spectral resolution, but with increasing scatter, suggesting there are larger systematic uncertainties at lower spectral resolution. 

\subsection{Fit robustness at different signal-to-noise ratios}
\label{sect:stat_err_vs_snr}

\begin{figure*}[h]
\begin{center}
\includegraphics[width=7in]{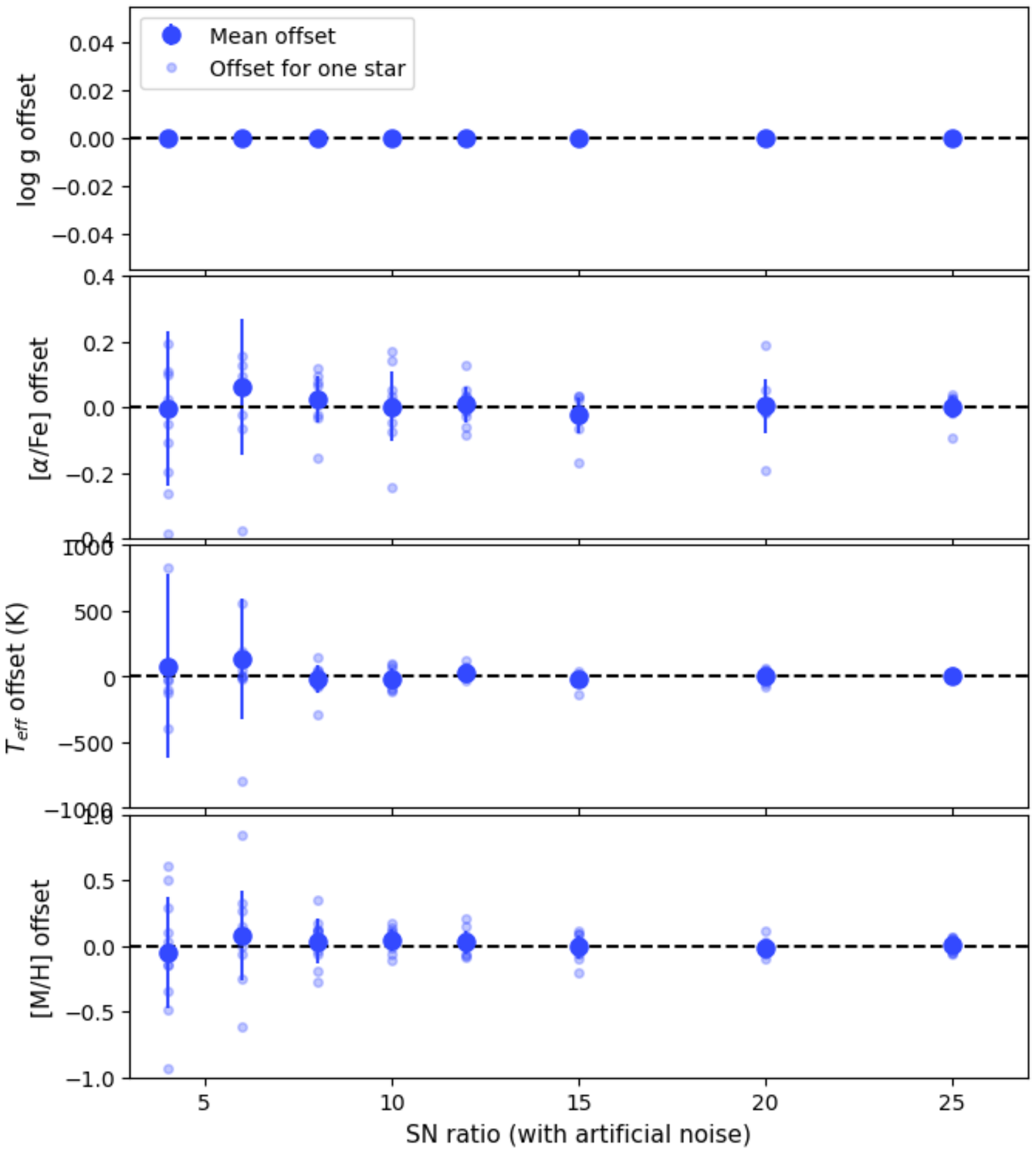}
\caption{\footnotesize We examine how the best-fit physical parameters from StarKit vary at lower SN ratios by artificially lowering the SN ratio by adding noise. We find that the scatter in the offsets from the values (individual offsets are transparent blue points, mean offsets are opaque blue points) without additional noise does not reach the scale of the systematic error until SN~10, indicating our fits are robust to a SN ratio of 10.}
\label{fig:mean_offset_vs_sn}
\end{center}
\end{figure*}

We find that the StarKit fits are robust at a signal-to-noise ratio (SNR) greater than $\sim10$. To determine this threshold, we performed a series of Monte Carlo simulations by adding artificial noise until the best-fit parameters become systematically different than without the extra noise. We added noise to the 11 calibrator spectra to reduce their SNR to between 2 and 25, and fit the new spectra with fixed log $g$, and the BOSZ grid. We find that the offsets do not diverge from the best-fit value without extra noise at SNR above 10 (Figure \ref{fig:mean_offset_vs_sn}). 

\begin{table*}[h]
\begin{center}
\caption{Mean offsets and standard deviations for fitting experiments}
\label{tab:mean_offsets}
\begin{tabular}{cccccc}
\hline\hline
Experiment & Grid & $T_{eff}$ & log $g$ & [$M/H$] & [$\alpha/Fe$] \\
 & & K & & &  \\
\hline
$v_{rad}$, log $g$ fixed & BOSZ & -117.0$\pm$180.0 & 0.0$\pm$0.0 & -0.16$\pm$0.09 & -0.2$\pm$0.15 \\
log $g$ fixed & BOSZ & -129.0$\pm$158.0 & 0.0$\pm$0.0 & -0.15$\pm$0.1 & -0.18$\pm$0.16 \\
$v_{rad}$ fixed & BOSZ & -172.0$\pm$149.0 & -0.27$\pm$0.67 & -0.21$\pm$0.16 & -0.19$\pm$0.15 \\
R fixed & BOSZ & -145.0$\pm$178.0 & -0.09$\pm$0.45 & -0.18$\pm$0.16 & -0.18$\pm$0.18 \\
R=4000 convolution & BOSZ & -103.0$\pm$91.0 & 0.16$\pm$0.54 & -0.03$\pm$0.15 & -0.19$\pm$0.18 \\
SN$\sim$12 & BOSZ & -93.0$\pm$120.0 & -0.08$\pm$0.61 & -0.04$\pm$0.2 & -0.16$\pm$0.16 \\
$S_{\lambda}$ mask & BOSZ & -90.0$\pm$120.0 & -0.11$\pm$0.5 & -0.07$\pm$0.19 & -0.16$\pm$0.15 \\
APOGEE spectrum & BOSZ & -130.0$\pm$139.0 & -0.09$\pm$0.68 & -0.2$\pm$0.32 & -0.15$\pm$0.16 \\
R fixed & PHOENIX & -147.0$\pm$250.0 & -1.22$\pm$1.07 & -0.2$\pm$0.29 & 0.19$\pm$0.24 \\
APOGEE spectrum & PHOENIX & -22.0$\pm$252.0 & -0.9$\pm$1.03 & -0.22$\pm$0.27 & 0.22$\pm$0.21 \\
\hline \hline
\end{tabular}
\end{center}
\end{table*}

\subsection{Spectral Atlas}
\label{sec:atlas}
\begin{figure}
\figurenum{18}
\plotone{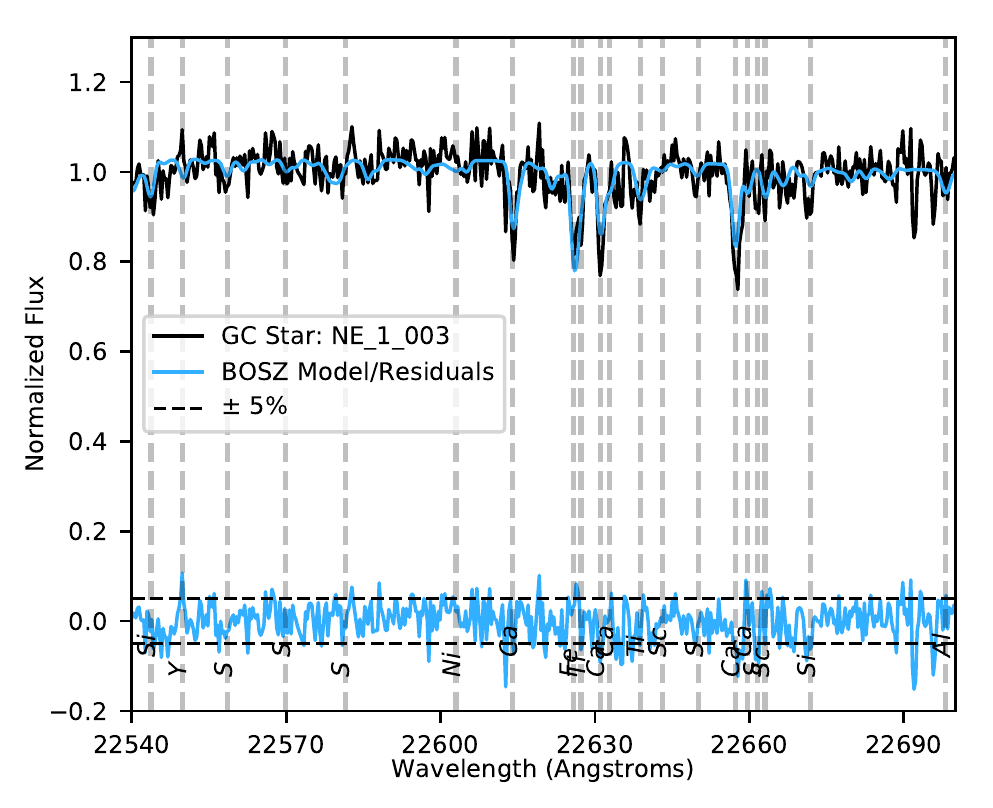}
\caption{Best-fit model (log $g$ fixed, BOSZ grid) compared to observed spectrum, with residuals.}
\label{fig:spec_example}
\end{figure}

An example of the  observed and best fit spectra for the Galactic center sources and all calibrations in each of the NIRSPAO orders is shown in Figure \ref{fig:spec_example}. See online journal for the complete spectral atlas.

\figsetstart
\figsetnum{18}
\figsettitle{The complete spectral atlas figure set (39 images) is available in the online journal}

\figsetgrpstart
\figsetgrpnum{18.1}
\figsetgrptitle{NE1-1 003 spectral order 34}
\figsetplot{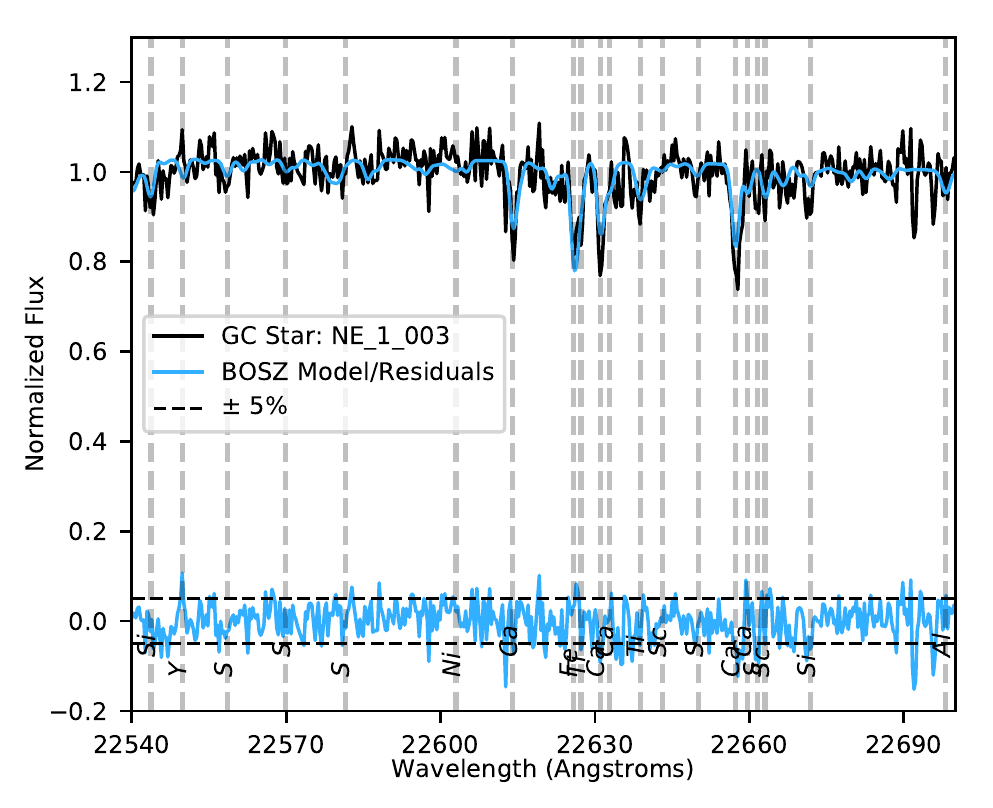}
\figsetgrpnote{Best-fit model (log $g$ fixed, BOSZ grid) compared to observed spectrum, with residuals.}
\figsetgrpend

\figsetgrpstart
\figsetgrpnum{18.2}
\figsetgrptitle{NE1-1 003 spectral order 35}
\figsetplot{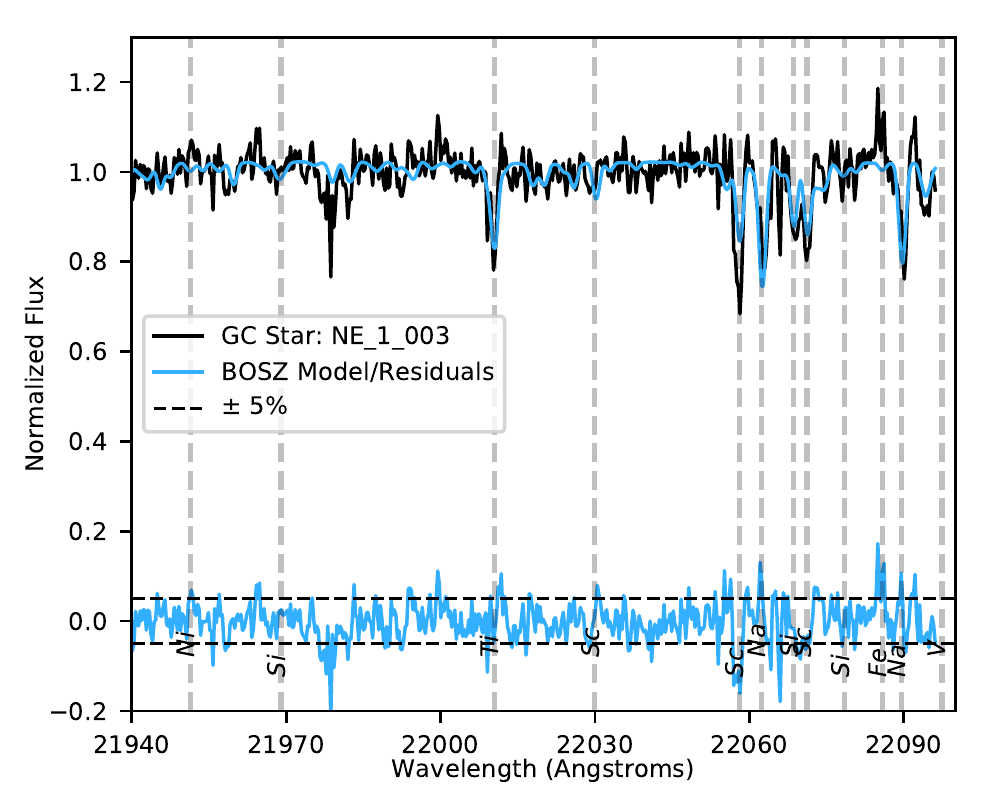}
\figsetgrpnote{Best-fit model (log $g$ fixed, BOSZ grid) compared to observed spectrum, with residuals.}
\figsetgrpend

\figsetgrpstart
\figsetgrpnum{18.3}
\figsetgrptitle{NE1-1 003 spectral order 36}
\figsetplot{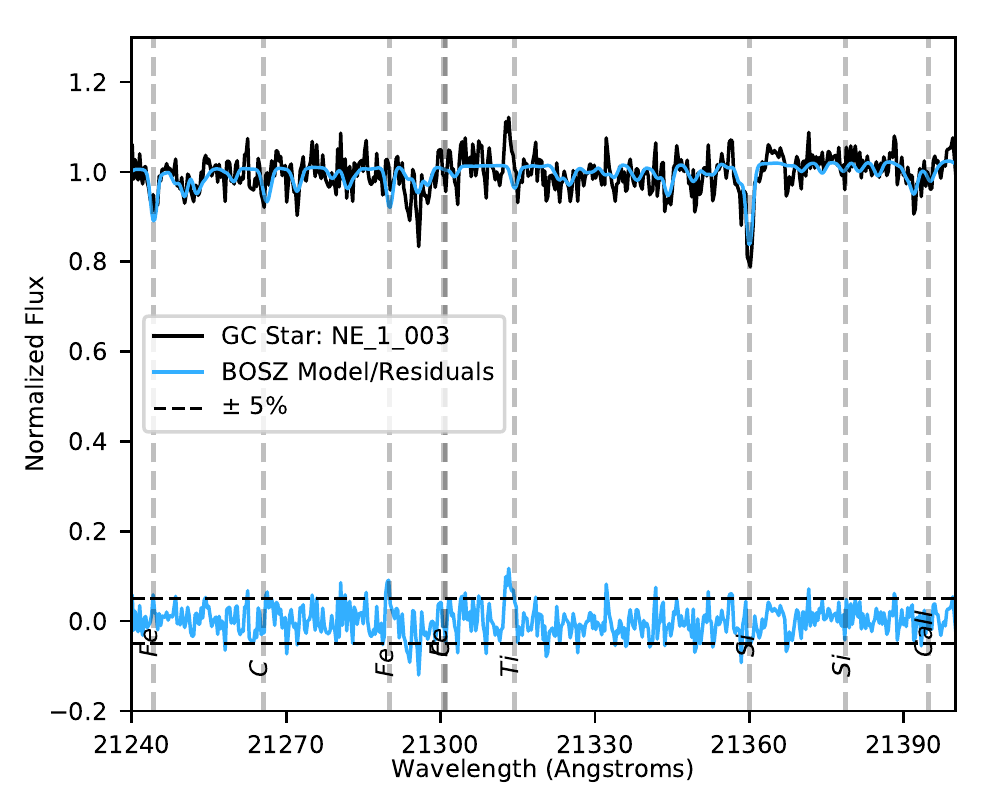}
\figsetgrpnote{Best-fit model (log $g$ fixed, BOSZ grid) compared to observed spectrum, with residuals.}
\figsetgrpend

\figsetgrpstart
\figsetgrpnum{18.4}
\figsetgrptitle{N2-1 002 spectral order 34}
\figsetplot{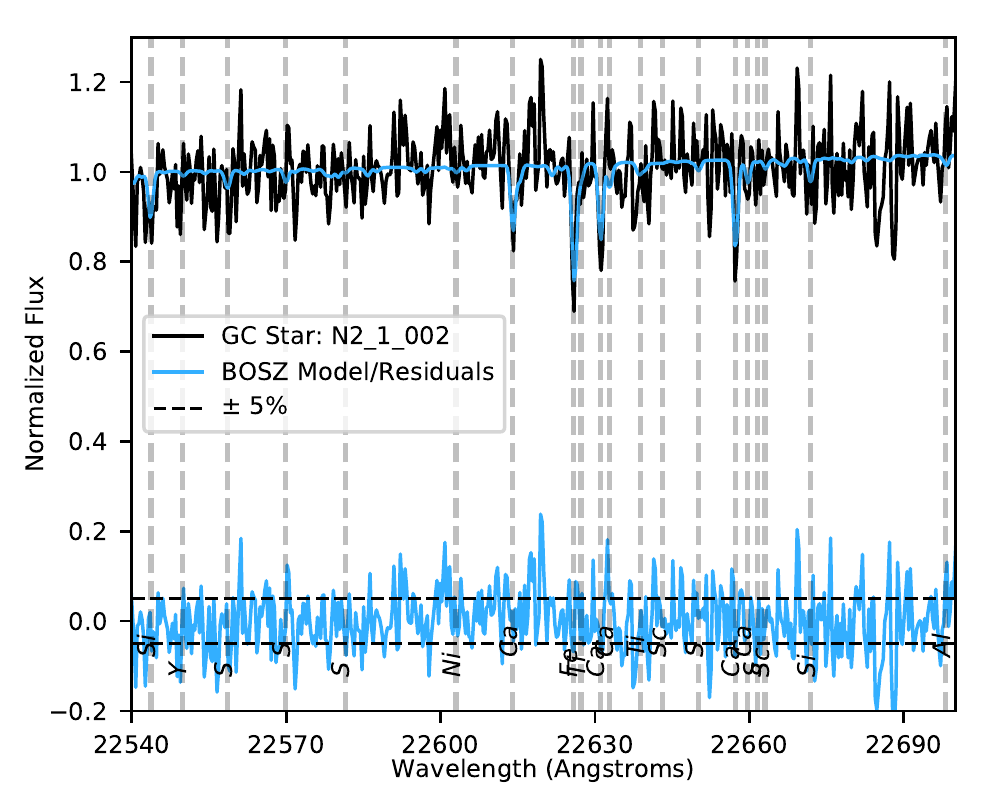}
\figsetgrpnote{Best-fit model (log $g$ fixed, BOSZ grid) compared to observed spectrum, with residuals.}
\figsetgrpend

\figsetgrpstart
\figsetgrpnum{18.5}
\figsetgrptitle{N2-1 002 spectral order 35}
\figsetplot{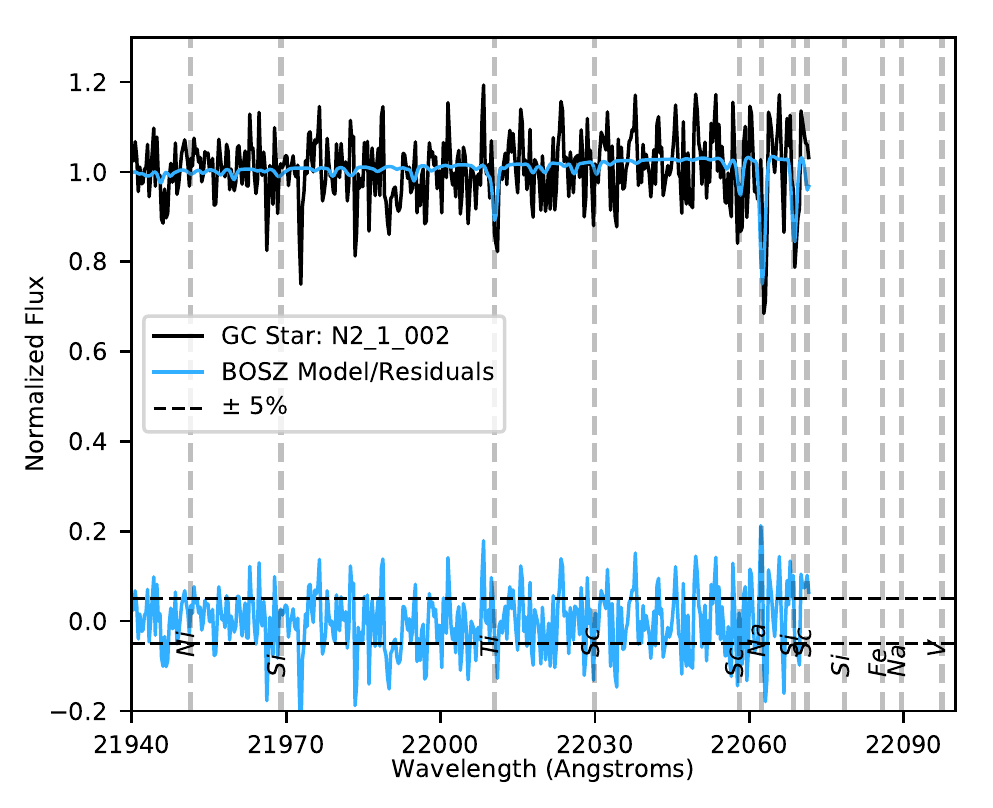}
\figsetgrpnote{Best-fit model (log $g$ fixed, BOSZ grid) compared to observed spectrum, with residuals.}
\figsetgrpend

\figsetgrpstart
\figsetgrpnum{18.6}
\figsetgrptitle{N2-1 002 spectral order 36}
\figsetplot{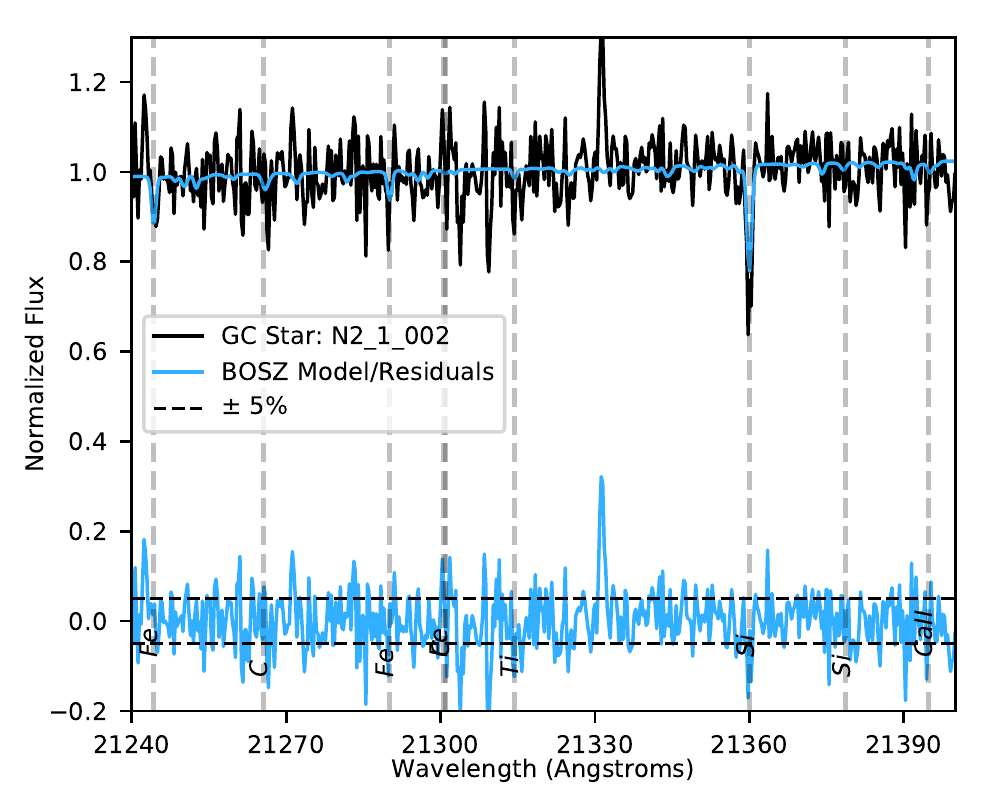}
\figsetgrpnote{Best-fit model (log $g$ fixed, BOSZ grid) compared to observed spectrum, with residuals.}
\figsetgrpend

\figsetgrpstart
\figsetgrpnum{18.7}
\figsetgrptitle{2M19205338+3748282 spectral order 34}
\figsetplot{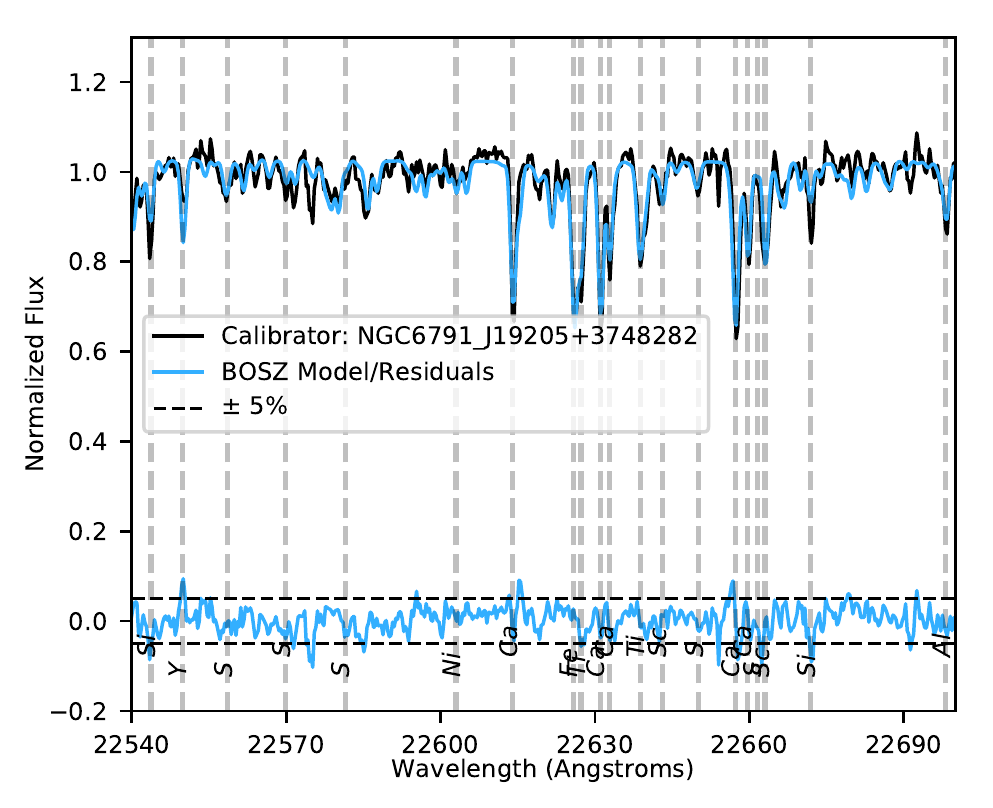}
\figsetgrpnote{Best-fit model (log $g$ fixed, BOSZ grid) compared to observed spectrum, with residuals.}
\figsetgrpend

\figsetgrpstart
\figsetgrpnum{18.8}
\figsetgrptitle{2M19205338+3748282 spectral order 35}
\figsetplot{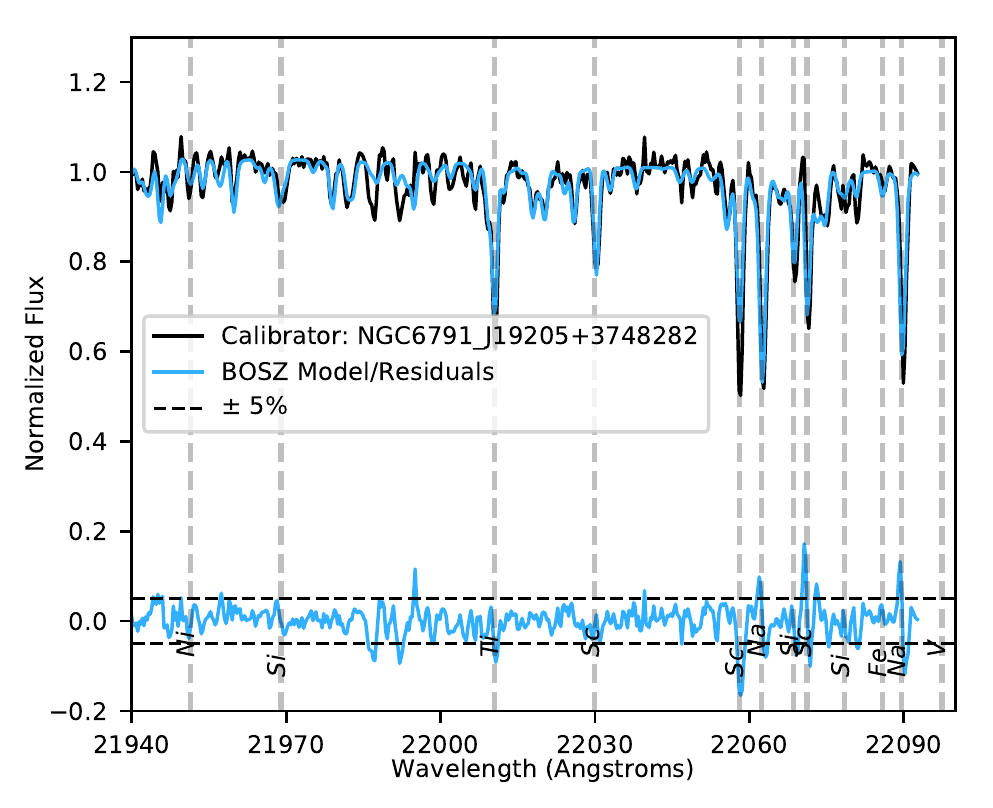}
\figsetgrpnote{Best-fit model (log $g$ fixed, BOSZ grid) compared to observed spectrum, with residuals.}
\figsetgrpend

\figsetgrpstart
\figsetgrpnum{18.9}
\figsetgrptitle{2M19205338+3748282 spectral order 36}
\figsetplot{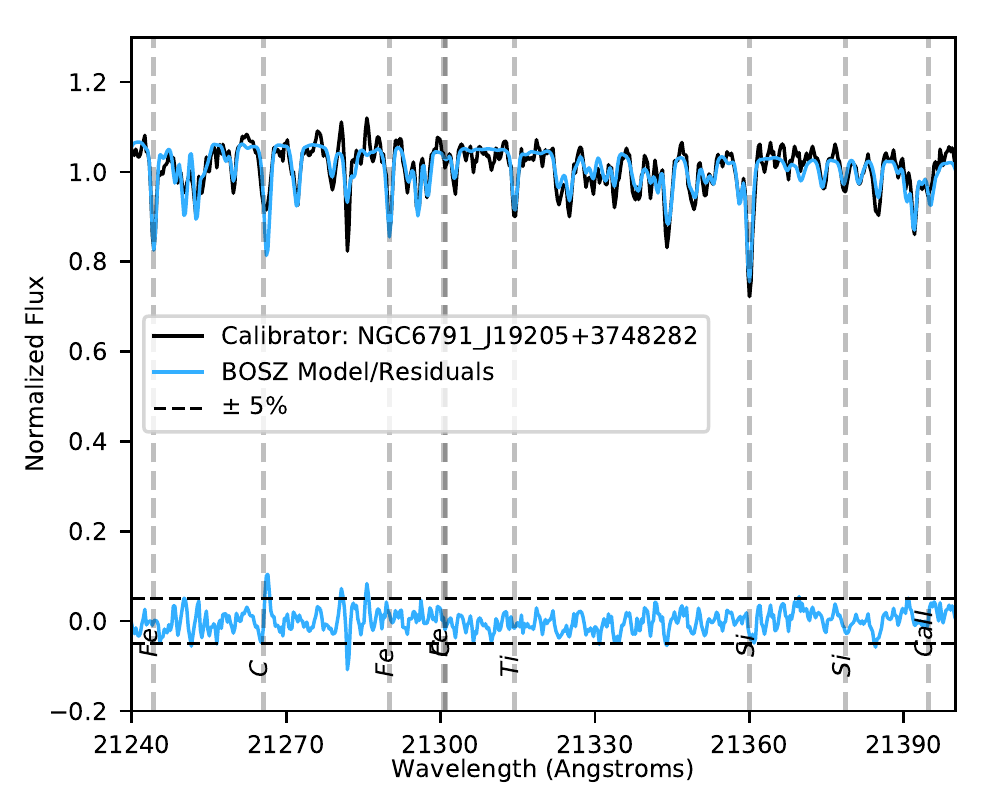}
\figsetgrpnote{Best-fit model (log $g$ fixed, BOSZ grid) compared to observed spectrum, with residuals.}
\figsetgrpend

\figsetgrpstart
\figsetgrpnum{18.10}
\figsetgrptitle{2M19411705+4010517 spectral order 34}
\figsetplot{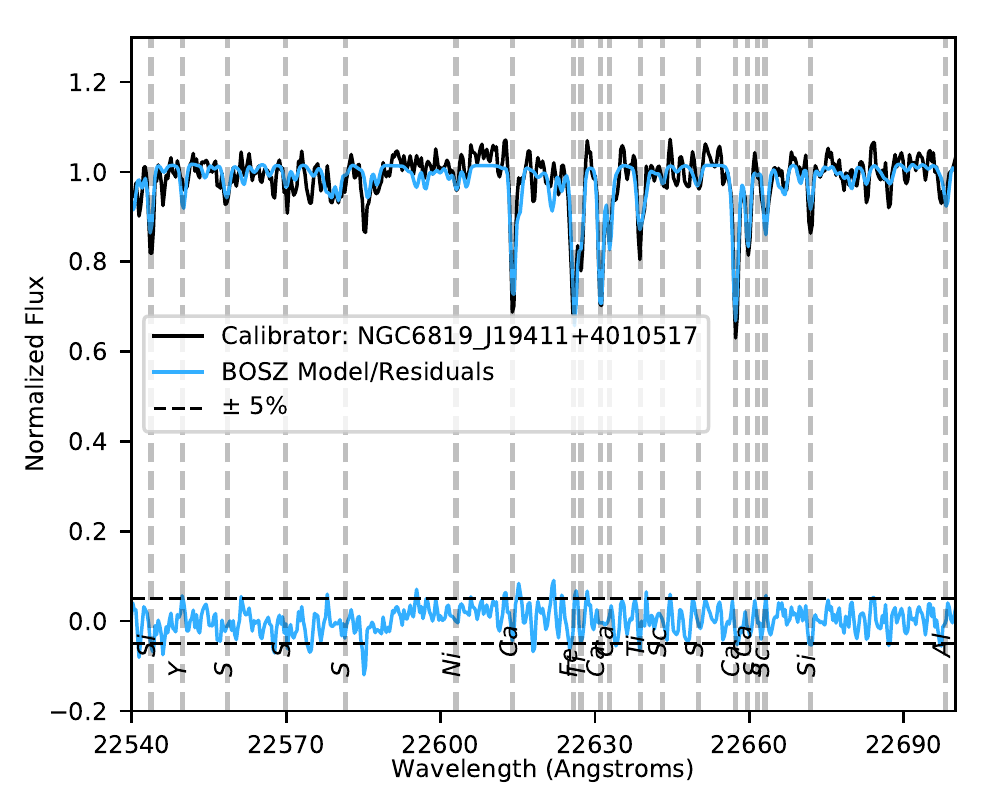}
\figsetgrpnote{Best-fit model (log $g$ fixed, BOSZ grid) compared to observed spectrum, with residuals.}
\figsetgrpend

\figsetgrpstart
\figsetgrpnum{18.11}
\figsetgrptitle{2M19411705+4010517 spectral order 35}
\figsetplot{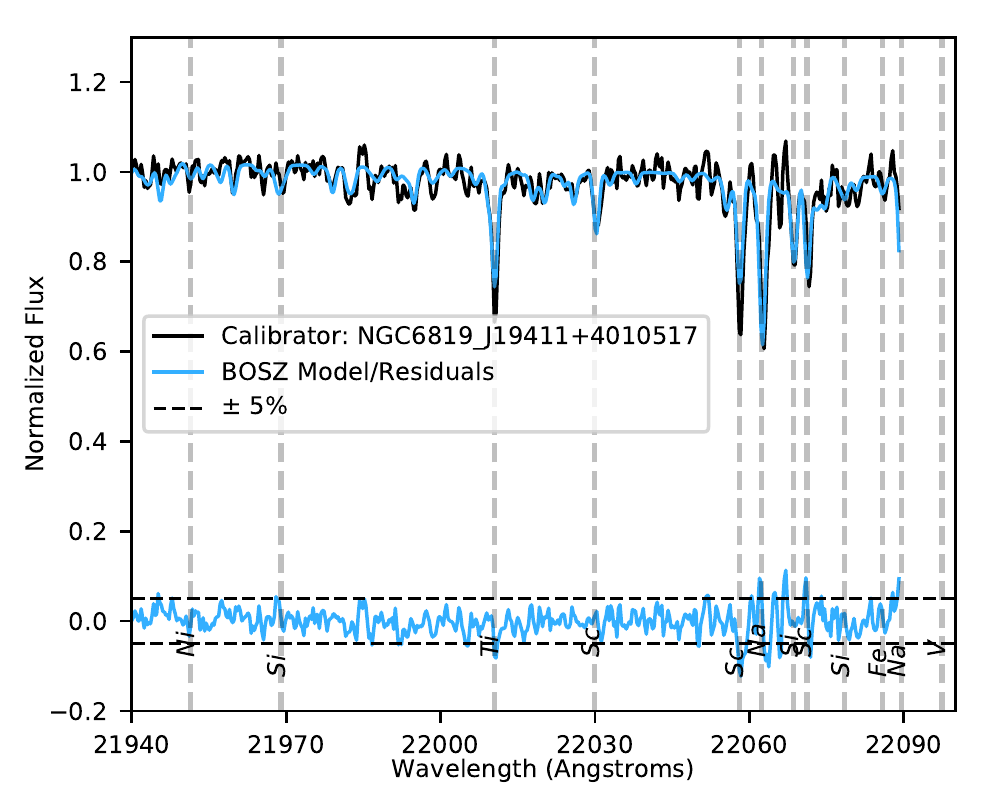}
\figsetgrpnote{Best-fit model (log $g$ fixed, BOSZ grid) compared to observed spectrum, with residuals.}
\figsetgrpend

\figsetgrpstart
\figsetgrpnum{18.12}
\figsetgrptitle{2M19411705+4010517 spectral order 36}
\figsetplot{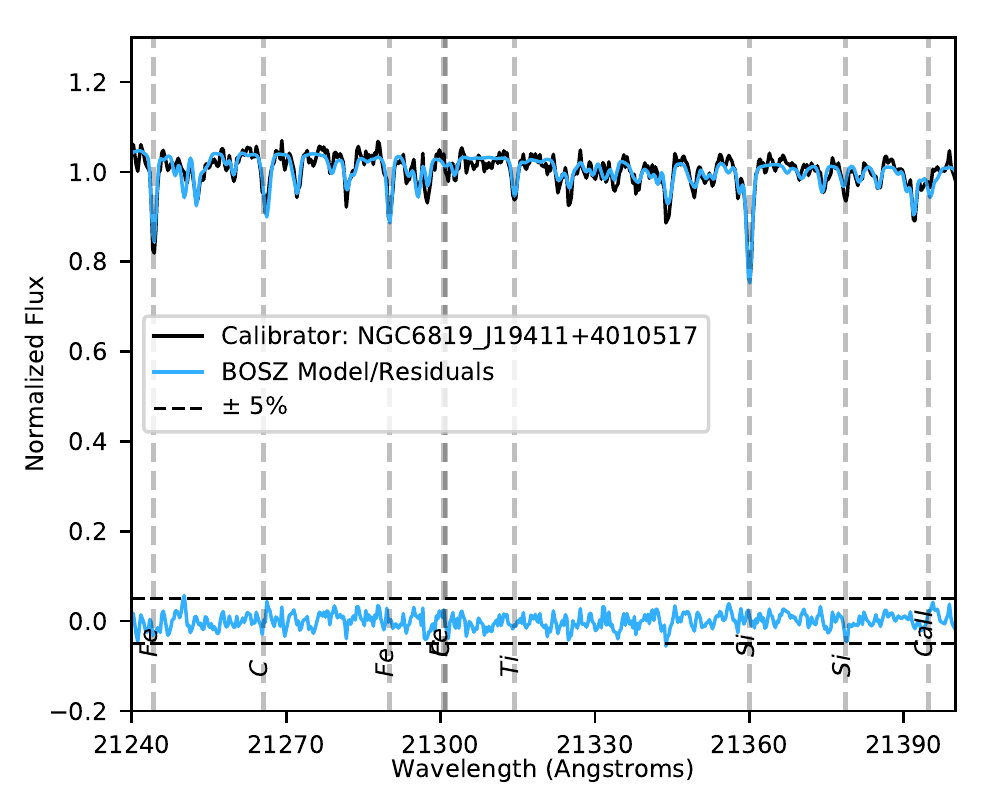}
\figsetgrpnote{Best-fit model (log $g$ fixed, BOSZ grid) compared to observed spectrum, with residuals.}
\figsetgrpend

\figsetgrpstart
\figsetgrpnum{18.13}
\figsetgrptitle{2M15190324+0208032 spectral order 34}
\figsetplot{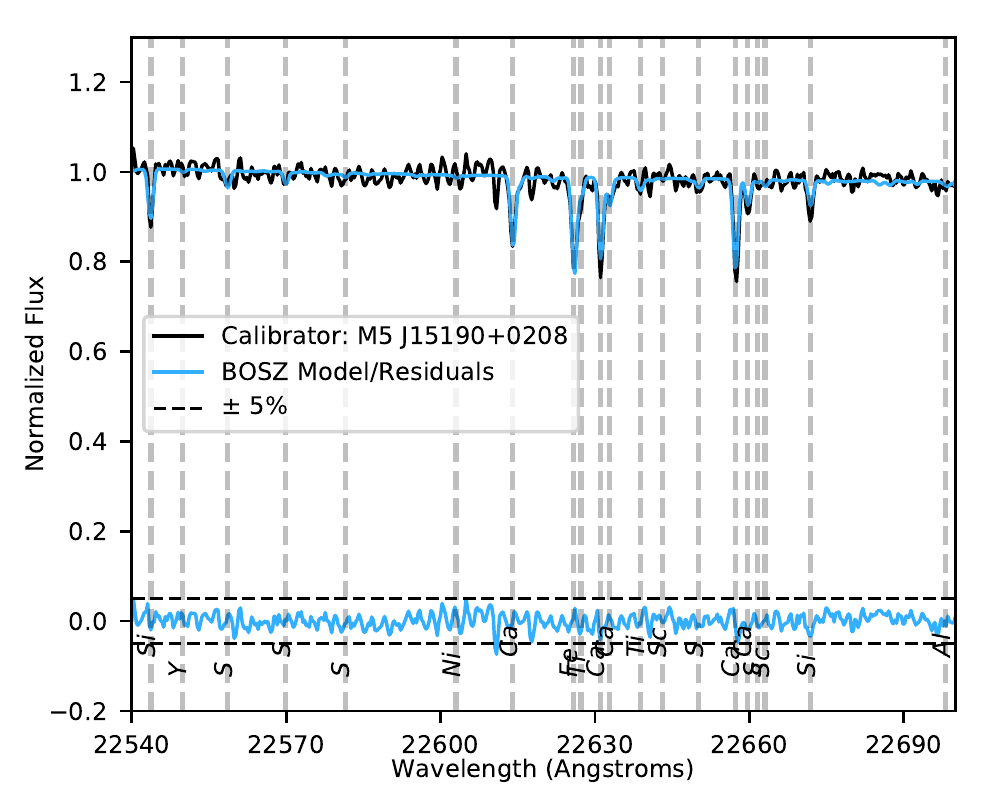}
\figsetgrpnote{Best-fit model (log $g$ fixed, BOSZ grid) compared to observed spectrum, with residuals.}
\figsetgrpend

\figsetgrpstart
\figsetgrpnum{18.14}
\figsetgrptitle{2M15190324+0208032 spectral order 35}
\figsetplot{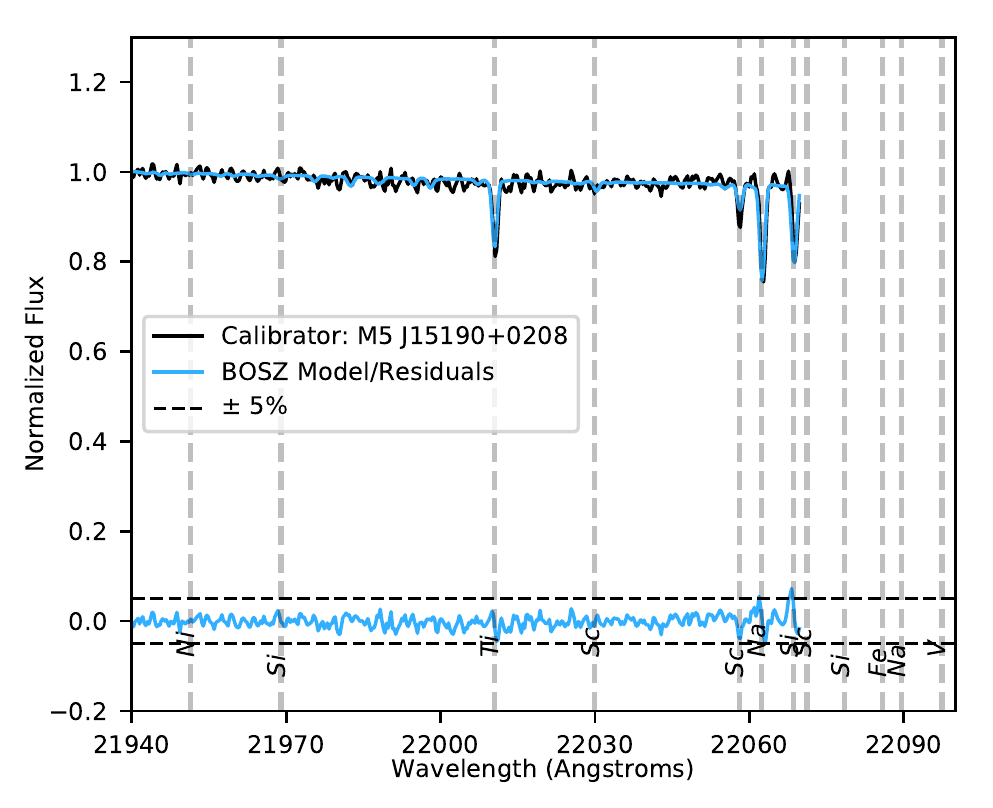}
\figsetgrpnote{Best-fit model (log $g$ fixed, BOSZ grid) compared to observed spectrum, with residuals.}
\figsetgrpend

\figsetgrpstart
\figsetgrpnum{18.15}
\figsetgrptitle{2M15190324+0208032 spectral order 36}
\figsetplot{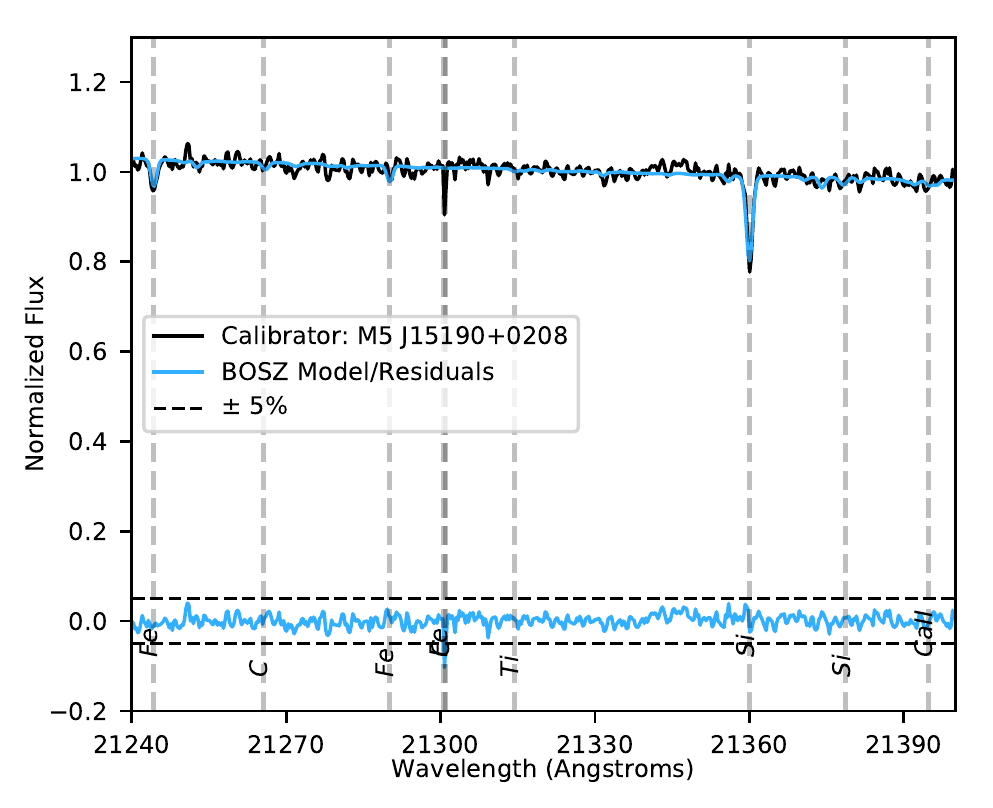}
\figsetgrpnote{Best-fit model (log $g$ fixed, BOSZ grid) compared to observed spectrum, with residuals.}
\figsetgrpend

\figsetgrpstart
\figsetgrpnum{18.16}
\figsetgrptitle{2M19213390+3750202 spectral order 34}
\figsetplot{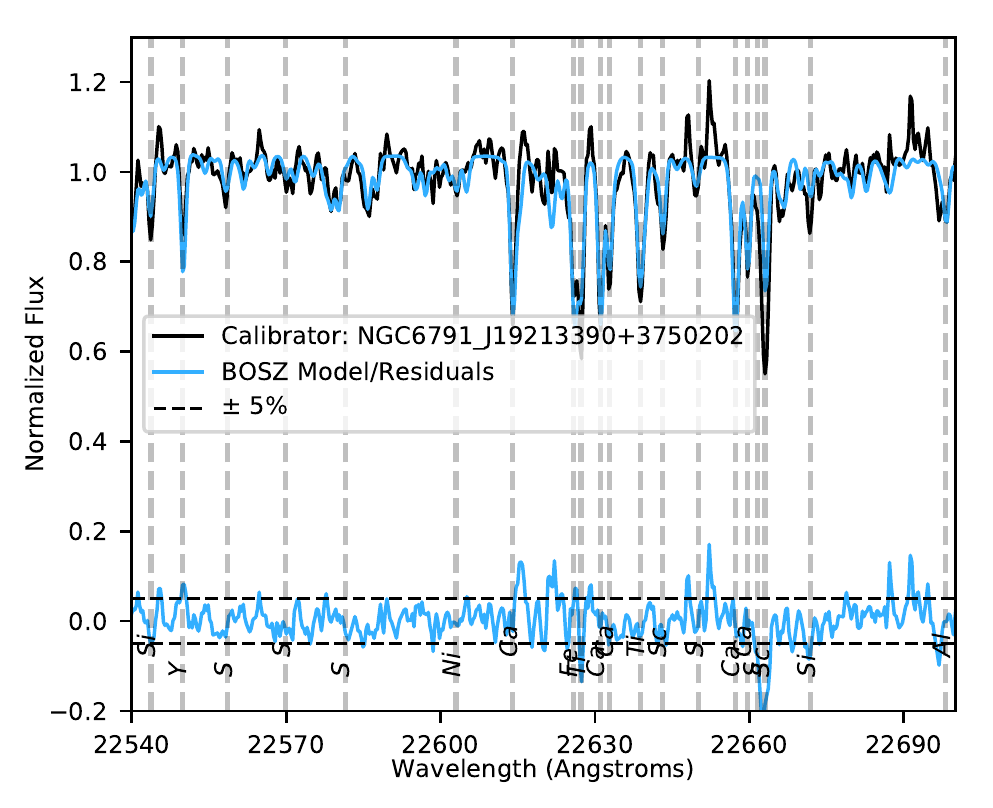}
\figsetgrpnote{Best-fit model (log $g$ fixed, BOSZ grid) compared to observed spectrum, with residuals.}
\figsetgrpend

\figsetgrpstart
\figsetgrpnum{18.17}
\figsetgrptitle{2M19213390+3750202 spectral order 35}
\figsetplot{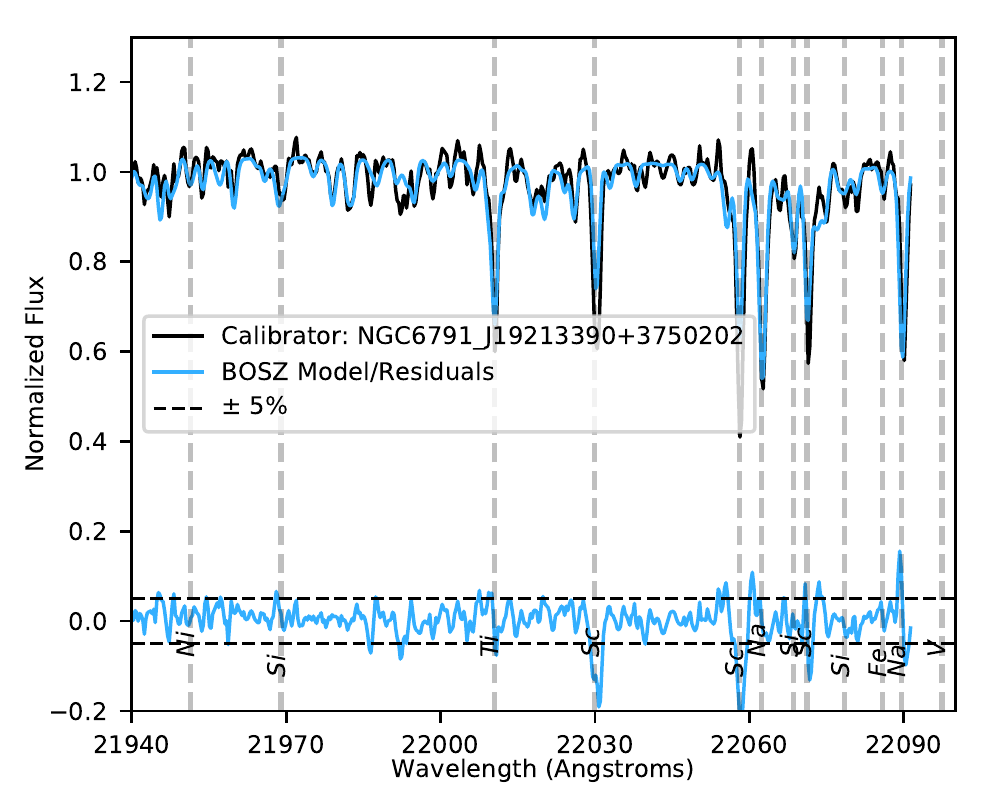}
\figsetgrpnote{Best-fit model (log $g$ fixed, BOSZ grid) compared to observed spectrum, with residuals.}
\figsetgrpend

\figsetgrpstart
\figsetgrpnum{18.18}
\figsetgrptitle{2M19213390+3750202 spectral order 36}
\figsetplot{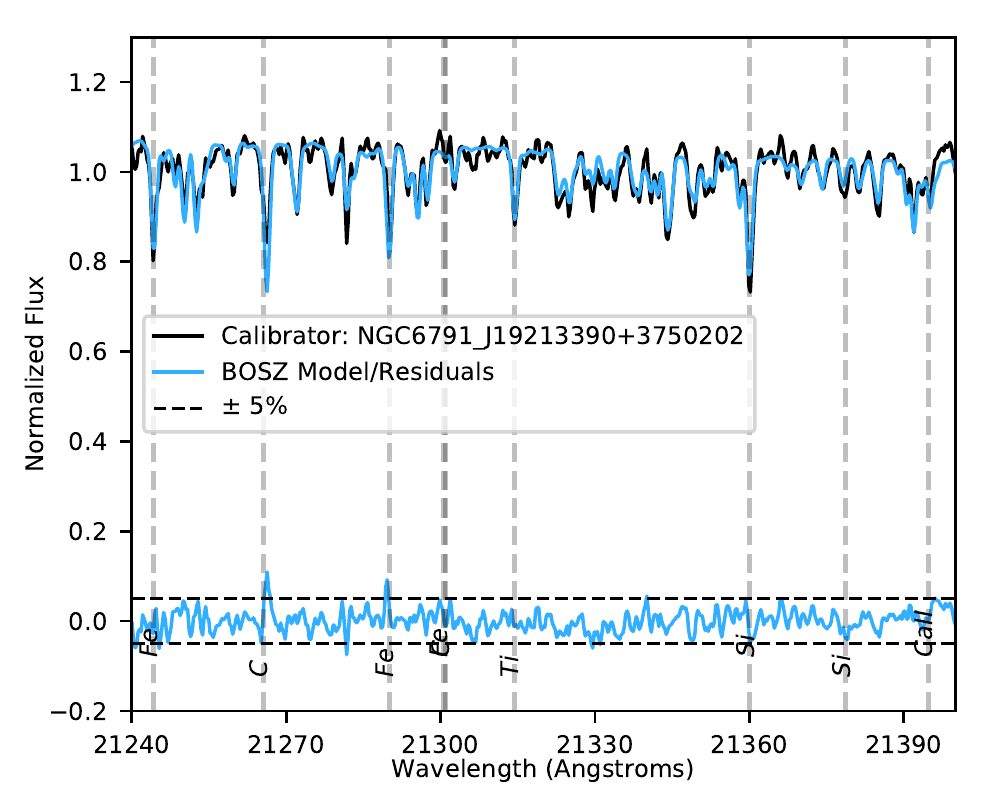}
\figsetgrpnote{Best-fit model (log $g$ fixed, BOSZ grid) compared to observed spectrum, with residuals.}
\figsetgrpend

\figsetgrpstart
\figsetgrpnum{18.19}
\figsetgrptitle{2M19413439+4017482 spectral order 34}
\figsetplot{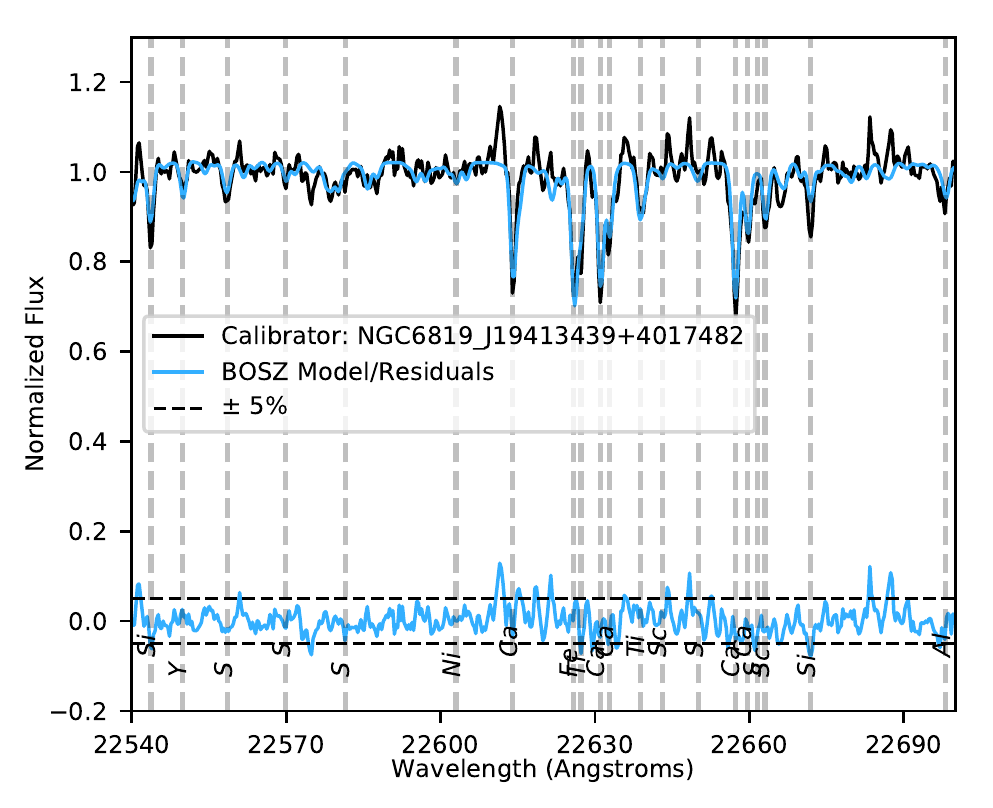}
\figsetgrpnote{Best-fit model (log $g$ fixed, BOSZ grid) compared to observed spectrum, with residuals.}
\figsetgrpend

\figsetgrpstart
\figsetgrpnum{18.20}
\figsetgrptitle{2M19413439+4017482 spectral order 35}
\figsetplot{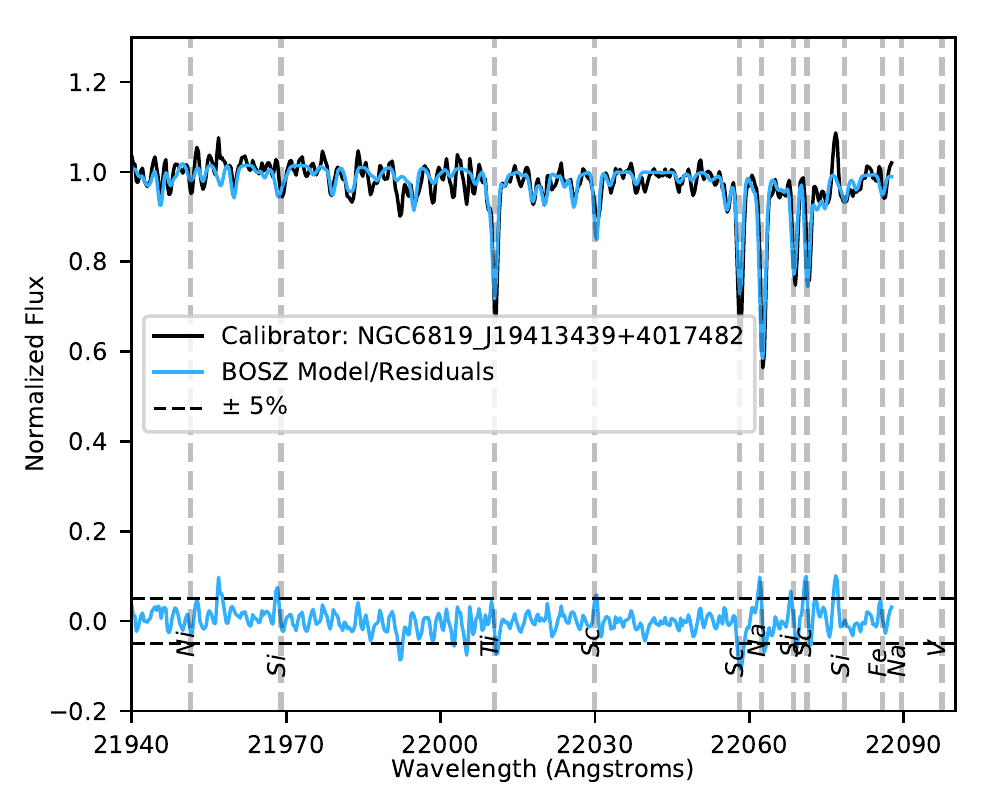}
\figsetgrpnote{Best-fit model (log $g$ fixed, BOSZ grid) compared to observed spectrum, with residuals.}
\figsetgrpend

\figsetgrpstart
\figsetgrpnum{18.21}
\figsetgrptitle{2M19413439+4017482 spectral order 36}
\figsetplot{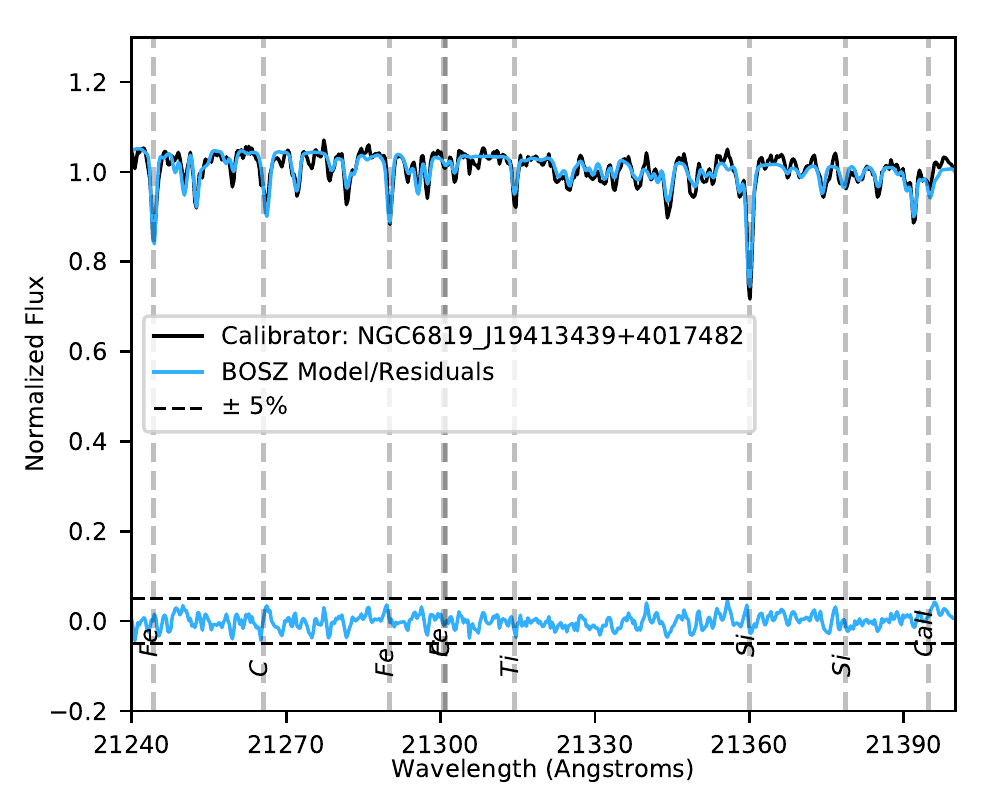}
\figsetgrpnote{Best-fit model (log $g$ fixed, BOSZ grid) compared to observed spectrum, with residuals.}
\figsetgrpend

\figsetgrpstart
\figsetgrpnum{18.22}
\figsetgrptitle{2M19534827+1848021 spectral order 34}
\figsetplot{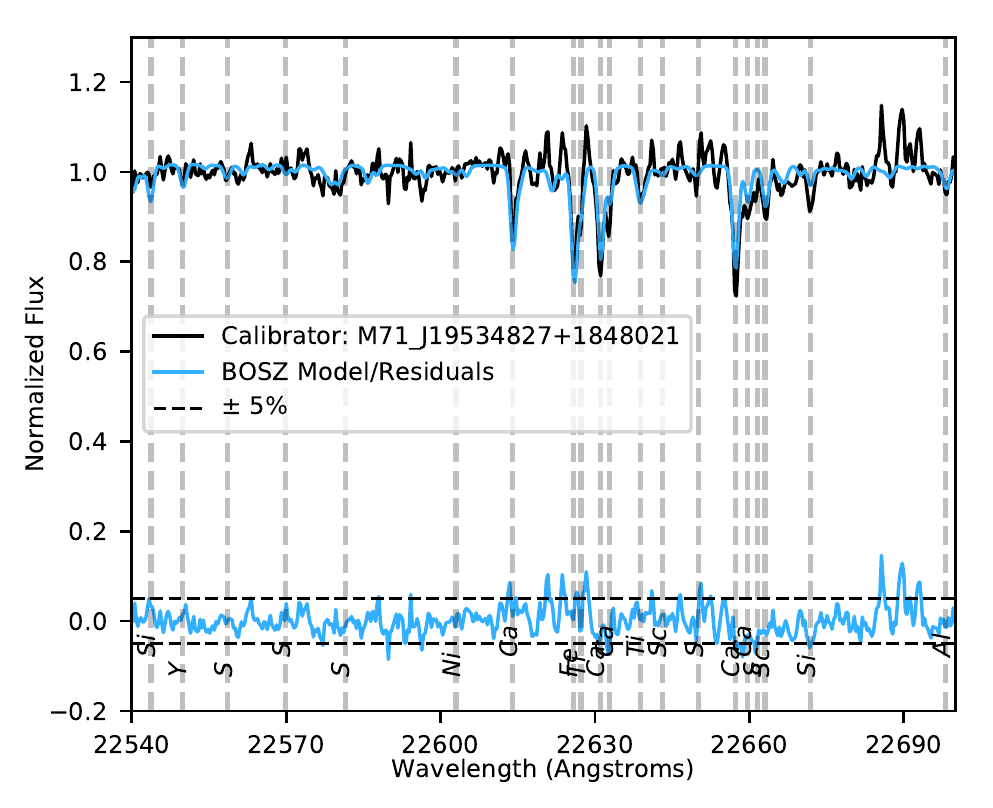}
\figsetgrpnote{Best-fit model (log $g$ fixed, BOSZ grid) compared to observed spectrum, with residuals.}
\figsetgrpend

\figsetgrpstart
\figsetgrpnum{18.23}
\figsetgrptitle{2M19534827+1848021 spectral order 35}
\figsetplot{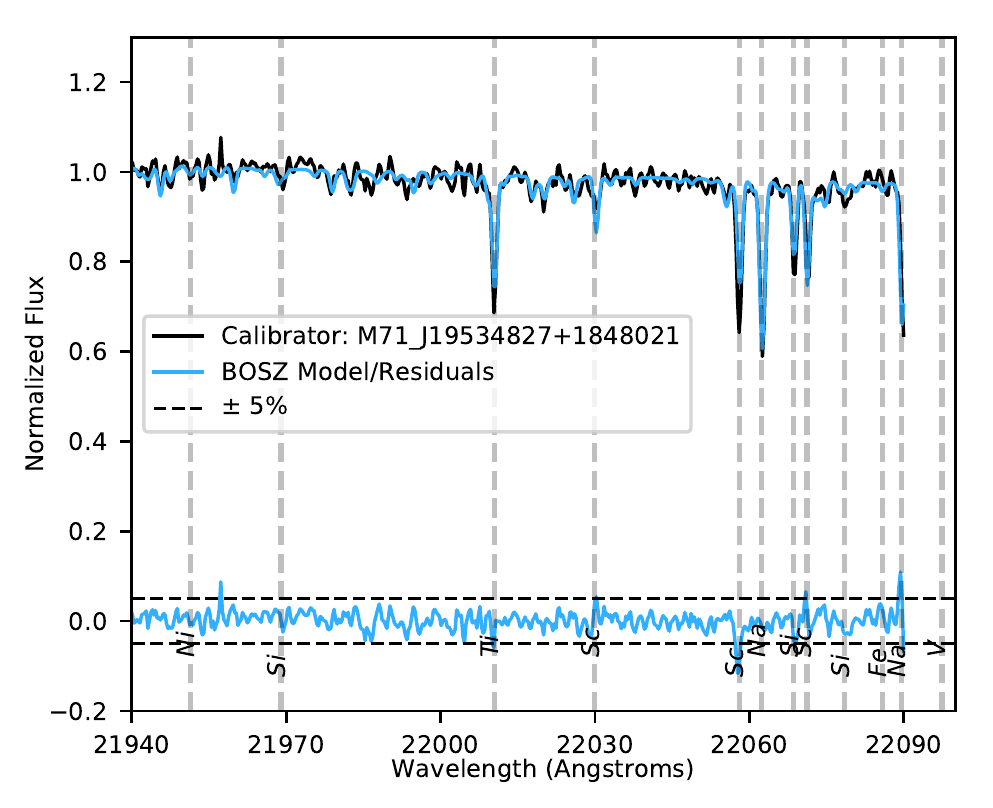}
\figsetgrpnote{Best-fit model (log $g$ fixed, BOSZ grid) compared to observed spectrum, with residuals.}
\figsetgrpend

\figsetgrpstart
\figsetgrpnum{18.24}
\figsetgrptitle{2M19534827+1848021 spectral order 36}
\figsetplot{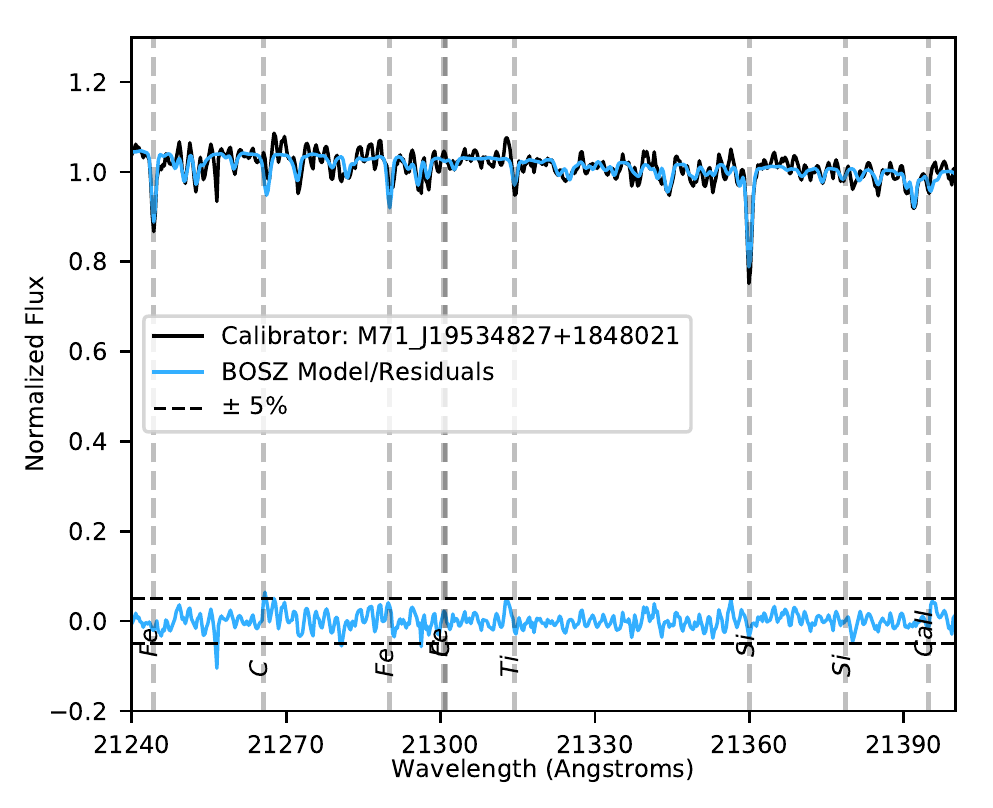}
\figsetgrpnote{Best-fit model (log $g$ fixed, BOSZ grid) compared to observed spectrum, with residuals.}
\figsetgrpend

\figsetgrpstart
\figsetgrpnum{18.25}
\figsetgrptitle{2M18482584+4828027 spectral order 34}
\figsetplot{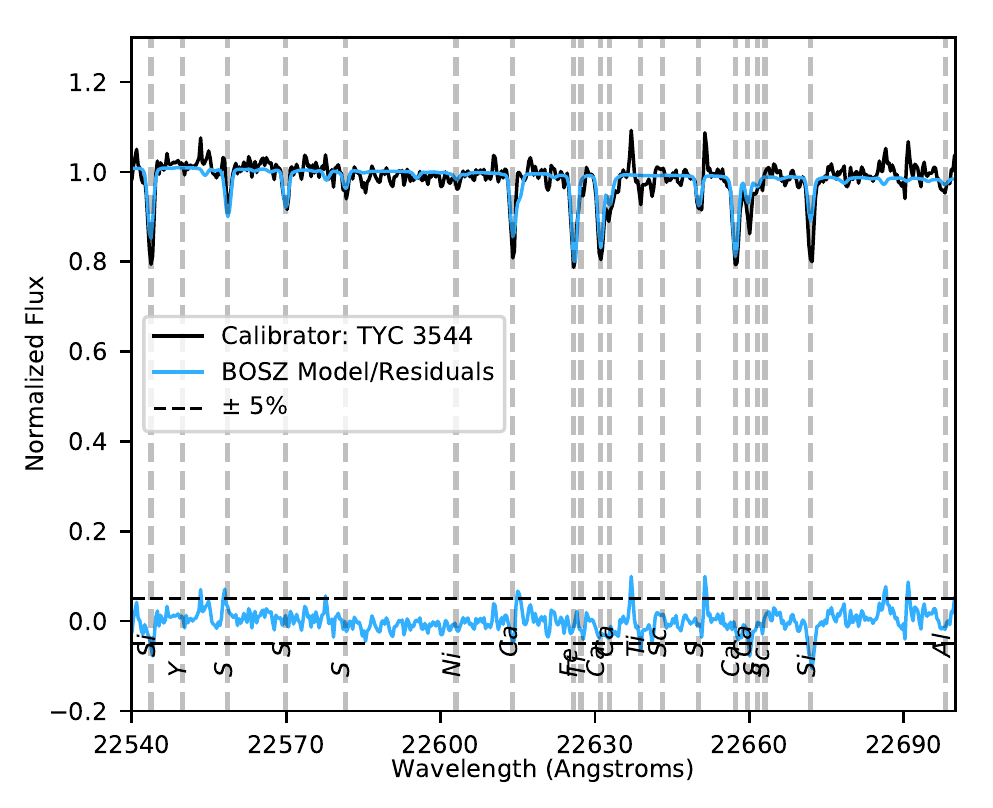}
\figsetgrpnote{Best-fit model (log $g$ fixed, BOSZ grid) compared to observed spectrum, with residuals.}
\figsetgrpend

\figsetgrpstart
\figsetgrpnum{18.26}
\figsetgrptitle{2M18482584+4828027 spectral order 35}
\figsetplot{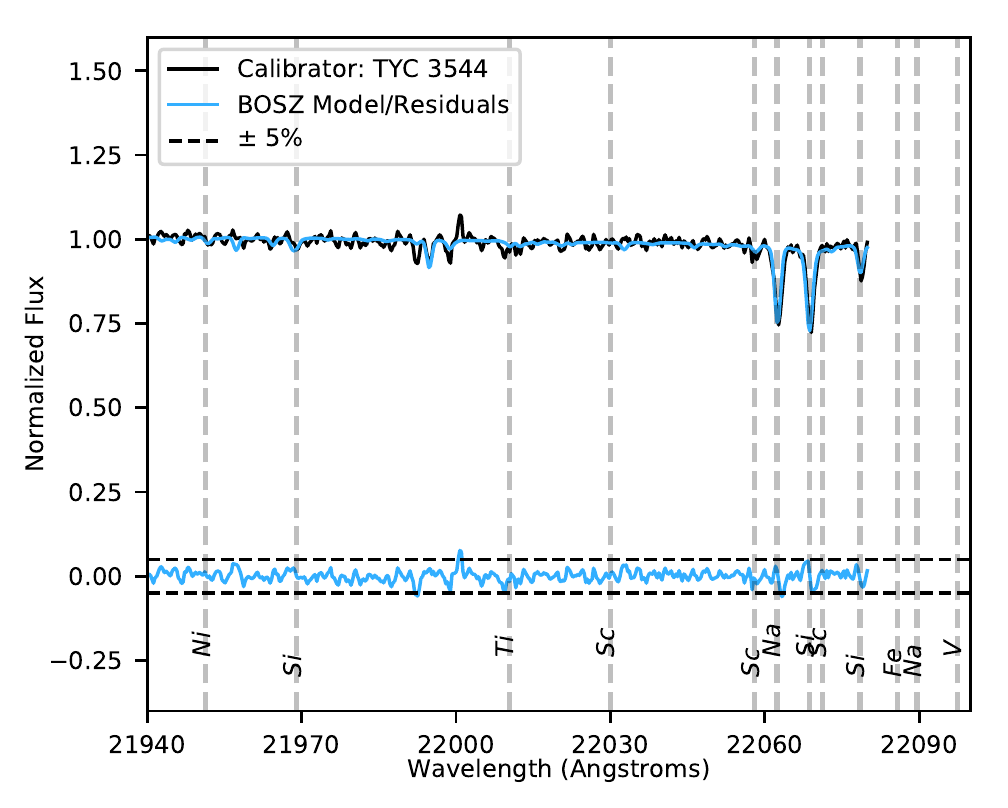}
\figsetgrpnote{Best-fit model (log $g$ fixed, BOSZ grid) compared to observed spectrum, with residuals.}
\figsetgrpend

\figsetgrpstart
\figsetgrpnum{18.27}
\figsetgrptitle{2M18482584+4828027 spectral order 36}
\figsetplot{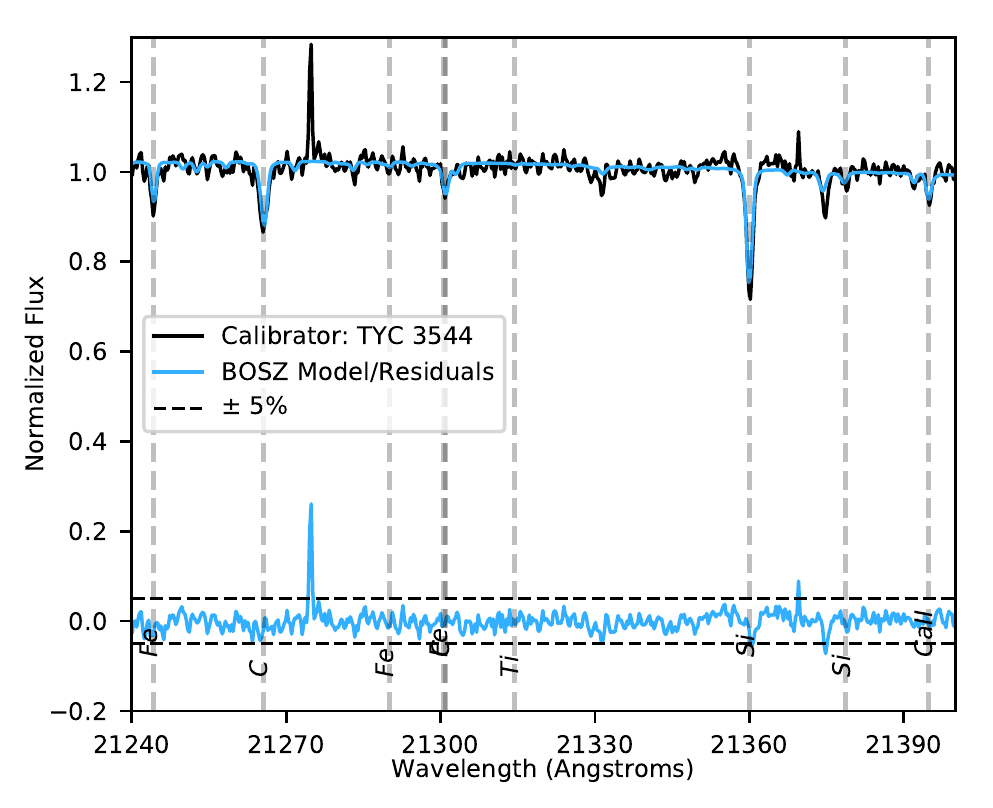}
\figsetgrpnote{Best-fit model (log $g$ fixed, BOSZ grid) compared to observed spectrum, with residuals.}
\figsetgrpend

\figsetgrpstart
\figsetgrpnum{18.28}
\figsetgrptitle{2M19534827+1848021 spectral order 34 (from Keck Observatory Archive)}
\figsetplot{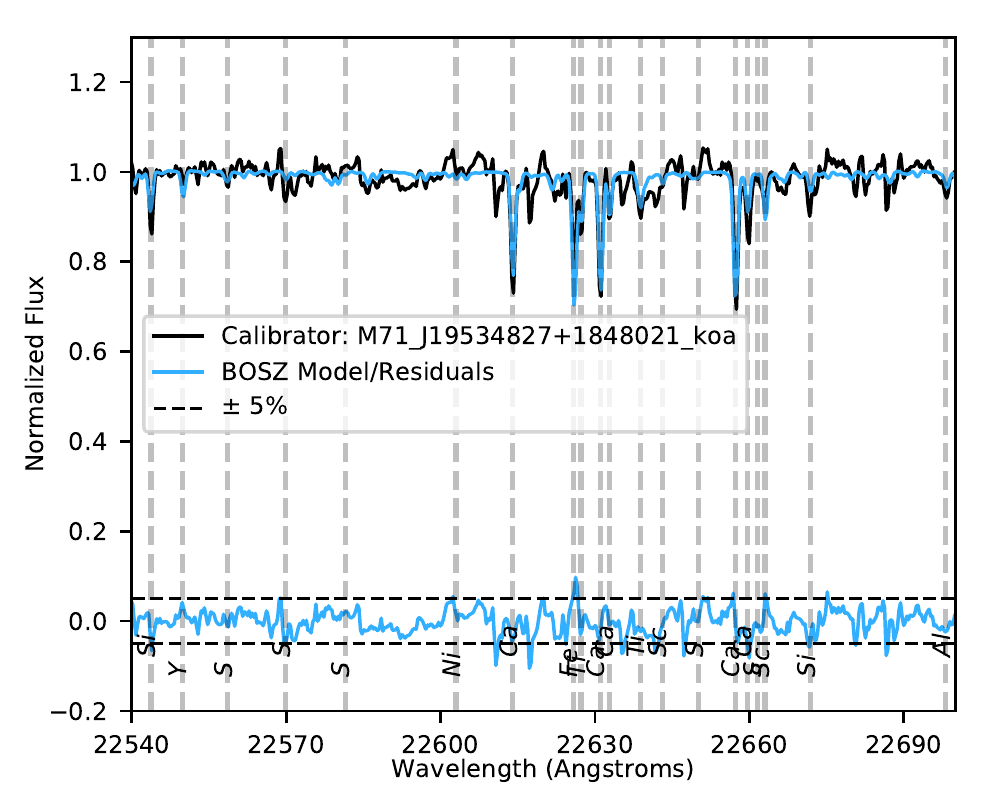}
\figsetgrpnote{Best-fit model (log $g$ fixed, BOSZ grid) compared to observed spectrum, with residuals.}
\figsetgrpend

\figsetgrpstart
\figsetgrpnum{18.29}
\figsetgrptitle{2M19534827+1848021 spectral order 35 (from Keck Observatory Archive)}
\figsetplot{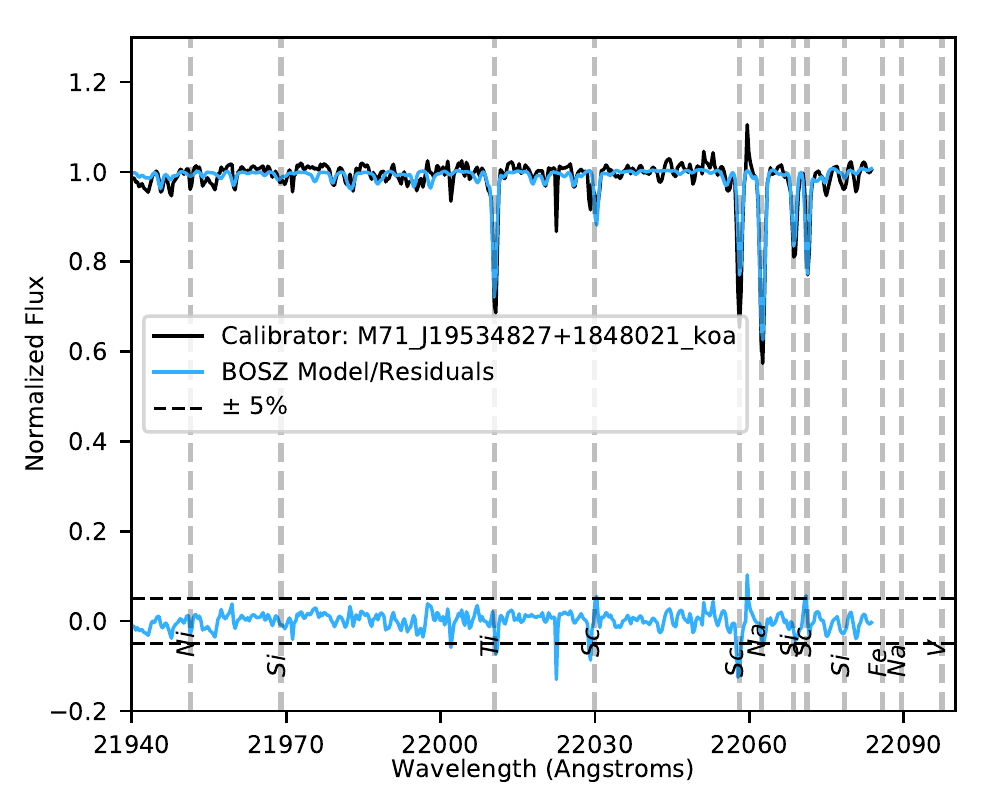}
\figsetgrpnote{Best-fit model (log $g$ fixed, BOSZ grid) compared to observed spectrum, with residuals.}
\figsetgrpend

\figsetgrpstart
\figsetgrpnum{18.30}
\figsetgrptitle{2M19534827+1848021 spectral order 36 (from Keck Observatory Archive)}
\figsetplot{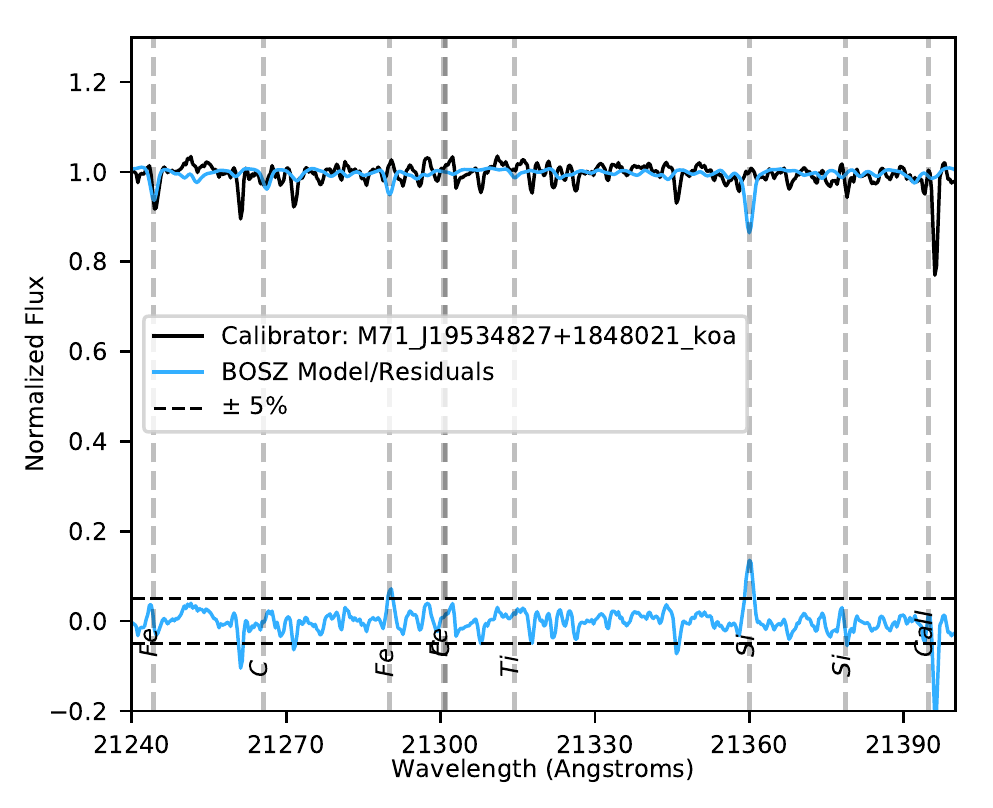}
\figsetgrpnote{Best-fit model (log $g$ fixed, BOSZ grid) compared to observed spectrum, with residuals.}
\figsetgrpend

\figsetgrpstart
\figsetgrpnum{18.31}
\figsetgrptitle{2M19534525+1846553 spectral order 34 (from Keck Observatory Archive)}
\figsetplot{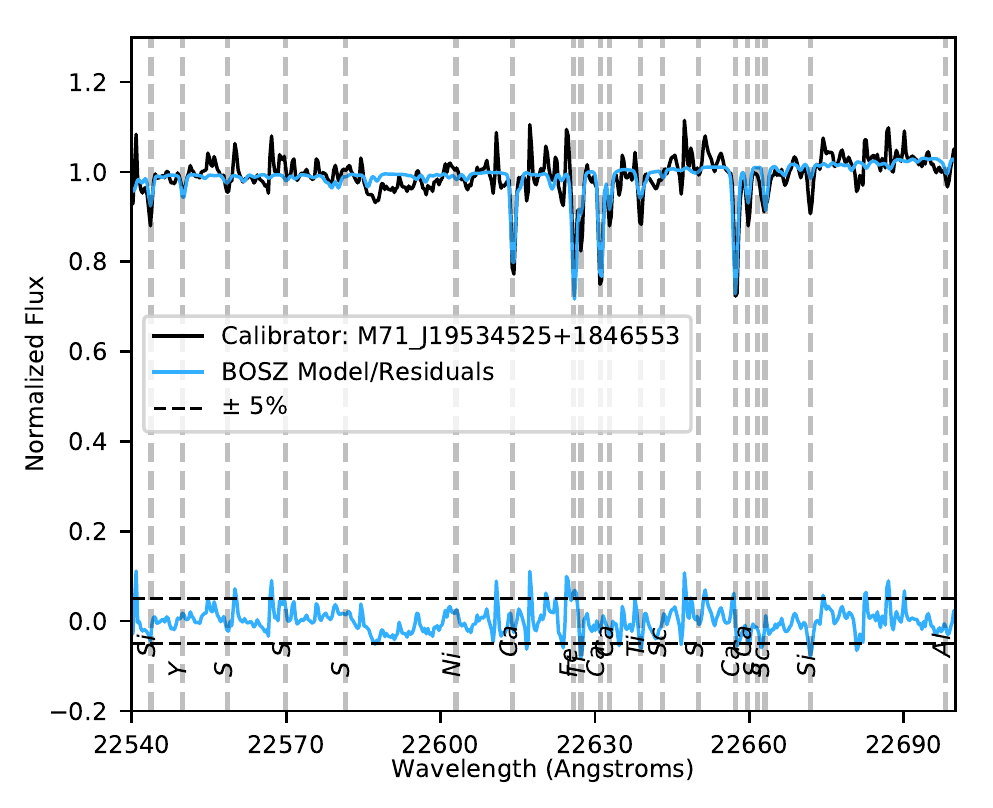}
\figsetgrpnote{Best-fit model (log $g$ fixed, BOSZ grid) compared to observed spectrum, with residuals.}
\figsetgrpend

\figsetgrpstart
\figsetgrpnum{18.32}
\figsetgrptitle{2M19534525+1846553 spectral order 35 (from Keck Observatory Archive)}
\figsetplot{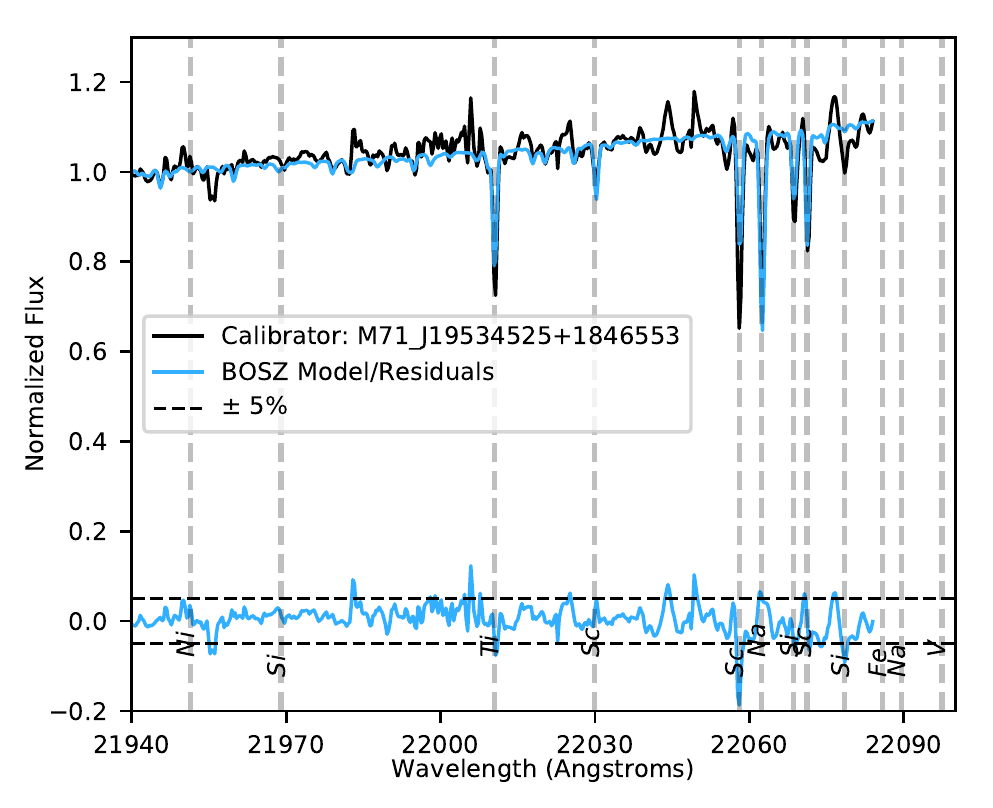}
\figsetgrpnote{Best-fit model (log $g$ fixed, BOSZ grid) compared to observed spectrum, with residuals.}
\figsetgrpend

\figsetgrpstart
\figsetgrpnum{18.33}
\figsetgrptitle{2M19534525+1846553 spectral order 36 (from Keck Observatory Archive)}
\figsetplot{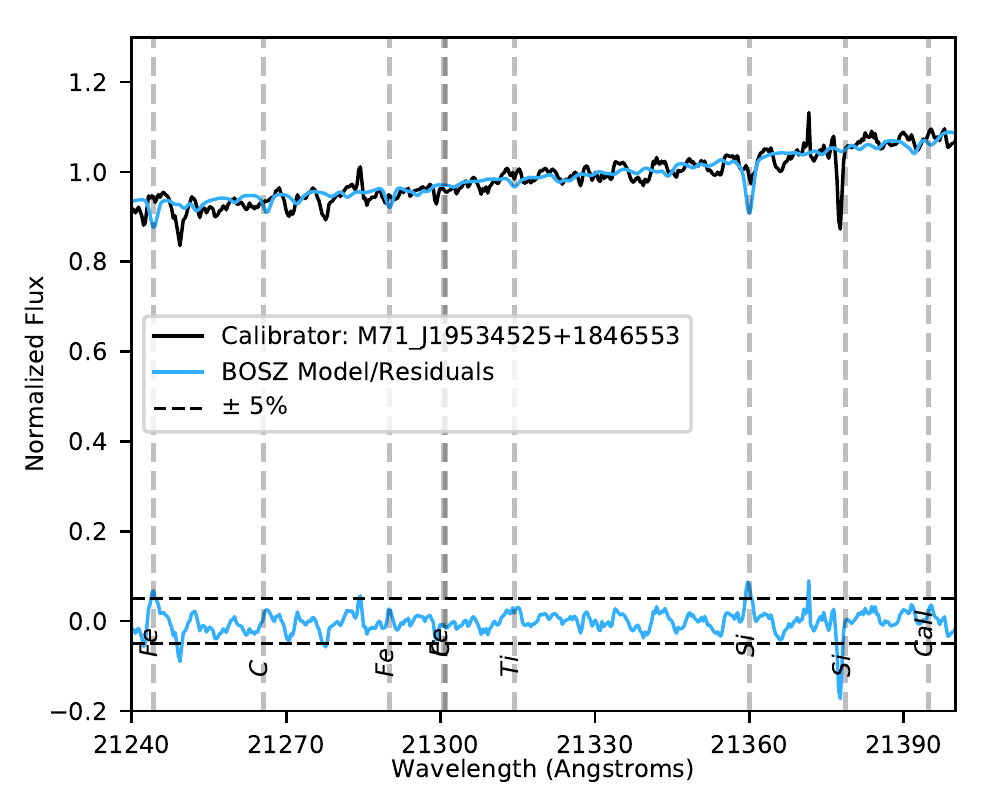}
\figsetgrpnote{Best-fit model (log $g$ fixed, BOSZ grid) compared to observed spectrum, with residuals.}
\figsetgrpend

\figsetgrpstart
\figsetgrpnum{18.34}
\figsetgrptitle{2M19535325+1846471 spectral order 34 (from Keck Observatory Archive)}
\figsetplot{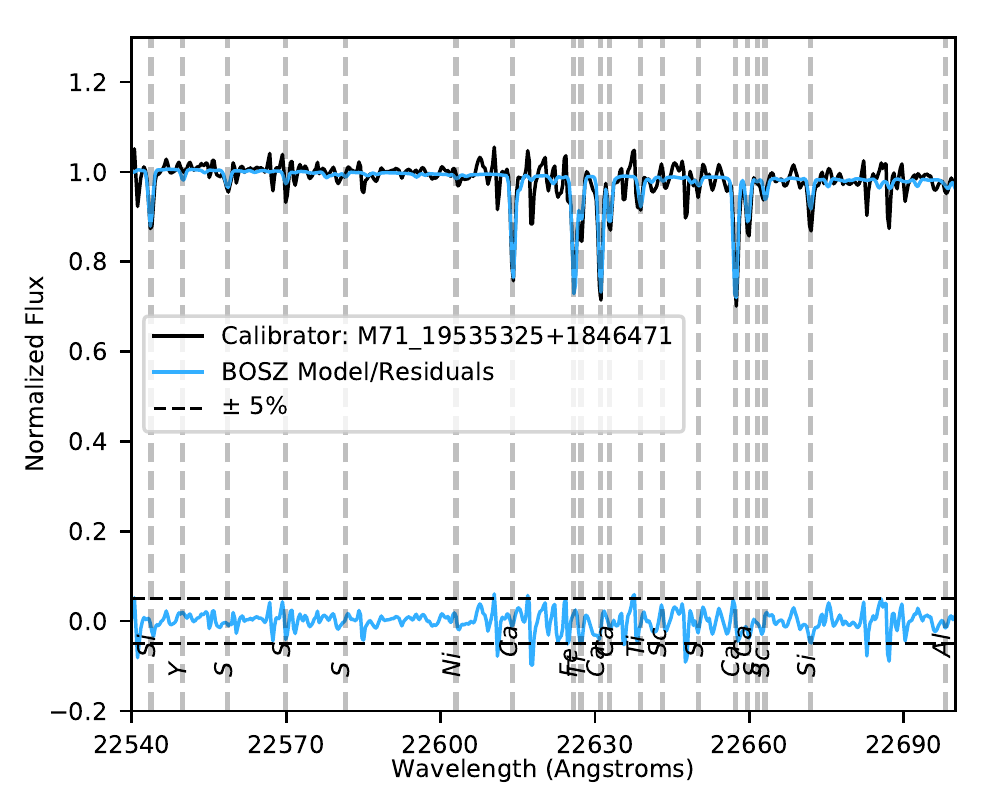}
\figsetgrpnote{Best-fit model (log $g$ fixed, BOSZ grid) compared to observed spectrum, with residuals.}
\figsetgrpend

\figsetgrpstart
\figsetgrpnum{18.35}
\figsetgrptitle{2M19535325+1846471 spectral order 35 (from Keck Observatory Archive)}
\figsetplot{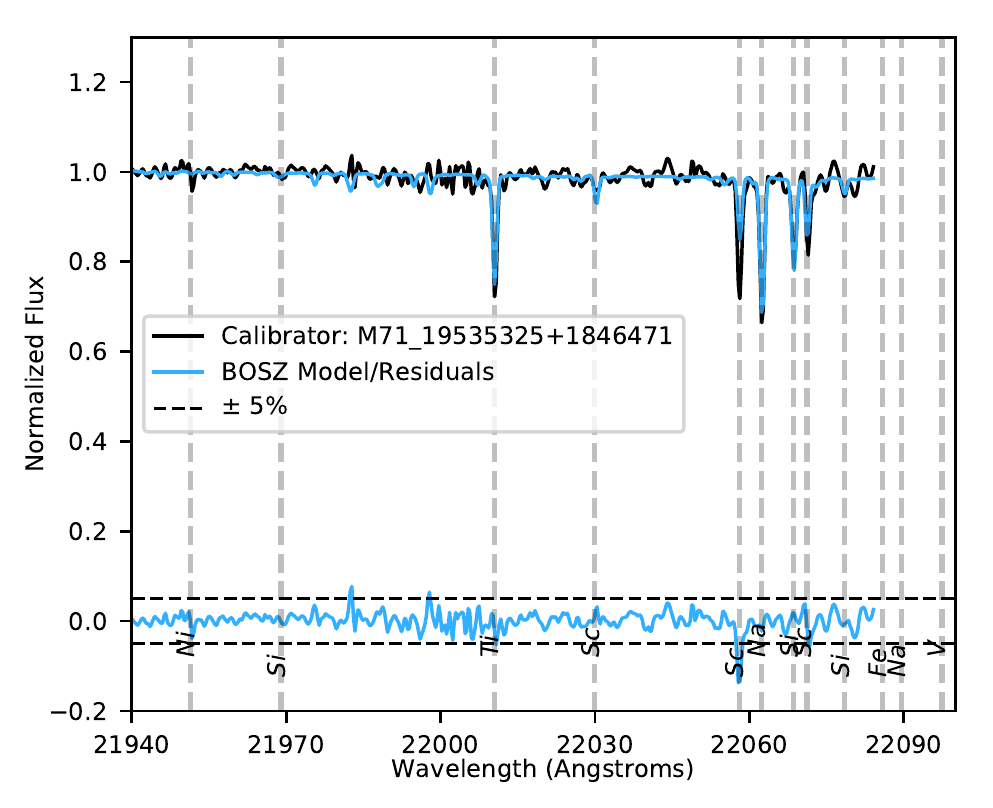}
\figsetgrpnote{Best-fit model (log $g$ fixed, BOSZ grid) compared to observed spectrum, with residuals.}
\figsetgrpend

\figsetgrpstart
\figsetgrpnum{18.36}
\figsetgrptitle{2M19535325+1846471 spectral order 36 (from Keck Observatory Archive)}
\figsetplot{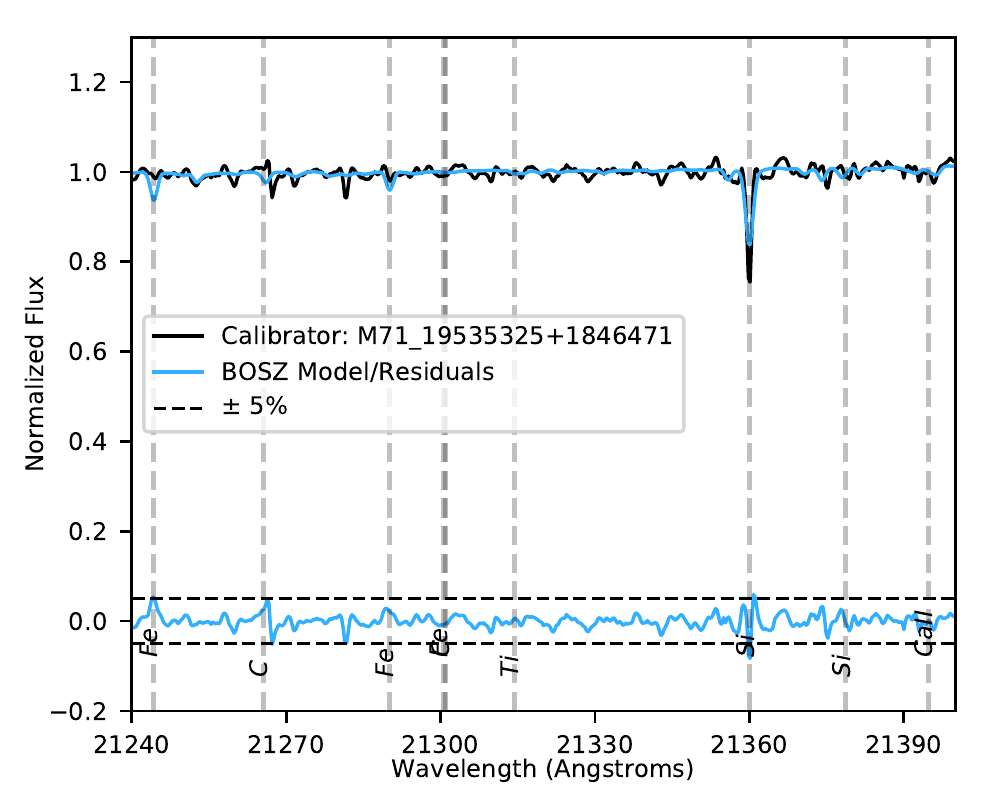}
\figsetgrpnote{Best-fit model (log $g$ fixed, BOSZ grid) compared to observed spectrum, with residuals.}
\figsetgrpend

\figsetgrpstart
\figsetgrpnum{18.37}
\figsetgrptitle{2M19533757+1847286 spectral order 34 (from Keck Observatory Archive)}
\figsetplot{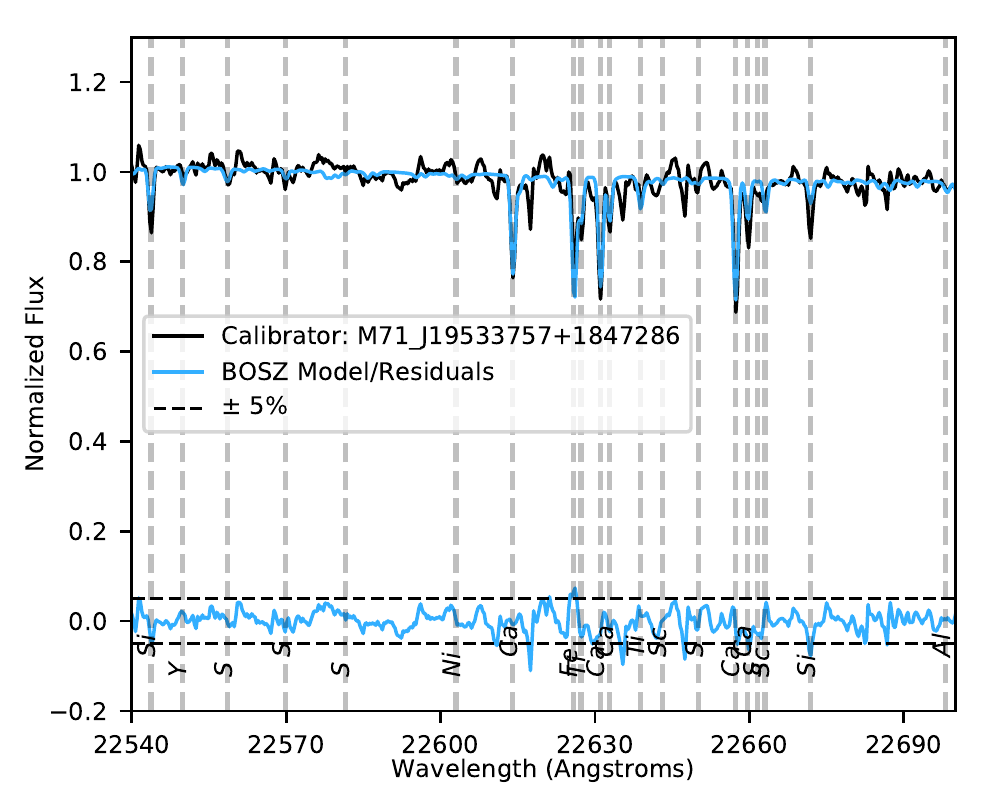}
\figsetgrpnote{Best-fit model (log $g$ fixed, BOSZ grid) compared to observed spectrum, with residuals.}
\figsetgrpend

\figsetgrpstart
\figsetgrpnum{18.38}
\figsetgrptitle{2M19533757+1847286 spectral order 35 (from Keck Observatory Archive)}
\figsetplot{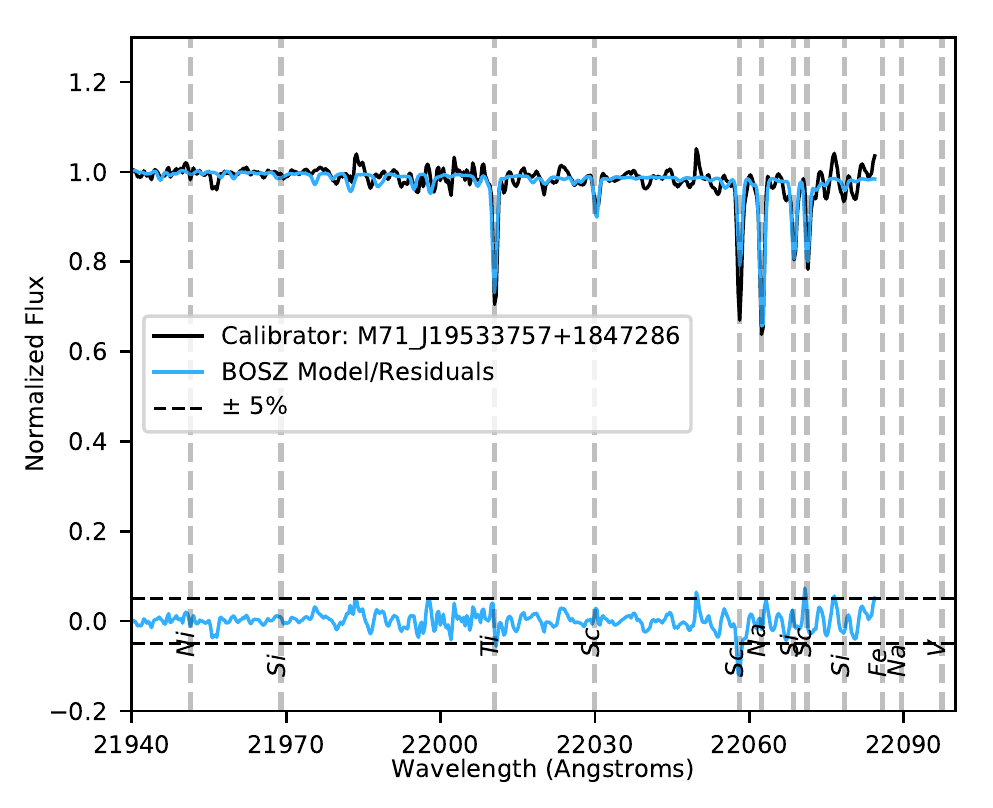}
\figsetgrpnote{Best-fit model (log $g$ fixed, BOSZ grid) compared to observed spectrum, with residuals.}
\figsetgrpend

\figsetgrpstart
\figsetgrpnum{18.39}
\figsetgrptitle{2M19533757+1847286 spectral order 36 (from Keck Observatory Archive)}
\figsetplot{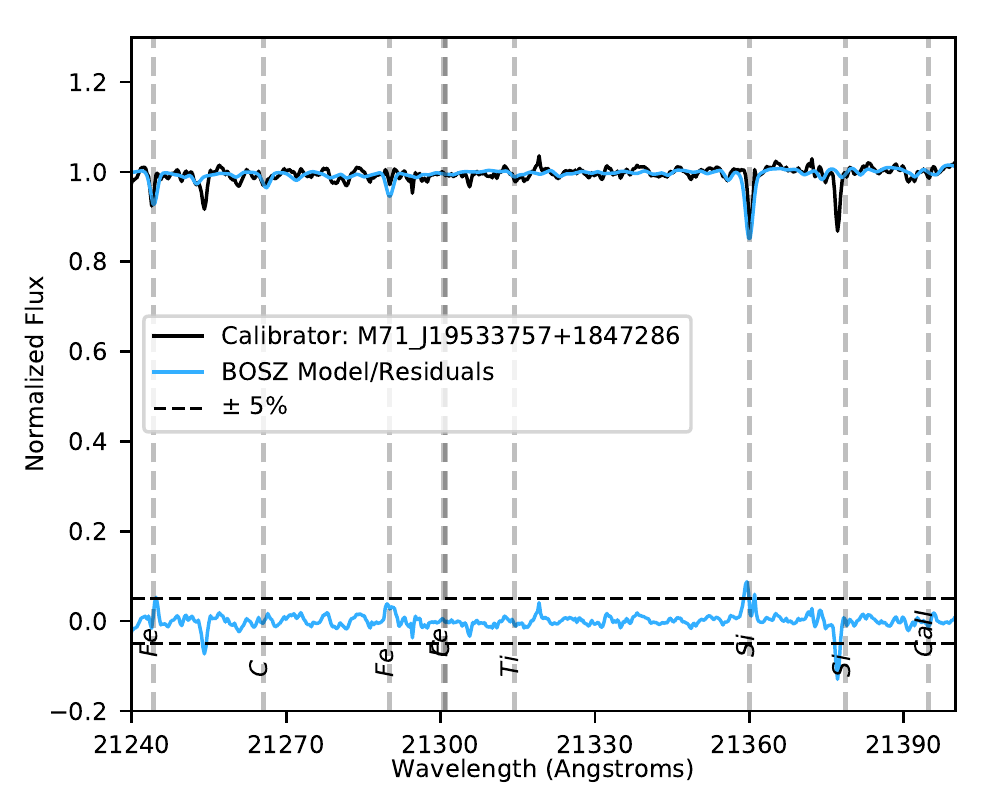}
\figsetgrpnote{Best-fit model (log $g$ fixed, BOSZ grid) compared to observed spectrum, with residuals.}
\figsetgrpend

\figsetend

\end{document}